\documentclass[a4paper,11pt]{article}
\usepackage{jheppub}

\usepackage{amsthm}

\theoremstyle{definition}

\usepackage[utf8]{inputenc}

\usepackage{amssymb,amsmath,amsbsy,mathtools}
\usepackage{braket}
\usepackage{graphicx}
\usepackage{xcolor}
\usepackage{braket}
\usepackage{multirow}
\usepackage{physics}
\usepackage{dcolumn}
\usepackage{hyperref}
\usepackage{subfig}
\usepackage[export]{adjustbox}
\usepackage{tikz-cd}

\usepackage{natbib}

\newcommand{\mM}{\mathcal{M}}

\newcommand{\mO}{\mathcal{O}}

\newcommand{\lb}{\left(}
\newcommand{\ep}{\epsilon}
\newcommand{\rb}{\right)}

\newcommand{\p}{\partial}

\newcommand{\nn}{\nonumber}
\newcommand{\ee}{\end{equation}}

\usepackage{graphicx}

\DeclareMathSymbol{\mlq}{\mathord}{operators}{``}
\DeclareMathSymbol{\mrq}{\mathord}{operators}{`'}

\usepackage{graphicx}
\usepackage{dcolumn}
\usepackage{bm}

\begin{document}

\preprint{APS/123-QED}

\title{A Lindbladian for exact renormalization of density operators in QFT}

\author{Samuel Goldman$^a$, Nima Lashkari$^b$, Robert G. Leigh$^a$}
\affiliation{$^a$ Department of Physics, University of Illinois,
1110 W. Green St., Urbana IL 61801-3080, U.S.A.}

\affiliation{$^b$ Department of Physics and Astronomy, Purdue University, West Lafayette, IN 47907, USA}

\abstract{In \cite{fliss2017unitary}, the authors extended the exact renormalization group (ERG) to arbitrary wave-functionals in quantum field theory (QFT). Applying this formalism, we show that the ERG flow of density matrices is given by a Lindblad master equation. The Lindbladian consists of a ``Hamiltonian" term which is the sum of a scaling and a coarse-graining (disentangling) operator, and a dissipative term with absorption and emission rates for each momentum mode. We consider as examples the flow of Gaussian states and the perturbative ground state of $\lambda \phi^4$ theory, and highlight the role of the dissipative terms in generating the correct flow of couplings. Integrating the Lindblad master equation, we find that a finite ERG flow of density matrices is described by a quantum channel. It follows from the data processing inequality that any distinguishability measure of states is an ERG monotone.}

\maketitle

\section{Introduction}

The exact renormalization group (ERG), also known as the functional renormalization group, is a powerful theoretical framework in statistical physics and quantum field theory (QFT) that uses path integral methods to implement the intuitive coarse-graining picture of Wilson and Kadanoff \cite{Kadanoff1966,Wilson1971I,Wilson1971II}. In a seminal work \cite{polchinski1984renormalization}, Polchinski developed the ERG formalism based on the central principle that the generating functional of QFT is independent of regularization. The ERG formalism has since found many applications in the study of strongly coupled QFTs.\footnote{The ERG formalism was further developed in \cite{wegner-1976, FJWegner_1974,Morris_1998, Latorre_2000}. See \cite{berges2002non, Rosten_2012} for a review and list of applications. For some more modern applications, see \cite{Cotler_2023, Berman2023BayesianR, Kline2023MultiRelevanceCB}.}

With the recent development of quantum information methods in many-body quantum physics, the ideas of Wilson and Kadanoff have inspired the development of novel numerical and analytic tools such as
entanglement renormalization and tensor networks \cite{vidal2007entanglement,orus2019tensor}. These tools have proven to be extremely effective in describing low energy wave-functions and finite temperature states of quantum spin systems with strongly interacting local Hamiltonians.
 Such methods have been extended to continuum Gaussian theories resulting in tensor networks such as the continuous multi-scale entanglement renormalization ansatz (cMERA) and continuous tensor network renormalization (cTNR) \cite{haegeman2013entanglement,hu2018continuous}. However, unlike the case of spin models, defining these continuous tensor networks away from the Gaussian limit has proven to be challenging.\footnote{See \cite{cotler2019entanglement,cotler2019renormalization} for attempts to generalize Gaussian cMERA in perturbation theory.} In spite of this, MERA and cMERA continue to provide a conceptual framework to explore the connection between entanglement and the emergence of geometry in holography \cite{swingle2012entanglement,nozaki2012holographic,milsted2018geometric}.

Motivated by applications of the ERG in holography \cite{heemskerk2011holographic,leigh2014holographic,leigh2015exact}, the authors of \cite{fliss2017unitary} extended the ERG formalism to arbitrary wave-functionals of QFT. They discovered that the ERG equations result in continuous  tensor networks, which as opposed to the cMERA and cTNR, systematically generalized to interacting models. In particular, by choosing regularization schemes that modify the dispersion relation, we recently showed the wave-functional ERG reproduces and generalizes cMERA tensor networks \cite{goldman2023exact}. 

In this work, we apply the ERG formalism developed in \cite{fliss2017unitary} to path integrals with two boundaries, and derive the ERG flow of density matrices in QFT. 
The theory is defined with an explicit cut-off scale $\Lambda$. The ERG is implemented by flowing to an effective scale $e^{-s}\Lambda$.
We find that the ERG flow of density matrices is described by a scale-dependent Lindblad master equation \cite{lindblad1976generators,gorini1976completely}:
\begin{equation}\label{densitymatrixERG}
    \frac{d}{ds}\rho(s) =\mathcal{L}_s(\rho(s)) = i[\hat{L}_s+\hat{K}, \rho(s)] + \mathcal{D}(\rho(s)).
\end{equation}
The Lindbladian above implies that the ERG is, in general, a non-unitary and irreversible process that erases the information of correlations at large momenta.\footnote{The erasure of the large momentum correlations in the ERG can be made precise using the language of quantum error correction codes \cite{Furuya:2020tzv,Furuya:2021lgx}. We come back to this in Section \ref{sec:monotones}.} 
 The ERG is expected to be irreversible, as is demonstrated by $C$-theorems in two, three and four spacetime dimensions \cite{Zamolodchikov:1986gt,casini2012renormalization,komargodski2011renormalization}. However, the non-unitarity of RG is novel.
We trace back the origin of this non-unitarity to the non-locality of the effective actions that ERG produces. Here, we point out some of the key properties of our ERG equation in (\ref{densitymatrixERG}):
\begin{itemize}
    \item {\bf The Hamiltonian term:}
As was previously observed in \cite{fliss2017unitary,goldman2023exact}, the ``Hamiltonian" piece is comprised of a scaling generator $\hat{L}_s$ (see \ref{rescaclingD}), and a squeezing generator $\hat{K}$ (see \ref{Kpart}).
In analogy with cMERA tensor networks, we refer to $\hat{K}$ as a ``disentangler". All terms in the Lindbladian are independent of $s$, except for the scaling operator $\hat{L}_s$ which we take to depend on $s$ to allow for a scale-dependent anomalous dimension.

\item {\bf The diffusion term:} The novel term is a diffusive piece $\mathcal{D}$ which for the ERG of the thermal state takes the form (see \ref{thermal-diffusion}):
\begin{align}
\begin{split}
 \mathcal{D}_\beta(\rho)&= \int_{\mathbf{p}}\frac{\Delta_\Sigma(\mathbf{p})}{\sinh(\beta\omega(\mathbf{p})/2)} \Bigg[e^{\beta \omega(\mathbf{p})/2}(a(\mathbf{p}) \rho a^\dagger(\mathbf{p})  - \frac{1}{2}\{a(\mathbf{p})^\dagger a(\mathbf{p}), \rho\})\\&\hspace{4cm}+e^{-\beta \omega(\mathbf{p})/2}(a(\mathbf{p})^\dagger \rho a(\mathbf{p}) - \frac{1}{2}\{a(\mathbf{p}) a(\mathbf{p})^\dagger, \rho\})\Bigg]
\end{split}
\end{align}
where $\Delta_\Sigma(\mathbf{p})$ defined in (\ref{deltasigma}) depends only on the ERG regulator.
The diffusive term above belongs to a class of open quantum dynamics that are commonly studied in quantum optics. Using the terminology of quantum optics, we call the  $a\rho a^\dagger$ and $a^\dagger \rho a$ the ``emission" and "absorption" terms, respectively. Their corresponding rates $e^{\beta \omega/2}$ and $e^{-\beta \omega/2}$ match the expectation from quantum detailed balance \cite{alicki2007quantum, carlen2017gradientflowentropyinequalities}. In the zero temperature limit, the absorption coefficient vanishes and the generator becomes (\ref{Drho})
\begin{equation}
    \mathcal{D}(\rho) = \int_\mathbf{p} \Delta_\Sigma (\mathbf{p}) \left[a(\mathbf{p})\rho a^\dagger(\mathbf{p}) - \frac{1}{2}\{\rho, a^\dagger(\mathbf{p})a(\mathbf{p})\}\right].
\end{equation}

\item {\bf A consistency condition:} There are two distinct ways to describe the ERG flow of a density matrix from initial UV scale $e^{-s_0}\Lambda$ to some lower IR scale $e^{-s}\Lambda$: 
\begin{description}
\item[\textit{Method 1:}] First, use the Polchinski equation to find an IR action $S(e^{-s}\Lambda)$, and then evaluate the path integral on a manifold with boundary to obtain the state $\rho(e^{-s}\Lambda)$.
\item[\textit{Method 2:}] First, ``slice'' the path integral to obtain the UV density matrix $\rho(e^{-s_0}\Lambda)$, and then use our ERG flow equation in (\ref{densitymatrixERG}) to obtain the IR state $\rho(e^{-s}\Lambda)$.
\end{description}
We show that the ERG flow equation in (\ref{densitymatrixERG}) is consistent in the sense that both methods give the same result for $\rho(e^{-s}\Lambda)$; see Figure \ref{fig:commutative-diagram}. In Section \ref{sec:gaussian-flows}, we show in a perturbative example that when using Method 2, the non-unitary terms of (\ref{densitymatrixERG}) are essential to capture the correct running of couplings from Method 1.
\end{itemize}

 \begin{figure}
 \[\begin{tikzcd}[row sep=1.5cm, column sep = 3cm]
 S[e^{-s_0}\Lambda]\arrow{r}{\text{Polchinski eq.}} \arrow[swap]{d}{\text{State preparation}} & S[e^{-s}\Lambda] \arrow{d}{\text{\textcolor{white}{State preparation}}} \\%
 \rho[e^{-s_0}\Lambda] \arrow{r}{\text{RG master eq.}}& \rho[e^{-s}\Lambda]
 \end{tikzcd}\]
 \caption{The RG master equation we derive is consistent with the Polchinski equation in the sense that the above diagram commutes.}
 \label{fig:commutative-diagram} 
 \end{figure}

As we discuss in Section \ref{sec:integrate}, an immediate application of our result is that integrating our Lindblad master equation results in a quantum channel. The data processing inequality then implies that any distinguishability measure of states  is non-increasing under the ERG. In other words, we naturally obtain a large class of ERG monotones.

In Section \ref{sec:prelim}, we review the ERG formalism. The derivation of our ERG Lindblad equation is in Section \ref{sec:ERGdensitymatrix}. In Section \ref{sec:gaussian-flows}, we work out examples of density matrix ERG in QFT for Gaussian states, and perturbatively for $\lambda \phi^4$. We comment on the role of the diffusive terms in reproducing the correct flow of the couplings. In Section \ref{sec:integrate} using methods from quantum optics, we integrate the ERG flow in the Heisenberg picture. In Section \ref{thermalERG}, we generalize our ERG flow to thermal density matrices. Finally, in Section \ref{sec:monotones}, we comment on connections between the ERG and approximate quantum error correction, in line with the recent work of \cite{Furuya:2020tzv,Furuya:2021lgx}. We also work out two examples of RG monotones, namely quantum fidelity of Gaussian vacuum states and the quantum relative entropy of thermal states.

\section{Review of the ERG formalism}\label{sec:prelim}

\subsection{Polchinski equation: ERG for effective actions}

We begin by reviewing Polchinski's formalism for the exact renormalization group (ERG) and the derivation of the Polchinski equation, which describes the ERG flow of the Wilsonian effective action \cite{polchinski1984renormalization}.\footnote{Polchinski's result is a special case of a more general formulation of RG known as the Wegner-Morris equation that, for completeness, we review in Appendix \ref{app:wegner}.}
To simplify the presentation, we use a modified DeWitt notation for spacetime integrals: if $K(x, y)$ is a bi-local kernel with $x,y\in \mathbb{R}^d$, then we write
\begin{equation}
    \left(K* \phi\right)(x) \equiv \int d^dy \: K(x, y)\phi(y).
\end{equation}
Since we will be considering path integrals with boundaries, we also write for integrals over a constant $t$ slice:
\begin{equation}
    \left(K\cdot \phi\right)(\mathbf{x}) \equiv \int d^{d-1}\mathbf{y} \: K(\mathbf{x}, \mathbf{y})\phi(\mathbf{y}),
\end{equation}
where $\mathbf{x}, \mathbf{y}\in \mathbb{R}^{d-1}$ are spatial coordinates on the constant $t = 0$ surface. 

\begin{figure}
    \centering
    \includegraphics[width = .6 \linewidth]{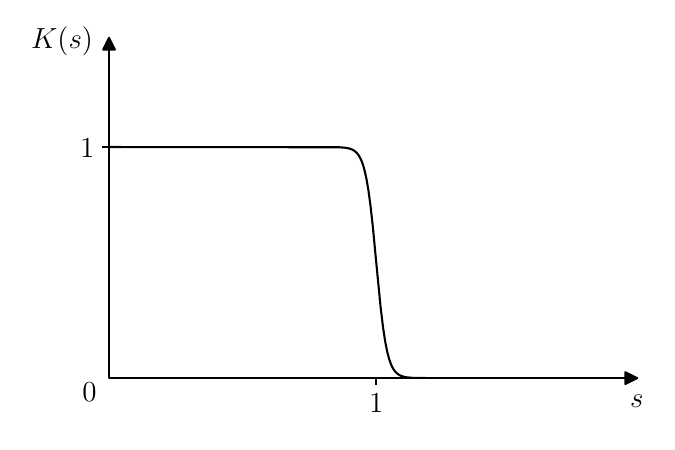}
    \caption{\small{An example of the smooth regulating function $K(s)$. When $\abs{p}\sim \Lambda$, the function $K(p^2/\Lambda^2)$ drops to zero quickly.}}
    \label{regulator-plot}
\end{figure}

The starting point of ERG is a general Euclidean action of the form 
\begin{equation}\label{actionterms}
    S_E = S_0 + S_I
\end{equation}
where $S_0$ is a quadratic (free) action, and $S_I$ is an arbitrary functional of the fields.\footnote{Note that, in what follows, we do not require $S_I$ to be small relative to $S_0$.} One usually regards $S_I$ as ``interaction terms,'' but here we will also view them as sources in the Euclidean path integral for preparing excited states. The Polchinski ERG is implemented by choosing $S_0$ to be a regulated free action with the following form,
\begin{eqnarray}\label{s0-def}
    S_0 = -\frac{1}{2} \phi * K^{-1}(-\partial^2/\Lambda^2) * \partial^2 \phi,\label{regaction}
\end{eqnarray}
where $\Lambda$ is a UV cutoff scale. The function $K(p^2/\Lambda^2)$ is a smooth regulator that approximates an indicator function on momenta below the cutoff (see Figure \ref{regulator-plot}). 
One then constructs the Euclidean path integral
\begin{equation}\label{Epathint}
    Z_E = \int \mathcal{D}\phi e^{-S_0 - S_I}\ .
\end{equation}
For momentum modes with support above the cutoff, $K^{-1}(p^2/\Lambda^2)$ in (\ref{regaction}) becomes large, suppressing the contribution of such configurations to $Z_E$. In perturbation theory, the function $K$ regulates UV divergences by modifying the propagator to
\begin{equation}
    G(p^2; \Lambda) = \frac{K(p^2/\Lambda^2)}{p^2}
\end{equation}
which smoothly cuts off the contribution of large momenta to loop integrals.

The RG condition is the invariance of the partition function under a variation of the momentum cutoff $\Lambda$:
\begin{equation}\label{rg-condition}
    -\Lambda\frac{d}{d\Lambda}Z_E = \int \mathcal{D}\phi \left(\Lambda\frac{d}{d\Lambda}S_0 + \Lambda\frac{d}{d\Lambda}S_I\right)e^{-S_E} =0\ .
\end{equation}
Writing $S_0$ in the form
\begin{equation}
    S_0 = \frac{1}{2} \phi * G^{-1} * \phi,
\end{equation}
one finds
\begin{align}\begin{split}\label{mdmS_0}
    \Lambda\frac{d}{d\Lambda}S_0 &= \frac{1}{2}\phi * \left(\Lambda\frac{d}{d\Lambda}G^{-1}\right)* \phi \\&= -\frac{1}{2}\phi * G^{-1} * \left(\Lambda\frac{d}{d\Lambda}G\right) * G^{-1}* \phi\\
    &= -\frac{1}{2} \frac{\delta S_0}{\delta \phi}* \left(\Lambda\frac{d}{d\Lambda} G\right) * \frac{\delta S_0}{\delta \phi}\end{split}
\end{align}
It is convenient to define a bilocal kernel
\begin{equation}\label{deltadef}
    G' \equiv \Lambda\frac{d}{d\Lambda}G
\end{equation}
which captures the logarithmic scale derivative of the regulated propagator. We emphasize that $G'$ is non-local in spacetime, and this will play a crucial role in our derivation of ERG flow for density matrices in Section \ref{sec:ERGdensitymatrix}.

Recall that for an arbitrary operator insertion $\mathcal{O}(x)$, the Schwinger-Dyson equation states that we can make the following replacement in the path integral,
\begin{equation}\label{schwinger-dyson}
    \mathcal{O}(x)\frac{\delta S_0}{\delta \phi(y)} = \frac{\delta \mathcal{O}(x)}{\delta\phi(y)} - \mathcal{O}(x)\frac{\delta S_I}{\delta \phi(y)}.
\end{equation}
Applying this twice to the expression in (\ref{mdmS_0}), we obtain
\begin{align}\label{schwingertwice}
    \begin{split}
        -\frac{1}{2} \frac{\delta S_0}{\delta \phi}*G'  * \frac{\delta S_0}{\delta \phi} &= -\frac{1}{2}\Tr\left(\frac{\delta^2 S_0}{\delta \phi\delta \phi}*G' \right) + \frac{1}{2}\frac{\delta S_0}{\delta \phi} * G' * \frac{\delta S_I}{\delta \phi}\\
        &= -\frac{1}{2}\Tr\left(\frac{\delta^2 S_0}{\delta \phi\delta \phi}*G' \right) + \frac{1}{2}\Tr(\frac{\delta^2 S_I}{\delta \phi \delta \phi}* G' ) - \frac{1}{2}\frac{\delta S_I}{\delta \phi} * G' * \frac{\delta S_I}{\delta \phi}
    \end{split}
\end{align}
Comparing with (\ref{rg-condition}), the RG condition is satisfied if $S_I$ obeys the {\it Polchinski equation}
\begin{equation}\label{polchinski-eq}
    \Lambda\frac{d}{d\Lambda}S_I = \frac{1}{2}\frac{\delta S_I}{\delta \phi}* G'* \frac{\delta S_I}{\delta \phi} - \frac{1}{2}\Tr\left(\frac{\delta^2 S_I}{\delta \phi \delta \phi}* G'\right)\ +\frac{1}{2}\Tr\left(\frac{\delta^2 S_0}{\delta \phi\delta \phi}*G' \right)
\end{equation}
where in our notation
\begin{eqnarray}
    \Tr(A)\equiv \int d^dx \: A(x,x)\ .
\end{eqnarray}
Notice that the final term in the Polchinski equation is field-independent, and thus has no bearing on local physics. It can be treated by paying attention to the normalization of the path integral: that is, this term can be dropped at the cost of scaling the partition function by an overall constant, so that the Polchinski equation is written solely in terms of $S_I$. 

An important feature of the Polchinski equation is that even if $S_I$ starts as a local polynomial of fields, the first term in (\ref{polchinski-eq}) generates non-local interactions, due to the non-locality of the kernel $G'$. In perturbative renormalization, these non-local terms are often attributed to irrelevant operators and neglected. However, non-locality is indispensable in the ERG formalism, 
and as we will see, these non-local terms are the origin of the non-unitarity of the ERG flow for density matrices.

\subsection{Path Integrals for Density Matrices}

\begin{figure}
    \centering
    \includegraphics[width=.9\linewidth]{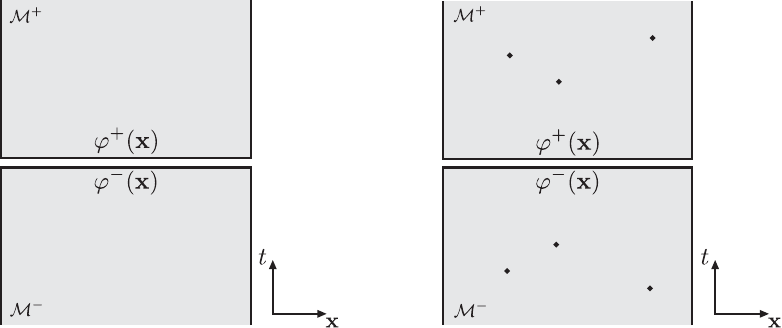}
    \caption{Schematic diagram of the path integrals considered for state preparation. There are two boundaries at $t=\pm \ep$. (Left) The path-integral that prepares the vacuum density matrix (Right) The Euclidean time-reflection symmetric insertions of operators in the path-integral prepare excited density matrices.}
    \label{fig:zero-temp-path-int}
\end{figure}

Here, we review the construction of density matrices using the path integral formalism. Consider a Euclidean spacetime $\mathcal{M}$ with a distinguished time coordinate $t \in (-\infty, \infty)$. In this section, we take the background to be $d$-dimensional Euclidean space, $\mathcal{M} = \mathbb{R}^d$; however, our discussion generalizes to more general background spacetimes.\footnote{In Section \ref{thermalERG}, we will discuss the case of a thermal cylinder, i.e., $\mM=\mathbb{R}^{d-1}\times S^1_\beta$.} We denote by  $\mathcal{M}^+$ ($\mathcal{M}^-$) the upper (lower) half-spaces $t\geq \epsilon$ $(t\leq -\epsilon)$ and their boundaries at $t=\ep$ ($t=-\ep)$ by $\Sigma^+$ ($\Sigma^-$), respectively. We denote coordinates on the constant time-slices $\Sigma^\pm$ using bold-face vectors $\mathbf{x}$. See Figure \ref{fig:zero-temp-path-int}. 

Let $S_E[\phi, \mathcal{M}]$ be the Euclidean action of a scalar field theory on $\mathcal{M}$, and let $\mathcal{F}[\phi]$ be some functional of the fields representing insertions in Euclidean time representing an excited state or a background source.\footnote{We limit our presentation to scalar field theories, but the general properties of our construction naturally carry over to fields with non-zero spin.} A density matrix in the basis of spatial field configurations $\varphi^\pm(\mathbf{x})$ on $\Sigma^\pm$ can be represented as the path-integral
\begin{equation}\label{densitymatpathint}
    \mel{\varphi^-}{\rho}{\varphi^+} = \frac{1}{Z}\int_{\phi(0^-, \mathbf{x}) = \varphi^-(\mathbf{x})}^{\phi(0^+, \mathbf{x} )= \varphi^+(\mathbf{x})} \mathcal{D}\phi\: \mathcal{F}[\phi] e^{-S_E[\phi, \mathcal{M}]},
\end{equation}
where the normalization $Z$ is given by
\begin{equation}
    Z = \int \mathcal{D}\phi\, \mathcal{F}[\phi] e^{-S_E[\phi, \mathcal{M}]}
\end{equation}
ensuring that $\tr(\rho) = 1$. Density matrices are positive semi-definite, i.e. $\rho=Q^\dagger Q$ for some operator $Q$, which implies that the Euclidean path-integral representing a density matrix must be reflection symmetric. Therefore, we require that both the action and the measure $\mathcal{D}\phi\,\mathcal{F}[\phi]$ be reflection symmetric.

We decompose the field $\phi(t, \mathbf{x})$ into its components on the two regions, which we denote as $\phi^+(t, \mathbf{x}) = \phi \vert_{\mathcal{M}^+}$ and $\phi^-(t, \mathbf{x}) = \phi \vert_{\mathcal{M}^-}$ so that
\begin{equation}
    \phi(t, \mathbf{x}) = \phi^-(t, \mathbf{x}) \Theta(-t) +  \phi^+(t, \mathbf{x}) \Theta(t)
\end{equation}
where $\Theta(t)$ is the Heaviside step function. In terms of these components, the path integral measure factors as
\begin{equation}
    \mathcal{D}\phi = \mathcal{D}\phi^+ \mathcal{D}\phi^-.
\end{equation}
In the special case where the action (including potential boundary terms) and the insertions $\mathcal{F}$ in (\ref{densitymatpathint}) factorize as
\begin{spreadlines}{6pt}
\begin{equation}
\begin{gathered}\label{factorization}
     S_E[\phi, \mathcal{M}] = S_+[\phi^+, \mathcal{M}^+] + S_-[\phi^-, \mathcal{M}^-],\\
     \mathcal{F}[\phi^+, \phi^-] = \mathcal{G}[\phi^+]\mathcal{G}[\phi^-],
\end{gathered}
\end{equation}
\end{spreadlines}
for some functional $\mathcal{G}$, the path-integral in (\ref{densitymatpathint}) represents a pure state:
\begin{equation}
\braket{\varphi^-}{\rho|\varphi^+}=\braket{\varphi^-}{\Psi}\braket{\Psi}{\varphi^+}.
\end{equation}
    The wavefunctional is defined by
 \begin{equation}
     \braket{\varphi^-}{\Psi} = \frac{1}{\sqrt{Z}}\int^{\phi^-(0, \mathbf{x}) = \varphi^-(\mathbf{x})} \mathcal{D}\phi^- \mathcal{G}[\phi^-]e^{-S_-[\phi^-, \mathcal{M}^-]}.
 \end{equation}
In general, due to the non-locality of the effective action in ERG, the factorization conditions (\ref{factorization}) fail and a pure density matrix is not guaranteed. One way to interpret this failure of factorization is via interactions that couple $\phi^+$ and $\phi^-$ in the path integral. To build intuition on how to use path integrals to prepare mixed states,  we work out two explicit examples of mixed states in quantum mechanics prepared using $0+1$-dimensional path integrals in Appendix \ref{app:mixed-states}.

\section{ERG flow of density matrices}\label{sec:ERGdensitymatrix}

In this section, we establish our central finding: we derive an equation for the ERG flow of density matrices and demonstrate that it is generated by a Lindbladian. The flow equation we derive is consistent with the Polchinski ERG in the sense of Figure \ref{fig:commutative-diagram}.

Our set up is a continuation of the line of work initiated in \cite{fliss2017unitary} that studies ERG flows for path integrals on manifolds with boundary. To have a well-defined path integral on a manifold with boundary, we need to ensure that the boundary value problem (BVP) associated with the action is well-posed. In Polchinski's scheme, the regulating function $K$ depends on the full Euclidean momentum $p^2$; in position space this amounts to adding time derivatives of arbitrarily high order into the action, resulting in an ill-posed BVP. It was shown in \cite{fliss2017unitary} that the replacement
\begin{equation}\label{spatial-regulator}
    K(p^2/\Lambda^2) \longrightarrow K(\mathbf{p}^2/\Lambda^2)
\end{equation}
with $\mathbf{p}$ the \textit{spatial} momentum is a sufficient modification to make the ERG well-defined on states. We refer the reader to \cite{fliss2017unitary} for details of the proof of this statement. In the following discussion, we will use (\ref{spatial-regulator}) to regulate the action.

There are two steps to ERG: lowering the cut-off and rescaling. We discuss these steps separately, and closely follow the manipulations of the path-integral in \cite{fliss2017unitary}.

In the first step of the ERG we lower the cut-off $\Lambda$. Consider a state $\rho$ prepared by the path integral
\begin{equation} \label{density-operator-def}
    \mel{\varphi^-}{\rho}{\varphi^+} = \frac{1}{Z}\int_{\phi(0^-, \mathbf{x}) = \varphi^-(\mathbf{x})}^{\phi(0^+, \mathbf{x}) = \varphi^+(\mathbf{x})} \mathcal{D}\phi \exp(-S_0 - S_I - S_{\textrm{bdry}})
\end{equation}
where $S_I$ is a general functional of the field $\phi$. The regulated action $S_0$ with a \textit{spatial} regulator is
\begin{equation}
    S_0 = -\frac{1}{2} \phi * K^{-1}(-\nabla^2/\Lambda^2) * \partial^2 \phi,
\end{equation}
and $S_{\textrm{bdry}}$ is an additional boundary term that ensures a consistent variational principle:
\begin{align}\begin{split}
    S_{\textrm{bdry}}&= \frac{1}{2} \varphi^-\cdot K^{-1}(-\nabla^2/\Lambda^2)\cdot \partial_t \phi^- - \frac{1}{2} \varphi^+\cdot K^{-1}(-\nabla^2/\Lambda^2)\cdot \partial_t \phi^+ \\
    &= \frac{1}{2} \varphi^-\cdot D_t\phi^- - \frac{1}{2} \varphi^+\cdot D_t \phi^+ .\end{split}
\end{align}
In the second line we have defined a non-local regulated derivative operator 
\begin{equation}
D_t \equiv K^{-1}(-\nabla^2/\Lambda^2)\partial_t.
\end{equation}
The RG condition is implemented by requiring the normalization of the state $Z$ is unchanged along the flow:
\begin{equation}\label{ERGeq}
    \Lambda\frac{d}{d\Lambda}Z = 0.
\end{equation}
To satisfy the renormalization condition above, we require that $S_I$ satisfies the Polchinski equation written in Eq. (\ref{polchinski-eq}) but with all instances of the regulator replaced with the spatial regulator used in $S_0$. To derive the flow of the density matrix we need to take the $\Lambda$ derivative of the path-integral in (\ref{density-operator-def}):
\begin{equation}
    -\Lambda\frac{d}{d\Lambda}\mel{\varphi^-}{\rho}{\varphi^+}= \frac{1}{Z}\int_{\phi(0^-, \mathbf{x}) = \varphi^-(\mathbf{x})}^{\phi(0^+, \mathbf{x}) = \varphi^+(\mathbf{x})} \mathcal{D}\phi \left(\Lambda\frac{d}{d\Lambda}S_0 + \Lambda\frac{d}{d\Lambda}S_I +\Lambda\frac{d}{d\Lambda}S_{\textrm{bdry}}\right)e^{-S_0-S_I- S_{\textrm{bdry}}}.
\end{equation}
In the presence of boundaries, the derivation of the ERG flow changes in two ways. First, the scale dependence of the boundary term $S_{\textrm{bdry}}$ results in new operator insertions in the path integral. Second, consider again the manipulations performed in (\ref{mdmS_0}) which involved the following steps:
\begin{equation}
    \phi * G^{-1} * (\cdots)= (G^{-1}*\phi) * (\cdots) = \frac{\delta S_0}{\delta \phi} * (\cdots).
\end{equation}
The kernel $G^{-1}$ is proportional to the Laplacian $\partial^2$ and in going to the second expression, we must integrate by parts twice. In the presence of boundaries, this results in boundary terms
that we denote by $\text{B.T.}$
\begin{eqnarray}\label{implicit-bt-def}
    \Lambda\frac{d}{d\Lambda}S_0 &&= -\frac{1}{2}(G^{-1}*\phi) *G'*(G^{-1}*\phi)+ \textrm{B.T.}\nn\\
    &&=-\frac{1}{2}\frac{\delta S_0}{\delta \phi}* G' * \frac{\delta S_0}{\delta \phi} + \textrm{B.T.}
\end{eqnarray}
The rest of the analysis remains the same: we repeat the steps of (\ref{schwingertwice})  using the Schwinger-Dyson equations, and apply the Polchinski equation (\ref{polchinski-eq}). All the bulk integrals cancel and the flow of $\rho$ reduces to a sum of boundary insertions in the path integral:
\begin{equation}
    -\Lambda\frac{d}{d\Lambda}\mel{\varphi^-}{\rho}{\varphi^+}= \frac{1}{Z}\int_{\phi(0^-, \mathbf{x}) = \varphi^-(\mathbf{x})}^{\phi(0^+, \mathbf{x}) = \varphi^+(\mathbf{x})} \mathcal{D}\phi \left(\Lambda\frac{d}{d\Lambda}S_{\textrm{bdry}}+\textrm{B.T}\right)e^{-S_0-S_I- S_{\textrm{bdry}}}.
\end{equation}

We start with the first term in the parentheses above.
The only $\Lambda$ dependence comes from the regulator, and we pick up a term corresponding to each boundary
\begin{spreadlines}{10pt}
\begin{align}\begin{split}\label{ldellSbound}
    \Lambda\frac{d}{d\Lambda}S_{\textrm{bdry}} &= -\frac{1}{2}\varphi^- \cdot K^{-1}\cdot \Lambda\frac{d}{d\Lambda}K\cdot K^{-1}\cdot \partial_t\phi^- + \frac{1}{2}\varphi^+ \cdot K^{-1}\cdot \Lambda\frac{d}{d\Lambda}K\cdot K^{-1}\cdot \partial_t \phi^+\\
    &=-\frac{1}{2}\varphi^-\cdot \Delta_\Sigma\cdot D_t\phi^- + \frac{1}{2}\varphi^+ \cdot \Delta_\Sigma \cdot D_t\phi^+.
    \end{split}
\end{align}
\end{spreadlines}
where we have defined the spatial kernel
\begin{equation}\label{deltasigma}
    \Delta_\Sigma(\mathbf{p};\Lambda) \equiv \Lambda\frac{d}{d\Lambda}\log(K(\mathbf{p}^2/\Lambda^2)) = K^{-1}(\mathbf{p}^2/\Lambda^2) \Lambda\frac{d}{d\Lambda}K(\mathbf{p}^2/\Lambda^2)\ .
\end{equation}
We will see in (\ref{SboundUnit}) that this term contributes to the unitary part of the evolution of the density matrix, and has the interpretation of a disentangler.

The non-unitary part of the evolution of the density matrix comes from the terms in $\text{B.T.}$ To compute these terms, we must explicitly go between the expressions 
\begin{equation}\label{ds0}
     -\frac{1}{2}\phi * G^{-1} * G'  * G^{-1}* \phi\ =-\frac{1}{2}(G^{-1}*\phi) *G'*(G^{-1}*\phi)+ \textrm{B.T.}
\end{equation}
The path integral has two boundaries located at $t =t_+$ and $t=t_-$. Therefore, the only boundary terms come from derivatives in the $t$-direction. As stated in the preliminaries, we take $t_\pm = \pm \epsilon$ with $\epsilon \to 0$ for the preparation of density matrices in flat space. As we will see, there are terms in $\text{B.T.}$ that couple the two boundaries at $t_+$ and $t_-$ which ultimately leads to  a non-unitary flow equation. To see how these boundary terms come about we apply a spatial Fourier transform to (\ref{ds0}) to obtain the expression
  \begin{align}\begin{split}\label{explainnonunit}
     &-\frac{1}{2}\phi * G^{-1}* G' * G^{-1}*\phi \\&\hspace{.7cm}= -\frac{1}{2}\int_{\mathbf{p}}\int dt dt' K^{-2}(\mathbf{p}^2/\Lambda^2)\left(\phi_{-\mathbf{p}}(t) (\mathbf{p}^2 - \partial_t^2) G'(t-t'; \mathbf{p})(\mathbf{p}^2 - \partial_{t'}^2)\phi_{\mathbf{p}}(t')  \right).
     \end{split}
 \end{align}
Here we have used a (time-dependent) spatial mode decomposition
 \begin{equation}
     \phi(\mathbf{x}, t) = \int \frac{d^{d-1}\mathbf{p}}{(2\pi)^{d-1}} \phi_{\mathbf{p}}(t) e^{-i\mathbf{p}\cdot \mathbf{x}}
 \end{equation}
 and the temporally non-local kernel is given by
 \begin{equation}
     G'(t-t';\mathbf{p}) = \frac{\Lambda\frac{d}{d\Lambda}K(\mathbf{p}^2/\Lambda^2)}{2\sqrt{\mathbf{p}^2}} e^{-\sqrt{\mathbf{p}^2}|t-t'|}\ .
 \end{equation}
 We also use the short hand for momentum space integrals
 \begin{equation}
     \int_{\mathbf{p}} \equiv \int \frac{d^{d-1}\mathbf{p}}{(2\pi)^{d-1}}
 \end{equation}
 Because the integrand in (\ref{explainnonunit}) is local in momenta, we can consider a single mode and focus on the integrals over $t$ and $t'$. For the mode of momentum $\mathbf{p}$, we have the expression
 \begin{equation}
     -\frac{1}{2}K^{-2}(\mathbf{p}^2/\Lambda^2) \int dt dt' \left(\phi_{-\mathbf{p}}(t) (\mathbf{p}^2 - \partial_t^2) G'(t, t'; \mathbf{p})(\mathbf{p}^2 - \partial_{t'}^2)\phi_{\mathbf{p}}(t')  \right).
 \end{equation}
After integrating by parts on the $t$ integral, we will obtain boundary terms at $t = t_\pm$. However, the $t'$ integral remains a ``bulk'' integral over the entire spacetime, and so B.T. appears to be a mix of boundary and bulk terms. To resolve this, we notice that $G'$ is proportional to the propagator, and so by integrating by parts twice on $t'$, we get a delta function at the cost of additional boundary terms. In Appendix \ref{app:computationboundary}, we show that after this manipulation, all bulk integrals in B.T. vanish. 

Schematically, each time we integrate by parts we get boundary terms of the form $(b_+ - b_-)$ where the subscript denotes the surface of support. Thus, after integrating by parts on both $t$ and $t'$ we formally obtain an expression with the structure 
\begin{eqnarray}
    (b_+ - b_-) \cdot \Delta \cdot (b_+ - b_-)
\end{eqnarray}
where $\Delta$ is a kernel and we recall the dot here denotes integration over the spatial coordinates. After multiplying these terms out, we obtain two types of boundary terms: those with insertions supported on the same boundary, and those that couple the two boundaries. The coupling of the $t_+$ and $t_-$ surfaces is incompatible with unitary evolution, and is therefore responsible for the non-unitarity of the ERG. Obtaining the terms denoted by $\textrm{B.T.}$ is straightforward but requires a careful treatment of the order of limits when approaching the boundary surfaces; this calculation is performed in Appendix \ref{app:computationboundary}. 
We cite the result here, which reads
\begin{align}\label{bt}
    \begin{split}\textrm{B.T.} =&  -\varphi^- \cdot \Delta_2\cdot \varphi^+ +\frac{1}{2} \varphi^- \cdot \Delta_2 \cdot \varphi^- +\frac{1}{2}  \varphi^+ \cdot \Delta_2\cdot \varphi^+ \\
    &- D_t\phi^- \cdot \Delta_0 \cdot D_t\phi^+ +\frac{1}{2}  D_t\phi^- \cdot \Delta_0 \cdot D_t\phi^-+\frac{1}{2} D_t\phi^+ \cdot \Delta_0 \cdot D_t\phi^+\\
    &-\varphi^-\cdot \Delta_1 \cdot D_t\varphi^+ + D_t\phi^- \cdot \Delta_1\cdot \varphi^+ \\
    & +\frac{1}{2}\left(   D_t\phi^- \cdot \Delta_1 \cdot \varphi^- -\varphi^- \cdot \Delta_1 \cdot D_t\phi^-+\varphi^+ \cdot \Delta_1\cdot D_t \phi^+ - D_t\phi^+\cdot \Delta_1 \cdot \varphi^+\right)
    \end{split}
\end{align}
where the three kernels $\Delta_0$, $\Delta_1$ and $\Delta_2$ expressed in momentum space are
\begin{eqnarray}
   && \Delta_0(\mathbf{p})= \frac{1}{2g(\mathbf{p})}\Delta_\Sigma(\mathbf{p}),\hspace{.6cm}
    \Delta_1(\mathbf{p}) = -\frac{1}{2} \Delta_\Sigma (\mathbf{p}),\hspace{.6cm}
    \Delta_2(\mathbf{p}) = -\frac{g(\mathbf{p})}{2}\Delta_\Sigma(\mathbf{p})\ .
\end{eqnarray}
with 
\begin{equation}
     g(\mathbf{p}; \Lambda) \equiv \frac{\abs{\mathbf{p}}}{K(\mathbf{p}^2/\Lambda^2)}\ .
 \end{equation} 
Although the full expression is messy, we can clearly see the claimed structure, namely the appearance of cross-terms which couple fields on the $t_+$ and $t_-$ surfaces. 

The final step of our computation is to convert the boundary terms in the path integral into operators acting on the state $\rho$. Insertions of $\varphi^+$ and $\varphi^-$ in the integrand correspond to the position space representation of the field operator $\hat{\varphi}$ inserted on the left or right of the density operator, respectively. For the momentum operator in the configuration basis, we have
\begin{gather}
    \mel{\varphi^-}{\hat{\pi}\rho}{\varphi^+} = -i\frac{\delta}{\delta \varphi^-}\rho(\varphi^-, \varphi^+),\hspace{1cm}
    \mel{\varphi^-}{\rho\hat{\pi}}{\varphi^+} = i\frac{\delta}{\delta \varphi^+}\rho(\varphi^-, \varphi^+).
\end{gather}
Thus, it follows from
\begin{equation}
    \frac{\delta}{\delta \varphi^\pm} e^{-S_{\textrm{bdry}}} = \pm D_t\phi^\pm
\end{equation}
that insertions of $D_t\phi^-$ correspond to applying $-i\hat{\pi}$ to the left of the state, while $D_t\phi^+$ corresponds to applying $-i\hat{\pi}$ to the right of the state.

We can now convert the terms of B.T. into operators acting on the state $\rho$, making sure to preserve the Euclidean time order discussed in Appendix \ref{app:computationboundary}. The result is
\begin{align}
\begin{split}
    \frac{1}{Z}\int &\mathcal{D}\phi\: (\textrm{B.T.}) e^{-S_0 - S_I - S_{\textrm{bdry}}}\\ &= \frac{1}{2}\int_{\mathbf{p}} \Delta_\Sigma(\mathbf{p})\Bigg[g(\mathbf{p})\left(\hat{\varphi}(-\mathbf{p})\rho \hat{\varphi}(\mathbf{p}) - \frac{1}{2}\{\rho, \hat{\varphi}(\mathbf{p})\hat{\varphi}(-\mathbf{p})\}\right)\\
    &\hspace{2.5cm}+\frac{1}{g(\mathbf{p})}\left(\hat{\pi}(-\mathbf{p})\rho \hat{\pi}(\mathbf{p}) - \frac{1}{2}\{\rho, \hat{\pi}(\mathbf{p})\hat{\pi}(-\mathbf{p})\}\right)\\
    &\hspace{2.5cm} -i\bigg(\hat{\varphi}(-\mathbf{p}) \rho \hat{\pi}(\mathbf{p}) -\hat{\pi}(-\mathbf{p})\rho \hat{\varphi}(\mathbf{p}) -  \frac{1}{2}\{\hat{\pi}(-\mathbf{p}) \hat{\varphi}(\mathbf{p}) - \hat{\varphi}(-\mathbf{p})\hat{\pi}(\mathbf{p}), \rho\}\bigg)\Bigg]
    \end{split}
\end{align}
To interpret the equations above, it is helpful to introduce scale-dependent annihilation operators
\begin{align}
    \begin{split}\label{adef}
        a(\mathbf{p}; \Lambda) = \sqrt{\frac{g(\mathbf{p}; \Lambda)}{2}}\hat{\varphi}(\mathbf{p}) + i\sqrt{\frac{1}{2g(\mathbf{p}; \Lambda)}} \hat{\pi}(\mathbf{p}).
    \end{split}
\end{align}
In terms of these operators, the contribution of the boundary terms corresponds to the action of the superoperator $\mathcal{D}$ below:
\begin{align}\label{Drho}
\begin{split}
    \mathcal{D}(\rho)&\equiv \frac{1}{Z}\int \mathcal{D}\phi\ (\textrm{B.T.}) e^{-S_0 - S_I - S_{\textrm{bdry}}}\\ &= \int_{\mathbf{p}} \Delta_\Sigma (\mathbf{p}) \left[a(\mathbf{p}) \rho a^\dagger(\mathbf{p}) - \frac{1}{2}\{\rho, a^\dagger(\mathbf{p}) a(\mathbf{p})\}\right]
\end{split}
\end{align}
where we have suppressed the $\Lambda$ dependence in $a(\mathbf{p})$ and $\Delta_\Sigma(\mathbf{p})$.
Next, we have the terms coming from $S_\textrm{bdry}$, which we recall are of the form 
\begin{equation}
    \frac{1}{Z}\int \mathcal{D}\phi \left(-\frac{1}{2}\varphi^-\cdot \Delta_\Sigma\cdot D_t\phi^- + \frac{1}{2}\varphi^+ \cdot \Delta_\Sigma \cdot D_t\phi^+\right) e^{-S_0 -S_I - S_{\textrm{bdry}}}
\end{equation}
Unlike terms in B.T which have a natural ordering for products of $\varphi$ and $D_t\phi$, the products appearing here are genuinely ``simultaneous.'' The simplest way to understand the operator order is to use the definition of the derivative directly in the path integral, so that we obtain
\begin{equation}
    \varphi^\pm(-\mathbf{p}) D_t\phi^\pm(\mathbf{p}) = \lim_{\epsilon \to 0^\pm}\left( \lim_{h \to 0^+}\phi_{-\mathbf{p}}(\epsilon) \frac{\phi_{\mathbf{p}}(\epsilon + h) - \phi_{\mathbf{p}}(\epsilon - h)}{2h}\right)\ .
\end{equation}
Putting everything into Euclidean time order, this definition results in symmetrized operator products:\footnote{Note that we are using a symmetric definition of the derivative to ensure the time-reflection symmetry of the Euclidean path integral.}
\begin{align}
\begin{split}
    \frac{1}{Z}\int \mathcal{D}\phi  \left(-\varphi^-(-\mathbf{p}) D_t\phi^-(\mathbf{p}) \right) e^{-S_0-S_I-S_{\textrm{bdry}}} &= \frac{i}{2}\left(\hat{\varphi}(-\mathbf{p}) \hat{\pi}(\mathbf{p}) + \hat{\pi}(\mathbf{p}) \hat{\varphi}(-\mathbf{p})\right)\rho\\
    \frac{1}{Z}\int \mathcal{D}\phi  \left(-\varphi^+(-\mathbf{p}) D_t\phi^+(\mathbf{p}) \right) e^{-S_0-S_I-S_{\textrm{bdry}}} &= \frac{i}{2}\rho\left(\hat{\varphi}(-\mathbf{p}) \hat{\pi}(\mathbf{p}) + \hat{\pi}(\mathbf{p}) \hat{\varphi}(-\mathbf{p})\right)
    \end{split}
\end{align}
Define the Hermitian ``disentangler'' operator
\begin{eqnarray}\label{Kpart}
     \hat{K} &&= \frac{1}{4}\int_{\mathbf{p}}\Delta_\Sigma(\mathbf{p})\left(\hat{\varphi}(-\mathbf{p}) \hat{\pi}(\mathbf{p}) + \hat{\pi}(\mathbf{p}) \hat{\varphi}(-\mathbf{p})\right)\nn\\
     &&=-\frac{i}{4}\int_{\mathbf{p}} \Delta_\Sigma(\mathbf{p}) \lb a(\mathbf{p})a(-\mathbf{p})-a^\dagger(\mathbf{p})a^\dagger(-\mathbf{p})\rb.
\end{eqnarray}
Then we see that
\begin{equation}\label{SboundUnit}
    \frac{1}{Z}\int \mathcal{D}\phi \left(\Lambda\frac{d}{d\Lambda}S_{\textrm{bdry}}\right)e^{-S_0-S_I-S_\textrm{bdry}} = i[\hat{K}, \rho]
\end{equation}

Combining the equation above with (\ref{Drho}) gives the  flow of the density operator under the first step of ERG as  a {\it scale-dependent} Lindblad equation
\begin{equation}\label{Lindbladevol}
    -\Lambda\frac{d}{d\Lambda}\rho = \mathcal{L}_\Lambda(\rho) = i[\hat{K}, \rho] + \mathcal{D}(\rho),
\end{equation}
whose ``Hamiltonian'' is described by the Hermitian operator $\hat{K}$. Note that both $\hat{K}$ and $\mathcal{D}$ depend on the cut-off $\Lambda$. We refer to the ERG flow equation above as the {\it bare flow} equation. 

In the second step of ERG, we perform the following Weyl transformation
\begin{eqnarray}
    \eta_{\mu\nu}\to e^{2s}\eta_{\mu\nu}, \qquad \phi(x)\to e^{-s(d-2)/2}\phi(x),
\end{eqnarray}
and bring $\Lambda$ back to its initial value \cite{fliss2017unitary}. 
Under an infinitesimal transformation above, the density matrix transforms as
\begin{gather}
    \rho \to \rho' = \rho + i\epsilon [\hat{L}, \rho]\\\label{rescaclingD}
     \hat{L} = -\frac{1}{2}\int d^{d-1}\mathbf{x}\left[((\mathbf{x}\cdot \nabla + \Delta_\phi) \hat{\varphi}(\mathbf{x}) )\hat{\pi}(\mathbf{x}) + h.c.\right]    
\end{gather}
where $\Delta_\phi \equiv (d-2)/2$ is the canonical scaling dimension of a scalar field in $d$ dimensions coming from the Weyl transformation and the $\mathbf{x}\cdot \nabla$ compensates for the  shift in the argument of $K(-\nabla^2/\Lambda^2)$.\footnote{See Appendix \ref{app:dilation} for a derivation.}

Putting the two steps of ERG together, the density matrix flows according to the equation 
\begin{equation}
    \frac{d}{ds}\rho = i[\hat{L}, \rho] + \mathcal{L}_\Lambda(\rho).
\end{equation}
We define a rescaled Lindbladian 
 \begin{equation}
     \mathcal{L}\equiv i[\hat{L}, \cdot ] + \mathcal{L}_\Lambda
 \end{equation}
which is independent of the scale $s$.\footnote{The cutoff $\Lambda$ after the rescaling is now a fixed parameter.} However, this conclusion is a little too quick. In a general interacting theory, under the ERG, the scaling dimension of the field $\phi$ does not remain the power-counting dimension $(d-2)/2$. Generally, the field acquires an anomalous dimension. Anomalous dimensions can be incorporated in the ERG formalism simply by choosing a scale-dependent scaling dimension $\Delta_\phi(s)=(d-2)/2+\gamma(s)$ in (\ref{rescaclingD}), resulting in the scale-dependent operator,
\begin{equation}\label{rescalingD2}
    \hat{L}_s =  -\frac{1}{2}\int d^{d-1}\mathbf{x}\left[((\mathbf{x}\cdot \nabla + \Delta_\phi(s)) \hat{\varphi}(\mathbf{x}) )\hat{\pi}(\mathbf{x}) + h.c.\right] 
\end{equation}
This discussion is standard in the ERG literature; for a more detailed discussion, we refer the reader to \cite{Rosten_2012}. In an ERG flow from a UV (to an IR) fixed point, the value of $\gamma(s)$ must be choosen so that it coincides with the anomalous scaling dimension at the fixed points at $s \to - \infty$ ($s\to \infty)$, respectively.

\subsection{The RG Lindbladian}

The effect of rescaling is to bring $\Lambda$ back to its original value, resulting in an effective scale which we keep track of with $s$. This adds an additional term to the generator of the flow in (\ref{Lindbladevol}), resulting in the flow equation
\begin{equation}\label{evoldensitymatrix}
    \frac{d}{ds}\rho =\mathcal{L}_s(\rho) = i[\hat{L}_s+\hat{K}, \rho] + \mathcal{D}(\rho).
\end{equation}
This flow equation belongs to a well-known class of evolutions in the study of open quantum systems. 
In this section, we would like to highlight some of the key properties of this equation.

\begin{figure}
    \centering
    \includegraphics[scale = .7]{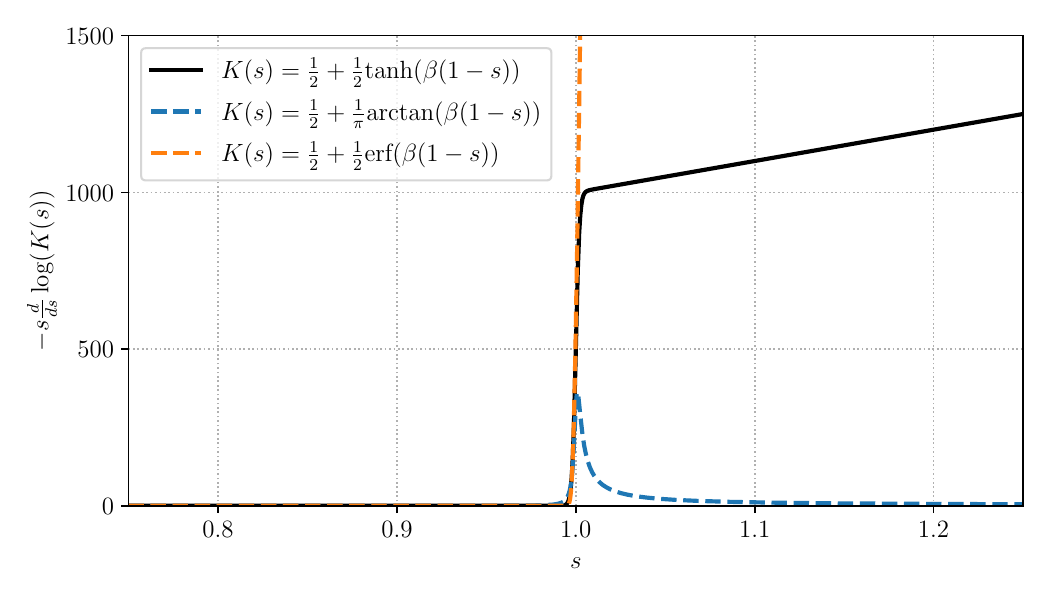}
    \caption{\small{Plots of the function $\Delta_\Sigma(s) = -s\frac{d}{ds}\log(K(s))$, where $s = \abs{\mathbf{p}}/\Lambda$, for three choices of regulator. The parameter $\beta = 500$ for all three curves. Each is nearly zero for values of $s < 1$, and grow sharply in the vicinity of the cutoff. The behavior above the cutoff is highly sensitive to the particular choice of function, but this should not affect any universal (low energy) behavior of the state.}}
    \label{fig:decay-rate}
\end{figure}

The evolution in (\ref{evoldensitymatrix}) is ``memory-less" (Markovian), meaning that the right-hand-side depends only on the state $\rho(s)$, and not its history. In the open quantum system literature, the generator of a Markovian flow, e.g. the superoperator on the right-hand-side, is called a Lindbladian. In finite dimensions, the Lindbladian can always be put into the form\footnote{Note that this form is highly non-unique.}
\begin{equation}\label{Lindbladian}
    \mathcal{L}_s(\rho) =-i[H(s), \rho] + \sum_{i = 1}^{N^2-1} \gamma_i(s) \left(L_i(s) \rho L_i(s)^\dagger - \frac{1}{2}\{\rho, L_i(s)^\dagger L_i(s)\}\right)
\end{equation}
where $N$ is the dimension of the Hilbert space on which $\rho$ acts as a matrix. $H(s)$ is the standard Hamiltonian time evolution of the state, while the $L_i(s)$ are known as Lindblad jump operators. Integrating (\ref{Lindbladian}) for a general time-dependent Lindbladian results in the ``time-ordered" flow
\begin{eqnarray}\label{integratedLindbladian}
    \Phi_{s,s_0}\equiv \mathcal{P} e^{\int_{s_0}^s ds'\mathcal{L}_{s'}}\ .
\end{eqnarray}
Since the Lindbladian in (\ref{evoldensitymatrix}) is manifestly traceless, the integrated map is trace-preserving. Moreover, if for all $s\geq s_0$ one has $\gamma_i(s)\geq 0$, then the integrated map $\Phi_{s,s_1}$ is completely-positive for all $s\geq s_1\geq s_0$, and hence a quantum channel \cite{breuer2016colloquium}. For our generator, the decay rates $\gamma_i(t)$ are set by $\Delta_\Sigma(\mathbf{p})$. As long as the regulator $K$ is a non-increasing function, the kernel $\Delta_\Sigma(\mathbf{p})\geq 0$ for all $\Lambda$ and $\mathbf{p}$. Hence, our integrated ERG flow $\Phi_{s,s_1}$ for all $s\geq s_1\geq 0$ is a quantum channel. See Figure \ref{fig:decay-rate} for plots of $\Delta_\Sigma$ for various choices of regulating function. We conclude that the Polchinski ERG, when applied to quantum states, is described by a one parameter family of quantum channels, whose scale-dependent Lindbladian is given by $\mathcal{L}_s$.

Identifying the ``Hamiltonian" term, the jump operators and the decay rates in our ERG equation allows us to interpret the flow equation in physical terms. The ``Hamiltonian" piece is made up of a coarse-graining generator $\hat{K}$ followed by a scaling generator $\hat{L}_s$. The operator $\hat{K}$ is the generator found in \cite{fliss2017unitary} in the context of pure state RG flows, where it was shown that it can be interpreted as a disentangler.\footnote{When the regulating function $K$ is taken to be a sharp momentum space cut off, it generates a coarse-graining local in momentum space, whereas for a more smooth momentum space cut-offs, it is more natural to view $\hat{K}_s$ as a disentangler in position-space.} In fact, as was established in \cite{goldman2023exact}, this unitary evolution generalizes the Multi-Scale Renormalization Ansatz to the continuum limit \cite{haegeman2013entanglement, zou2019magic}.

Now consider the dissipative term $\mathcal{D}(\rho)$ in the Lindblad equation. Generators which are discrete versions of (\ref{Drho}) are commonly studied in modeling the coupling of a cavity mode to a thermal bath \cite{alicki2007quantum}. In this context, the Lindblad coefficients $\gamma_i$ are proportional to decay rates of the associated jump process. The jump operators for us are simply given by $a(\mathbf{p})$ for each value of $\mathbf{p}$, representing an emission process for the field mode with the corresponding momentum. The corresponding decay rate is set by the function $\Delta_\Sigma(\mathbf{p})$. Turning again to Figure \ref{fig:decay-rate}, we can investigate the decay rates for various choices of regulating function. One can see that essentially no decay occurs for modes well below the cutoff scale $\Lambda$. As we approach modes with magnitude comparable to $\Lambda$, the value of $\Delta_\Sigma$ becomes large with a value that diverges in the limit of a sharp cutoff. For momenta above the cutoff, the behavior of $\Delta_\Sigma$ depends on the choice of regulator. This is a manifestation of the fact that the renormalization group flow, away from a fixed point, is scheme-dependent. However, the dependence on the scheme does not affect the qualitative behavior of solutions when we flow from a UV fixed point deep into the infrared, since the scale ordering of the integrated channel guarantees that modes above the cutoff are always removed at a prior RG step.

 For an alternative point of view, let us re-express the bare flow equation in the field configuration basis. In this basis, the density matrix has components
 \begin{equation}
     \mel{\varphi^-}{\rho}{\varphi^+} \equiv \rho(\varphi^-, \varphi^+)
 \end{equation}
 and the action of the bare generator can be expressed as a functional differential equation:
 \begin{align}
 \begin{split}
 \label{matrix-convection-diffusion}
     -\Lambda \partial_\Lambda \rho(\varphi^-, \varphi^+) &= \frac{1}{2}\int_{\mathbf{p}}\Delta_\Sigma \Bigg[\frac{1}{g}\left(\frac{\delta^2}{\delta \varphi^-\delta \varphi^+} + \frac{1}{2}\frac{\delta^2}{\delta \varphi^-\delta \varphi^-}+\frac{1}{2}\frac{\delta^2}{\delta \varphi^+\delta \varphi^+}\right)\rho(\varphi^-, \varphi^+)\\
     &\hspace{3cm} + \left(\frac{\delta}{\delta \varphi^-}+\frac{\delta}{\delta \varphi^+}\right)((\varphi^-+\varphi^+)\rho(\varphi^-, \varphi^+) )\\
     &\hspace{3cm} -\frac{g}{2}\left(\varphi^--\varphi^+\right)^2 \rho(\varphi^-,\varphi^+)\Bigg].
 \end{split}
 \end{align}
 Note that this takes the form of a functional convection diffusion equation. The first line implements a diffussion process while the second line represents a drift term. Interestingly, if one regards 
 \begin{equation}
     P(\phi) = \frac{1}{Z}e^{-S_0 - S_I}
 \end{equation}
 as a probability distribution on the space of field configurations, then the Polchinski equation can also be recast as a convection-diffusion equation for $P(\phi)$ \cite{Rosten_2012}. The novel feature of the density matrix flow equation is the final line, which represents a sink term for the matrix elements. Notice that the sink is stronger the further off the ``diagonal'' one gets in the configuration basis, and so this term has a tendency to suppress the contribution of off-diagonal matrix elments in the density matrix. In this sense, the last term can be regarded as a dephasing effect. In summary, the RG flow, when regarded as a differential equation for the matrix elements of the state, is a combination of a formal diffusion of matrix elements and dephasing, both of which are irreversible.

\section{Two examples}\label{sec:gaussian-flows}

\subsection{Example 1: Gaussian states}

As a first example of our formalism, we consider the case of Gaussian density matrices, for which the flow equation is exactly solvable. 
In the configuration basis, a general Gaussian density matrix takes the form
\begin{align}\label{gaussian-state}
    \mel{\varphi^-}{\rho}{\varphi^+} = \rho(\varphi^-, \varphi^+)= \mathcal{N}\exp(-\frac{1}{2}\varphi^-\cdot c_1 \cdot \varphi^- - \frac{1}{2}\varphi^+ \cdot c_1 \cdot \varphi^+ + \varphi^- \cdot c_2 \cdot \varphi^+).
\end{align}
Here we specialize to translation invariant states where $c_1 = c_1(\mathbf{p})$ and $c_2 = c_2(\mathbf{p})$ are local functions in momentum space. Hermiticity requires that $c_1(\mathbf{p})$ and $c_2(\mathbf{p})$ be symmetric under reflections $\mathbf{p} \to -\mathbf{p}$. The constant $\mathcal{N}$ is a normalization factor. Taking the $\Lambda$ derivative of the state, we have
\begin{align}
\begin{split}
    -\Lambda \frac{d}{d\Lambda}\rho(\varphi^-, \varphi^+) &= \Bigg(\frac{1}{2}\varphi^-\cdot \left(\Lambda\frac{d}{d\Lambda}c_1\right)\cdot \varphi^- + \frac{1}{2}\varphi^+\cdot \left(\Lambda\frac{d}{d\Lambda}c_1\right)\cdot \varphi^+ \\&\hspace{2cm}- \varphi^- \cdot \left(\Lambda\frac{d}{d\Lambda}c_2 \right)\cdot \varphi^+- \frac{\Lambda}{\mathcal{N}}\frac{d}{d\Lambda}\mathcal{N}\Bigg)\rho(\varphi^-, \varphi^+)
\end{split}
\end{align}
The Lindblad flow we derived in (\ref{Lindbladevol}) and (\ref{Drho}) for the ERG is diagonal in momentum space and preserves the Gaussianity of density matrices. Therefore, the evolution can be expressed as flow equations for the functions $c_1(\mathbf{p})$ and $c_2(\mathbf{p})$, and $\mathcal{N}$. However, because the flow equation is trace-preserving, the flow of $\mathcal{N}$ is redundant, and we only need to compute terms which contribute to the flow of $c_1$ and $c_2$. Using Eq. (\ref{matrix-convection-diffusion}) and collecting in powers of $\varphi^\pm$, we obtain the following flow equations, 
\begin{align}\label{coarse-grain-gaussian-flow}
    \begin{split}
    \Lambda\frac{d}{d\Lambda}c_1 = -\Delta_\Sigma \cdot (c_1-c_2)  - \frac{1}{2}\Delta_\Sigma \cdot g + \frac{1}{2}(c_1-c_2)\cdot \Delta_\Sigma \cdot g^{-1}\cdot (c_1-c_2)\\
    \Lambda\frac{d}{d\Lambda}c_2 = -\Delta_\Sigma \cdot (c_2-c_1) - \frac{1}{2}\Delta_\Sigma \cdot g -  \frac{1}{2}(c_1-c_2)\cdot \Delta_\Sigma \cdot g^{-1}\cdot (c_1-c_2)
    \end{split}
\end{align}
Now define the functions $c_\pm=c_1\pm c_2$. Then, using the relation
\begin{eqnarray}
\Lambda\frac{d}{d\Lambda}g=-\Delta_\Sigma\cdot g
\end{eqnarray}
we find that in momentum space, $c_\pm$ satisfy the decoupled flow equations
\begin{align}
    \Lambda\frac{d}{d\Lambda}c_+ =\Lambda\frac{d}{d\Lambda}\lb \frac{g^2}{c_-}\rb&= \Lambda\frac{d}{d\Lambda}g.
\end{align}
Integrating these equations with initial scale $\Lambda_0$ and final scale $\Lambda$, results in
\begin{align}
    c_+(\mathbf{p};\Lambda) &= c_+(\mathbf{p};\Lambda_0)  +g(\mathbf{p};\Lambda)-g(\mathbf{p};\Lambda_0)\\
    \frac{g^2(\mathbf{p};\Lambda)}{c_-(\mathbf{p};\Lambda)} &=\frac{g^2(\mathbf{p};\Lambda_0)}{c_-(\mathbf{p};\Lambda_0)}+g(\mathbf{p};\Lambda)-g(\mathbf{p};\Lambda_0)\ .
\end{align}
The solution above can be used to construct solutions to the fully rescaled flows,
\begin{equation}
    \frac{d}{ds}\rho = i[\hat{L}, \rho] - \Lambda\frac{d}{d\Lambda}\rho.
\end{equation}
Acting with the dilatation operator on the Gaussian density matrix and once again collecting powers $\varphi^\pm$, the result for the flow of $c_1$ and $c_2$ reads
\begin{align}\label{scaled-gaussian-flow}
\begin{split}
     \frac{d}{ds}c_1&= -(\mathbf{p}\cdot \nabla_{\mathbf{p}} + \gamma -1)c_1-\Lambda \frac{d}{d\Lambda}c_1, \\
     \frac{d}{ds}c_2&= -(\mathbf{p}\cdot \nabla_{\mathbf{p}} + \gamma -1)c_2-\Lambda \frac{d}{d\Lambda}c_2.
     \end{split}
\end{align}
If we now consider free fields, we can set the anomalous dimension $\gamma = 0$. 
If $c_{1, 2}(\mathbf{p},\Lambda)$ are solutions to (\ref{coarse-grain-gaussian-flow}), then the functions $c_{1,2}(s) = e^s c_{1,2} (e^{-s}\mathbf{p}, e^{-s} \Lambda)$ solve
the scaled equation above.

As a concrete example, suppose at scale $\Lambda_0$ we have the massive free field action
\begin{equation}
    S[\Lambda_0] = S_0[\Lambda_0] + \frac{m^2}{2}\int d^{d}x \phi(x)^2
\end{equation}
Preparing the ground state, the initial values of $c_\pm$ are given by
\begin{align}
    c_+(\mathbf{p};\Lambda_0) = c_-(\mathbf{p};\Lambda_0) = \frac{\sqrt{\mathbf{p}^2 + K(\mathbf{p}^2/\Lambda_0^2) m^2}}{K(\mathbf{p}^2/\Lambda_0^2)} =g(\mathbf{p};\Lambda_0)\frac{\omega_m(\mathbf{p};\Lambda_0)}{\omega_0(\mathbf{p})}
\end{align}
where in going to the last expression we have introduced the massive dispersion relation $\omega_m(\mathbf{p};\Lambda) = \sqrt{\mathbf{p}^2 + K(\mathbf{p}^2/\Lambda^2)m^2}$. Plugging in the initial condition and solving for the kernels, the solution at a lower scale $\Lambda$ is given by
\begin{align}\label{lowercs}
\begin{split}
    c_+(\mathbf{p};\Lambda) &= g(\mathbf{p};\Lambda)+g(\mathbf{p};\Lambda_0)\lb \frac{\omega_m(\mathbf{p};\Lambda_0)}{\omega_0(\mathbf{p})}-1\rb\\
    c_-(\mathbf{p};\Lambda) &= g(\mathbf{p};\Lambda)\lb 1+\frac{g(\mathbf{p};\Lambda_0)}{g(\mathbf{p};\Lambda)}\lb \frac{\omega_0(\mathbf{p})}{\omega_m(\mathbf{p};\Lambda_0)}-1\rb\rb^{-1}\ .
\end{split}
\end{align}
For the massive free field ground state with $\gamma=0$, we conclude
\begin{align}
\begin{split}
    c_+(s) &= \frac{\sqrt{\mathbf{p}^2}}{K(\mathbf{p}/\Lambda)} \left(1  +\frac{K(\mathbf{p}/\Lambda)}{K(e^{-s}\mathbf{p}/\Lambda)}\left(\frac{\sqrt{\mathbf{p}^2 + K(e^{-s}\mathbf{p}/\Lambda)e^{2s}m^2}}{\sqrt{\mathbf{p}^2}}-1\right)\right)\\
    c_-(s) &= \frac{\sqrt{\mathbf{p}^2}}{K(\mathbf{p}/\Lambda)} \left(1  -\frac{K(\mathbf{p}/\Lambda)}{K(e^{-s}\mathbf{p}/\Lambda)}\left(1-\frac{\sqrt{\mathbf{p}^2}}{\sqrt{\mathbf{p}^2 + K(e^{-s}\mathbf{p}/\Lambda)e^{2s}m^2}}\right)\right)^{-1}
\end{split}
\end{align}
where $s = 0$ is the initial condition. The first observation is that in the expressions above, the mass $m^2$ always appears with an additional factor of $e^{2s}$ as expected from naive power counting. To further interpret these solutions, it is helpful to take the limit that $K(s)$ is nearly a Heaviside function. As long as $K$ goes to zero quickly enough at $\abs{\mathbf{p}} \approx \Lambda$, one can approximate\footnote{As a concrete example, one can consider $K_\beta(s) = \frac{1}{2} + \frac{1}{2}\tanh(\beta(1-s))$ and take the limit $\beta \to \infty$. }
\begin{equation}
    \frac{K(\mathbf{p}/\Lambda)}{K(e^{-s}\mathbf{p}/\Lambda)} \approx K(\mathbf{p}/\Lambda).
\end{equation}
In this limit our solution becomes
\begin{align}\label{piecewise-sol}
    c_+(s) \approx c_-(s)&\approx \begin{cases}
        \sqrt{\mathbf{p}^2 + e^{2s} m^2} & |\mathbf{p}| < \Lambda \\
        \frac{\sqrt{\mathbf{p}^2}}{K(\mathbf{p} / \Lambda)} & |\mathbf{p}| > \Lambda
    \end{cases}
\end{align}
A Gaussian density matrix is pure if and only if $c_+=c_-$. Therefore, in the sharp momentum cut-off limit, the flowed state is well-approximated by the pure ground state of a field theory with mass $m(s) = e^s m(0)$, as expected. The behavior above the cutoff is divergent in this limit, which simply enforces the cutoff by suppressing the contribution of these modes to the state.

Let us now compare directly with the effective action of the Polchinski equation. For an initial action which is quadratic at scale $\Lambda_0$, it can be shown that at a lower scale $\Lambda$ the effective action takes the form
\begin{equation}
    S[\Lambda] = S_0[\Lambda] + \frac{1}{2}\phi * B(\Lambda)* \phi.
\end{equation}
Here $B(\Lambda)$ is a bilocal function and the initial condition for the massive case is $B(x, y;\Lambda_0) = m^2 \delta^{(d)}(x-y)$. From the Polchinski equation, one finds that $B$ satisfies the flow equation
\begin{equation}
    -\Lambda \frac{d}{d\Lambda}B=-B * G' * B.
\end{equation}
In Appendix \ref{app:massive-flow} we solve this equation for translation invariant $B$. Then, because the action is quadratic, we can use a field redefinition in the path integral to derive the ``ground state'' density matrix, i.e., the density matrix associated with the flowed path integral with no operator insertions. The result is in precise agreement with (\ref{lowercs}), as claimed. 

\subsection{Example 2: $\lambda \phi^4$ ground state}\label{subsec:interaction}

As a second example, we consider the interacting ERG flow with initial condition
\begin{equation}\label{phi4-action}
    S[\Lambda_0] = S_0[\Lambda_0] + \frac{\lambda}{4!}\int d^dx \phi(x)^4.
\end{equation}
We take the initial state $\rho(\Lambda_0)$ to be the ground state of $S[\Lambda_0]$. Because the initial action is local, the corresponding ground state is pure, and we denote it by $\rho(\Lambda_0) = \ket{\Omega_\lambda}\bra{\Omega_\lambda}$.
Our goal in this example, besides demonstrating the validity and computational power of our formalism, is to provide intuition as to how specific terms in the Polchinski equation are reflected in our master equation. 

In the presence of interactions, the non-linear terms in the Polchinski equation generate all possible polynomial terms in the action. To perform a calculation, we will have to resort to a perturbative expansion in $\lambda$. Our main goal will be to highlight how the non-linear behavior of the Polchinski equation is captured by the dissipative terms $\mathcal{D}$ of our generator. As we will see, these non-linear contributions do not appear at first order in $\lambda$, and so in what follows we will work to second order.

To compare the Polchinski equation and our Lindbladian, we will view the state as a function of $S_I$ as it flows under the Polchinski equation. Formally, we have $\rho = \rho(S_I[\Lambda], \Lambda)$, and its total derivative is given by
\begin{equation}\label{fullderivative}
    -\Lambda \frac{d}{d\Lambda}\rho = \delta_{\scriptscriptstyle{S_I}}\rho - \Lambda\partial_\Lambda \rho.
\end{equation}
The total derivative of the state from the path integral is
\begin{align}\label{perturbative-polchinski-state-flow}
\begin{split}
    -\Lambda \frac{d}{d\Lambda}\rho\Bigg\vert_{\Lambda_0} &= \int \mathcal{D}\phi \left(\Lambda \frac{d}{d\Lambda}S_I\right)\Bigg\vert_{\Lambda_0} e^{-S_0[\Lambda_0] - S_I[\Lambda_0] - S_{\textrm{bdry}}[\Lambda_0]}\\&\hspace{3cm} - \Lambda \frac{\partial}{\partial \Lambda}\left(\int\mathcal{D}\phi e^{-S_0[\Lambda] - S_I[\Lambda_0] - S_{\textrm{bdry}}[\Lambda]}\right)\Bigg\vert_{\Lambda_0}.
\end{split}
\end{align}
and we identify
\begin{align}
    -\Lambda \partial_\Lambda \rho\Big\vert_{\Lambda_0} &\equiv -\Lambda \frac{\partial}{\partial \Lambda}\left(\int\mathcal{D}\phi e^{-S_0[\Lambda] - S_I[\Lambda_0] - S_{\textrm{bdry}}[\Lambda]}\right)\Bigg\vert_{\Lambda_0},\\
    \delta_{\scriptscriptstyle{S_I}}\rho\Big\vert_{\Lambda_0} &\equiv \int \mathcal{D}\phi \left(\Lambda \frac{d}{d\Lambda}S_I\right)\Bigg\vert_{\Lambda_0} e^{-S_0[\Lambda_0] - S_I[\Lambda_0] - S_{\textrm{bdry}}[\Lambda_0]}.
\end{align}
These definitions highlight the fact that $\delta_{\scriptscriptstyle{S_I}}\rho$ captures the ``dynamical'' contribution of the Polchinski equation while the partial derivative simply captures the \textit{parametric} dependence of the state on $\Lambda$. In Appendix \ref{app:perturbative}, we show that the second term in (\ref{fullderivative}) takes the form
\begin{eqnarray}\label{defineoperatorS}
  -\Lambda \p_\Lambda \rho(S_I;\Lambda)= i [\hat{K},\rho] - \mathcal{S}(\rho)\ .
\end{eqnarray}
and so the unitary evolution of the state is entirely captured by the parametric dependence on $\Lambda$. The definition of the operator $\mathcal{S}$ is given in the Appendix, as we will not use it here. Now equating (\ref{fullderivative}) with the Lindbladian $\mathcal{L}_\Lambda$, we have
\begin{equation}\label{dissipative-to-polchinski}
    \mathcal{D}(\rho) = \delta_{\scriptscriptstyle{S_I}}\rho - \mathcal{S}(\rho).
\end{equation}
This equation establishes explicitly that the ERG flow of the effective action in the Polchinski equation is \textit{entirely} captured by the non-unitary generator $\mathcal{D}$. In our derivation in Section \ref{sec:ERGdensitymatrix} we used the Schwinger-Dyson equation to convert the bulk term $\delta_{\scriptscriptstyle{S_I}}\rho$ to boundary terms in $\mathcal{D}(\rho)$. One of our main objectives in this example is to explicitly study this step in more detail in a perturbative expansion for $\lambda$. 

The initial interaction term in (\ref{phi4-action}) is
\begin{equation}
    S_I[\Lambda_0] = \frac{\lambda}{4!}\int d^dx \phi(x)^4.
\end{equation}
Under the Polchinski equation, the action flows to $S_I[\Lambda'] = S_I[\Lambda_0] - \epsilon\left(\Lambda\frac{d}{d\Lambda}S_I\right)\big\vert_{\Lambda_0}$ with
\begin{align}
    -\Lambda \frac{d}{d\Lambda}S_I\Bigg\vert_{\Lambda_0}
    &=- \frac{1}{2} \frac{\delta S_I[\Lambda_0]}{\delta \phi}* G' * \frac{\delta S_I[\Lambda_0]}{\delta \phi} + \frac{1}{2}\Tr\left(G' * \frac{\delta^2S_I[\Lambda_0]}{\delta \phi\delta \phi}\right)\\
    &\label{pol-perturbative}= \frac{\lambda}{4}\int d^dx \phi(x)^2 G'(0)- \frac{\lambda^2}{2(3!)^2} \int d^dx d^dy \left(\phi(x)^3 G'(x-y) \phi(y)^3\right)
\end{align}
The linear term in $\lambda$ comes from the trace and describes the first order correction to the propagator. As we claimed earlier, the non-linear term of the Polchinski equation contributes at $O(\lambda^2)$ and describes a non-local six point interaction. It is this term which we will focus on. Let us suppress the spatial coordinates for a moment and simply analyze the time integrals. We can break these integrals up into three terms:
\begin{equation}
    -\frac{\lambda^2}{2(3!)}\left[\int_{-}\int_{-}dtdt' + \int_{+}\int_{+}dtdt' +2 \int_{-}\int_{+} dtdt'\right]\phi(t)^3 G'(t-t') \phi(t')^3
\end{equation}
where we are using the shorthand
\begin{equation}
    \int_{\pm} dt \equiv \pm \int_0^{\pm \infty} dt
\end{equation}
and in the last term we have taken advantage of the reflection symmetry of the kernel $G'$. Because insertions in negative Euclidean time appear to the left of the state while insertions at positive Euclidean time appear to the right, we can see that the first two terms correspond to inserting six field operators to either the right or the left of the state, while the final term corresponds to having three field insertions on both the left and the right. This implies that this term contributes to matrix elements with odd particle number:
\begin{equation}\label{non-local-state}
    \frac{\lambda^2}{(3!)^2} \int_{\mathbf{x}, \mathbf{y}}\int_{-\infty}^0 \int_{0}^\infty dt dt' G'(t-t',\mathbf{x}-\mathbf{y}) \left(e^{t\hat{H}} \hat{\varphi}(\mathbf{x})^3\right)\ket{\Omega_0}\bra{\Omega_0}\left(\hat{\varphi}(\mathbf{y})^3 e^{-t'\hat{H}}\right),
\end{equation}
After performing Wick contractions, the expression above leads to both 3-particle and 1-particle matrix elements of the state. In fact, the term above is the \textit{only} contribution to the odd particle matrix elements of the flowed density matrix at this order.\footnote{It is straightforward to check that no other terms in $\delta_{\scriptscriptstyle{S_I}}\rho$ can contribute matrix elements with odd particle number. Moreover, the operator $\mathcal{S}$ in (\ref{defineoperatorS}) cannot change the particle number of a given term in the perturbative expansion, and so $\mathcal{S}(\ket{\Omega_\lambda}\bra{\Omega_\lambda})$ never has odd particle matrix elements, simply by the parity symmetry of the action.}
Hence, we can isolate this term in the Polchinski equation by considering matrix elements in the odd particle number sector. Define the three particle wave functions
\begin{equation}
    \ket{\psi_{1, 2}} = \int_{\mathbf{p}_1, \mathbf{p}_2, \mathbf{p}_3} \psi_{1, 2}(\mathbf{p}_1, \mathbf{p}_2, \mathbf{p}_3) \ket{\mathbf{p}_1, \mathbf{p}_2, \mathbf{p}_3}.
\end{equation}
Then we have
\begin{equation}
    -\mel{\psi_1}{\left(\Lambda \frac{d}{d\Lambda}\rho\right)}{\psi_2}\Big\vert_{\Lambda_0} = \mel{\psi_1}{\delta_{\scriptscriptstyle{S_I}}\rho}{\psi_2}\Big\vert_{\Lambda_0}
\end{equation}
where we are able to drop the $\mathcal{S}$ term as it does not contribute. We can now investigate explicitly how the Lindbladian accounts for this matrix element. Going to momentum space, the expression in (\ref{non-local-state}) becomes 
\begin{align}
\begin{split}
    &\frac{\lambda^2}{(3!)}\int_{\mathbf{p}_i, \mathbf{q}_i, \mathbf{p}} \int_{-\infty}^0 \int_{0}^\infty dt dt' \delta^{(d-1)}\left(\sum_{i=1}^3 \mathbf{p}_i - \mathbf{p}\right)\delta^{(d-1)}\left(\sum_{i=1}^3 \mathbf{q}_i + \mathbf{p}\right) \frac{e^{-\omega(\mathbf{p})(t'-t)}}{2\omega(\mathbf{p})}\\
&\hspace{1.4cm}\times\Delta_\Sigma(\mathbf{p})K(\mathbf{p}/\Lambda^2) \left[e^{t\hat{H}} \hat{\varphi}(\mathbf{p}_1)\hat{\varphi}(\mathbf{p}_2)\hat{\varphi}(\mathbf{p}_3)\ket{\Omega_0}\bra{\Omega_0}\hat{\varphi}(\mathbf{q}_1)\hat{\varphi}(\mathbf{q}_2)\hat{\varphi}(\mathbf{q}_3)e^{-t'\hat{H}}\right]
\end{split}
\end{align}
Expanding the field operators in creation and annihilation operators and performing Wick contractions, we can collect terms which contribute to the three particle sector. The final expression takes the form
\begin{align}
\begin{split}\label{nonlocal-explicit}
    \delta_{\scriptscriptstyle{S_I}}\rho\Big\vert_{\textrm{3-particle}}
    &= \lambda^2\int_\mathbf{p} \hat{\delta}(\mathbf{p}) \frac{1}{E_1 E_2}\Delta_\Sigma(\mathbf{p}) \frac{K(\mathbf{p}^2/\Lambda^2)}{2\omega(\mathbf{p})}  \\
    &\hspace{.8cm}\times\left(\prod_{i=1}^3 \sqrt{\frac{K(\mathbf{p}_i^2/\Lambda^2)}{2\omega(\mathbf{p}_i)}}a^\dagger(\mathbf{p}_i)\right)\ket{\Omega_0}\bra{\Omega_0}\left(\prod_{i=1}^3 \sqrt{\frac{K(\mathbf{q}_i^2/\Lambda^2)}{2\omega(\mathbf{q}_i)}}a(\mathbf{q}_i)\right)
\end{split}
\end{align}
where we have defined
\begin{gather}
    \hat{\delta}(\mathbf{p})\equiv \delta^{(d-1)}\left(\sum_{i=1}^3 \mathbf{p}_i - \mathbf{p}\right)\delta^{(d-1)}\left(\sum_{i=1}^3 \mathbf{q}_i + \mathbf{p}\right)
\end{gather}
and 
\begin{gather}
    E_1\equiv \omega(\mathbf{p}) + \sum_{i=1}^3 \omega(\mathbf{p}_i),\hspace{2cm}
    E_2 \equiv  \omega(\mathbf{p}) + \sum_{i=1}^3 \omega(\mathbf{q}_i)
\end{gather}

In Appendix \ref{app:perturbative} we discuss, in detail, a diagramatic approach to the perturbation theory in $\lambda$ of density matrices $\rho$. There are two types of propagators:
\begin{equation}\label{prop}
    \includegraphics[valign = c, width = 2.5cm]{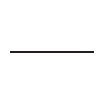} = \begin{cases}
        \frac{K(\mathbf{p}^2/\Lambda^2)}{2\omega(\mathbf{p})} & \textrm{internal}\vspace{.2cm}\\
        \sqrt{\frac{K(\mathbf{p}^2/\Lambda^2)}{2\omega(\mathbf{p})}} a^\dagger(\mathbf{p};\Lambda) & \textrm{external}
    \end{cases}
\end{equation}
Each external leg corresponds to a creation operator which acts on the free vacuum state. For internal legs, the propagator evaluates to a scalar coefficient corresponding to a ``spatial'' Green's function. Diagrams with external legs are therefore operator-valued, with $n$ external legs generating an $n$-particle state when acting on the vacuum. To evaluate diagrams, every propagator is assigned a (spatial) momentum and all momenta are integrated over. As usual, vertices carry a factor of the interaction $\lambda$ as well as a momentum conserving delta function. There are some additional details in determining an overall energy factor associated to every diagram which is described in the Appendix which we will not emphasize here.\footnote{These factors are always $\Lambda$ independent, and so do not affect any of our manipulations in this section.}

Note that, in the perturbative expansion for $\ket{\Omega_\lambda}$, diagrams always appear to the left of the state. In constructing the density operator, we must compute $\bra{\Omega_\lambda}$ which contains conjugates of the diagrams in its perturbative expansion, placing them to the right of the unperturbed state. For convenience, in what follows there will always be an implicit complex conjugation associated to diagrams appearing to the right of the unperturbed density operator $\ket{\Omega_0}\bra{\Omega_0}$.

Combining the diagrammatic rules from Appendix \ref{app:perturbative} with the Polchinski equation\footnote{We refer to this perturbative expansion as the Feynman-Polchinski diagrams.} (\ref{pol-perturbative}), the expression in Equation (\ref{nonlocal-explicit}) can be summarized  as\footnote{Matching propagators and momentum conserving delta functions is straightforward, but the energy factors of $E_1$ and $E_2$ require some more care to understand. See the Appendix for details.}
\begin{equation}\label{nonlocal-pol-diagram}
    \mel{\psi_1}{\delta_{\scriptscriptstyle{S_I}}\rho}{\psi_2} = \bra{\psi_1}\left[\includegraphics[scale=1, valign = c]{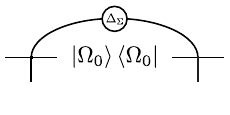}\right]\ket{\psi_2}
\end{equation}
where we have introduced the modified propagator 
\begin{equation}\label{prop-deriv}
    \includegraphics[valign = c, width = 2.5cm]{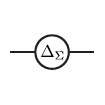} = \begin{cases}
        \Delta_\Sigma(\mathbf{p})\frac{K(\mathbf{p}^2/\Lambda^2)}{2\omega(\mathbf{p})} & \textrm{internal}\vspace{.2cm}\\
        \frac{1}{2}\Delta_\Sigma(\mathbf{p})\sqrt{\frac{K(\mathbf{p}^2/\Lambda^2)}{2\omega(\mathbf{p})}} a^\dagger(\mathbf{p};\Lambda) & \textrm{external}
    \end{cases}
\end{equation}

In the remainder of this section, we reproduce the diagram in (\ref{nonlocal-pol-diagram}) by computing the matrix elements of our Lindbladian generator $\mathcal{D}$:
\begin{equation}
    \mel{\psi_1}{\mathcal{D}(\ket{\Omega_\lambda}\bra{\Omega_\lambda})}{\psi_2}.
\end{equation}
From the parity symmetry of the initial action, the state $\ket{\Omega_\lambda}$ only has diagrams with even particle number. Consider the explicit expression for $\mathcal{D}$,
\begin{equation}
    \mathcal{D}(\rho) = \int_{\mathbf{p}}\Delta_\Sigma(\mathbf{p} \left[a(\mathbf{p})\rho a^\dagger(\mathbf{p}) - \frac{1}{2}\{a^\dagger(\mathbf{p})a(\mathbf{p}), \rho\}\right]
\end{equation}
The anticommutator term $\{a^\dagger(\mathbf{p})a(\mathbf{p}), \rho\}$ does not change the particle number of a diagram, and so the only contribution to the matrix element must come from
\begin{equation}
    \int_{\mathbf{p}}\Delta_\Sigma(\mathbf{p})\mel{\psi_1}{a(\mathbf{p})\ket{\Omega_\lambda}\bra{\Omega_\lambda}a^\dagger(\mathbf{p})}{\psi_2}
\end{equation}
The creation/annihilation operators lower the particle number of any diagram in the state by one, and so for a term in the expansion of $\ket{\Omega_\lambda}\bra{\Omega_\lambda}$ to contribute to a three particle matrix element, it must have a four particle diagram on both sides of the density operator. If one inspects all of the terms in the second order expression for the state (\ref{second-order-full}), we see that at second order there is only one term which satisfies this requirement, namely
\begin{equation}
    \left[\includegraphics[scale = .6, valign = c]{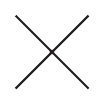} \right]\ket{\Omega_0}\bra{\Omega_0}\left[\includegraphics[scale = .6, valign = c]{figs/firstorder4pt.pdf}\right]
\end{equation}
where the explicit expression for the diagram appearing above is
\begin{equation}
    \includegraphics[valign = c]{figs/firstorder4pt.pdf} = -\frac{\lambda}{4!}\int_{\mathbf{p}_1, \mathbf{p_2}, \mathbf{p}_3, \mathbf{p}_4} \frac{\delta^{(d-1)}(\sum_{i=1}^4 \mathbf{p}_i)}{\sum_{i=1}^4\omega(\mathbf{p}_i)} \left(\prod_{i=1}^4\sqrt{\frac{K(\mathbf{p}_i^2/\Lambda^2)}{2\omega(\mathbf{p}_i)}} a^\dagger(\mathbf{p}_i;\Lambda)\right)
\end{equation}
We can commute the jump operators past the expression for the diagram, which precisely reproduces the equation (\ref{nonlocal-explicit}), or as a diagrammatic manipulation,
\begin{align}
    \int_\mathbf{p} \Delta_\Sigma(\mathbf{p})a(\mathbf{p}) \left[\includegraphics[scale = .6, valign = c]{figs/firstorder4pt.pdf} \right]\ket{\Omega_0}\bra{\Omega_0}\left[\includegraphics[scale = .6, valign = c]{figs/firstorder4pt.pdf}\right] a^\dagger(\mathbf{p}) = \includegraphics[valign = c]{figs/nonlocal-modified.pdf}
\end{align}
where we can interpret the creation and annihilation operators from $\mathcal{D}$ as contracting one of the open legs from each diagram. Thus, we conclude that at second order, the three-particle matrix elements satisfy
\begin{equation}
    \mel{\psi_1}{\mathcal{D}(\ket{\Omega_\lambda}\bra{\Omega_\lambda})}{\psi_2} = \mel{\psi_1}{\delta_{\scriptscriptstyle{S_I}}\rho}{\psi_2}\Big\vert_{\Lambda_0}
\end{equation}
We now see that the non-linear contributions of the Polchinski equation are accurately accounted for  by the dissipative terms of the Lindbladian. This is important, for example, for using our flow equation to accurately compute the second order beta function for $\lambda$. 

\section{Integrating the ERG flow}\label{sec:integrate}

In this section, we will explore the Heisenberg evolution induced by the RG master equation. This allows us to exactly integrate the ERG flow by solving the flow of arbitrary correlation functions implicitly in terms of the correlation functions at the UV scale. Recall that, for a quantum channel $\Phi$, the \textit{dual map} $\Phi^*$ is defined by the adjoint relation
\begin{equation}
    \tr(\Phi(\rho)X) = \tr(\rho \Phi^*(X))
\end{equation}
where $\rho$ is a density matrix and $X\in \mathcal{B}(\mathcal{H})$ is any bounded operator on the Hilbert space.\footnote{A quantum channel is a compeletely positive trace-preserving map, and its dual is a completely-positive map that preserves the identity operator (a unital map) \cite{Furuya:2020tzv}.} As a Lindbladian $\mathcal{L}$ may be regarded as the generator of an ``infinitesimal channel,'' it also has a corresponding dual, $\mathcal{L}^*$. For a general Lindbladian of the form
\begin{equation}
    \mathcal{L}(\rho) = -i[H, \rho] + \sum_{i}\gamma_i \left(L_i \rho L_i^\dagger - \frac{1}{2}\{\rho, L_i^\dagger L_i\}\right),
\end{equation}
the dual generator can be given explicitly as 
\begin{equation}
    \mathcal{L}^*(X) = i[H, X] + \sum_i \gamma_i \left(L_i^\dagger X L_i - \frac{1}{2}\{X, L_i^\dagger L_i\}\right).
\end{equation}
Thus, for example, the dual generator for the dissipative term $\mathcal{D}$ is given by
\begin{equation}
    \mathcal{D}^*(X) =  \int_{\mathbf{p}} \Delta_\Sigma(\mathbf{p}) \left[a^\dagger(\mathbf{p}) X a(\mathbf{p}) - \frac{1}{2}\{X, a^\dagger(\mathbf{p})a(\mathbf{p})\}\right],
\end{equation}
and the dual RG Lindbladian can be written as
\begin{equation}
    \mathcal{L}^*(X) = -i[\hat{L}+ \hat{K}, X]  + \mathcal{D}^*(X)
\end{equation}

\subsection{Correlator generating function}

We will now use the dual flow equation to implicitly solve the ERG master equation for arbitrary correlation functions.\footnote{The discussion is inspired by well-known methods in the study of open quantum system dynamics \cite{alicki2007quantum}.} Suppose we have an initial state $\rho_0$ defined at a UV scale $\Lambda$. Introduce the operator
\begin{equation}
    D(\alpha_1, \alpha_2) = \exp(i\int_\mathbf{p}\alpha_1(-\mathbf{p}) \hat{\varphi}(\mathbf{p}))\exp(-i\int_{\mathbf{p}}\alpha_2(-\mathbf{p}) \hat{\pi}(\mathbf{p})).
\end{equation}
Then, consider the characteristic function
\begin{equation}\label{characteristic}
    F(\alpha_1, \alpha_2) = \tr(\rho_0 D(\alpha_1,\alpha_2))
\end{equation}
which up to a phase matches the Wigner quasi-probability distribution. Similar to the Wigner distibution, we can use the characteristic function in (\ref{characteristic}) as a generating functional for correlation functions of the form $\expval{\varphi^n \pi^m}_{\Lambda}$ at the cutoff scale.\footnote{This set of correlators along with the canonical commutation relations is sufficient to reproduce any correlation function. The choice of generating function one uses always induces a preferred ordering of operators. Which one is used is a matter of convenience. See Appendix \ref{app:normal} for a calculation using normal ordered products.} If we now flow the state along the RG, the value of the correlator flows in the Schr\"odinger picture to
\begin{equation}
    F_s(\alpha_1, \alpha_2) = \tr(\rho(s) D(\alpha_1, \alpha_2)).
\end{equation}
where $\rho(s)$ is the flowed state,
\begin{equation}
    \rho(s) = \Phi_{s}(\rho_0)
\end{equation}
Equivalently, we may write 
\begin{equation}
    F_s(\alpha_1, \alpha_2) = \tr(\rho_0 \Phi_s^* (D(\alpha_1, \alpha_2))),
\end{equation}
and study the flow of the generating function by solving the ERG flow in the Heisenberg picture,
\begin{equation}
    \frac{d}{ds}X(s) = \mathcal{L}^* X(s)
\end{equation}
with the initial condition $X(0) = D(\alpha_1, \alpha_2)$. As before, we solve this equation by first solving the bare flow equation
\begin{equation}
    -\Lambda \partial_\Lambda X(\Lambda) = \mathcal{L}_\Lambda^*(X(\Lambda))
\end{equation}
and then we find a solution to the rescaled equations by making an appropriate redefinition. Consider the ansatz
\begin{align}
    X(\Lambda) &= e^{\delta(\Lambda)}D(\beta_1(\Lambda),\beta_2(\Lambda)) \\&=  e^{\delta(\Lambda)}\exp(i\int_\mathbf{p}\beta_1(-\mathbf{p};\Lambda) \hat{\varphi}(\mathbf{p}))\exp(-i\int_{\mathbf{p}}\beta_2(-\mathbf{p};\Lambda) \hat{\pi}(\mathbf{p})).
\end{align}
That is, we keep the functional form of $D$ up to a normalization, but where we have introduced scale dependent functions $\beta_{1, 2}(\Lambda)$ and $\delta(\Lambda)$ satisfying the initial condition
\begin{align}
    \delta(\Lambda_0) = 0,\qquad
    \beta_{1, 2}(\Lambda_0) = \alpha_{1, 2}.
\end{align}
The explicit derivative of $X$ is given by
\begin{equation}
   -\Lambda \partial_\Lambda X = \dot{\delta} X + i\int_{\mathbf{p}}\left[\dot{\beta}_1(\mathbf{p}) \hat{\varphi}(\mathbf{p})X - \dot{\beta}_2(\mathbf{p})X\hat{\pi}(-\mathbf{p}))\right] . 
\end{equation}
where we have implemented a dot notation for the differential operator $-\Lambda\partial\Lambda$. 

We need to compute the action of the dual generator on $X$. First, let us write the diffusive term in a more convenient form. Written in terms of $\hat{\varphi}$ and $\hat{\pi}$, it can be shown that 
\begin{align}
\begin{split}
    \mathcal{D}^*(X) &= -\frac{1}{2}\int_{\mathbf{p}}\Delta_\Sigma(\mathbf{p}) \Bigg(\frac{g(\mathbf{p})}{2} [\hat{\varphi}(-\mathbf{p}), [\hat{\varphi}(\mathbf{p}), X]] + \frac{1}{2 g(\mathbf{p})} [\hat{\pi}(-\mathbf{p}), [\hat{\pi}(\mathbf{p}), X]]\\&\hspace{5cm} + i \hat{\varphi}(-\mathbf{p}) [\hat{\pi}(\mathbf{p}), X] - i \hat{\pi}(-\mathbf{p}) [\hat{\varphi}(\mathbf{p}), X]\Bigg)
\end{split}
\end{align}
Then, it follows that
\begin{align}
\begin{split}
    \mathcal{D}^*(X) &= -\frac{1}{2}\int_{\mathbf{p}} \Delta_\Sigma(\mathbf{p}) \Bigg( \left(\frac{g(\mathbf{p})}{2} \beta_2(-\mathbf{p})\beta_2(\mathbf{p}) + \frac{1}{2g(\mathbf{p})} \beta_1(-\mathbf{p})\beta_1(\mathbf{p})\right)X \\&\hspace{3cm}+ i\beta_1(\mathbf{p}) \hat{\varphi}(-\mathbf{p}) X - i\beta_2(\mathbf{p}) X \hat{\pi}(-\mathbf{p}) -i\beta_1(-\mathbf{p})\beta_2(\mathbf{p})X\Bigg)\ .
\end{split}
\end{align}
Next we need the unitary term, which is given by
\begin{align}
    -i[\hat{K}, X] &= -\frac{i}{2}\int_{\mathbf{p}}\Delta_\Sigma(\mathbf{p})\left[\beta_1(\mathbf{p})\hat{\varphi}(-\mathbf{p})X + \beta_2(\mathbf{p})X \hat{\pi}(-\mathbf{p})\right]
\end{align}
Matching terms with our ansatz, we obtain the system of equations
\begin{gather}
\begin{align}
    \dot{\beta}_1(\mathbf{p}) &= -\Delta_\Sigma(\mathbf{p})\beta_1(\mathbf{p})\\
    \dot{\beta}_2(\mathbf{p}) &= 0\end{align}\\ 
    \dot{\delta} = -\frac{1}{4}\int_\mathbf{p} \Delta_\Sigma(\mathbf{p}) \left[g(\mathbf{p})\beta_2(\mathbf{p})\beta_2(-\mathbf{p}) + \frac{1}{g(\mathbf{p})}\beta_1(\mathbf{p})\beta_1(-\mathbf{p}) - 2i\beta_1(\mathbf{p})\beta_2(-\mathbf{p})\right] 
\end{gather}
The equations for $\beta_1$ and $\beta_2$ are easily solved. Using the initial condition, one finds
\begin{align}
    \beta_1(\mathbf{p};\Lambda) &= \frac{K(\mathbf{p}^2/\Lambda^2)}{K(\mathbf{p}^2/\Lambda_0^2)} \alpha_1(\mathbf{p})\\
    \beta_2(\mathbf{p};\Lambda) &= \alpha_2(\mathbf{p}).
\end{align}
Plugging these solutions back into the equation for $\delta$, we can directly integrate to obtain
\begin{equation}
    \delta(\Lambda) = \frac{1}{4}\int_\mathbf{p}\left[g\left(\frac{K}{K_0} - 1\right) \alpha_2^2 + \frac{1}{3g_0}\left(\frac{K^3}{K_0^3} - 1\right)\alpha_1^2 - 2 i \alpha_1\alpha_2 \left(\frac{K}{K_0} - 1\right)\right]
\end{equation}
We have dropped the dependence of functions on $\mathbf{p}$ and $\Lambda$ above to keep the expression concise. The subscripted variables $K_0$ and $g_0$ are evaluated at the initial scale $\Lambda_0$, while $K$ and $g$ are evaluated at $\Lambda$. 

Now, consider the rescaled flow equation. We must include the action of the dilation generator on $X$, which reads
\begin{equation}
    -i[\hat{L}, X] = i \int_\mathbf{p} \left(\left(\mathbf{p}\cdot \nabla_\mathbf{p} + \frac{d-2}{2}\right)\beta_1(\mathbf{p} )\hat{\varphi}(-\mathbf{p}) X - \left(\mathbf{p}\cdot \nabla_\mathbf{p} + \frac{d}{2}\right)\beta_2(\mathbf{p}) X \hat{\pi}(-\mathbf{p})\right)
\end{equation}
Now suppose that $\beta_1(\mathbf{p};\Lambda)$, $\beta_2(\mathbf{p};\Lambda)$ and $\delta(\Lambda)$ are solutions to the bare flow equation. We see that we can construct solutions to the rescaled equation by defining
\begin{align}
    \beta_1(s) &= e^{\frac{d-2}{2}s}\beta_1(e^s\mathbf{p}; e^{-s}\Lambda_0) \\
    \beta_2(s) &= e^{\frac{d}{2}s}\beta_2(e^s\mathbf{p}; e^{-s}\Lambda_0) \\
    \delta(s) &= \delta(e^{-s}\Lambda_0).
\end{align}
Unlike in the Schr\"odinger picture, the momenta and cutoff do not scale in the same manner, which may seem odd at first glance. However, once we insert these functions back into $X(s)$, we can perform a change of variables as
\begin{align}
\begin{split}
    X(s) &= e^{\delta(s)}\exp(i\int_\mathbf{p} e^{\frac{d-2}{2}s}\beta_1(e^{s}\mathbf{p}; e^{-s}\Lambda)\hat{\varphi}(-\mathbf{p})) \exp(-i\int_\mathbf{p}e^{\frac{d}{2}s}\beta_2(e^{s}\mathbf{p}; e^{-s}\Lambda)\hat{\pi}(-\mathbf{p}))\\
    &= e^{\delta(s)}\exp(i\int_\mathbf{p} e^{-\frac{d}{2}s}\beta_1(-\mathbf{p}; e^{-s}\Lambda)\hat{\varphi}(e^{-s}\mathbf{p})) \exp(-i\int_\mathbf{p}e^{-\frac{d-2}{2}s}\beta_2(-\mathbf{p}; e^{-s}\Lambda)\hat{\pi}(e^{-s}\mathbf{p}))
\end{split}
\end{align}
Plugging in the scaled solutions, we obtain the final result
\begin{equation}
X(s)=e^{\delta(s)}\exp(i\int_\mathbf{p} e^{-\frac{d}{2}s} \frac{K(e^{2s}\mathbf{p}^2/\Lambda^2)}{K(\mathbf{p}^2/\Lambda^2)}\alpha_1(-\mathbf{p})\hat{\varphi}(e^{-s}\mathbf{p})) \exp(-i\int_\mathbf{p}e^{-\frac{d-2}{2}s}\alpha_2(-\mathbf{p})\hat{\pi}(e^{-s}\mathbf{p}))
\end{equation}
The scaling factors in the exponents match what one expects from power counting of the field variables. This expression may be
 used to construct the generating function $F_s(\alpha_1, \alpha_2)$. Note that $\delta(s)$ depends on $\alpha_1$ and $\alpha_2$ and must be included when taking derivatives to compute correlators. 

\section{Thermal ERG}
\label{thermalERG}

\begin{figure}
    \centering
    \includegraphics[width=0.45\linewidth]{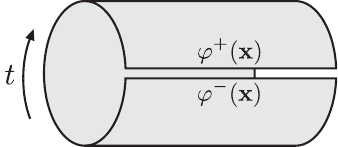}
    \caption{The path integral used for state preparation in Section \ref{thermalERG}. Note that the resulting density matrix need not be the thermal state, due to the presence of sources in the path integral.}
    \label{fig:thermal-path-int}
\end{figure}

Our derivation of the ERG flow density matrices as a Lindblad equation in Section \ref{sec:ERGdensitymatrix} applies to any background spacetime $\mathcal{M}$ with two spacelike boundaries $\Sigma^+$ and $\Sigma^-$. However, the explicit form of the Lindblad operator depends on the propagator in this spacetime. In computing the ``Hamiltonian" piece in (\ref{Kpart}) and the diffusive piece in (\ref{Drho}) we specialized to the vacuum propagator in $\mathbb{R}^d$, which after Fourier transforming in the spatial directions becomes
\begin{eqnarray}
    G(t-t';\mathbf{p})=K(\mathbf{p}^2/\Lambda^2) \frac{e^{-\omega |t-t'|}}{2\omega}
\end{eqnarray}
with $\omega=|\mathbf{p}|$. Another physically important background is the Euclidean cylinder, i.e. $\mathbb{R}^{d-1}\times S^1_\beta$, depicted in Figure \ref{fig:thermal-path-int}. In this background, to satisfy the Polchinski equation for the trace of the density matrix, we must use the thermal propagator at inverse temperature $\beta$:
\begin{align}\label{replacementprop}
    G_\beta(t-t';\mathbf{p}) = \frac{K(\mathbf{p}^2/\Lambda^2)}{2 \omega }\sum_{n=-\infty}^\infty e^{-|t-t'+n\beta|\omega}=K(\mathbf{p}^2/\Lambda^2)\frac{\cosh((|t-t'|-\beta/2)\omega)}{2\omega\sinh(\beta \omega/2)}
\end{align}
where $|t-t'| \in [0, \beta)$ and $\omega = \abs{\mathbf{p}}$. Using the thermal propagator above in our derivation in Section \ref{sec:ERGdensitymatrix} only changes the variation of the bulk term in (\ref{implicit-bt-def}) and B.T., whereas the variation of $S_{\textrm{bdry}}$ in (\ref{SboundUnit}) remains unchanged. Physically, this means that the new background only affects the dissipative terms. 

We repeat the integration by parts performed in  Appendix \ref{app:computationboundary} that led to the derivation of B.T. The result takes the same form as in (\ref{bt})
but now with temperature dependent kernels $\Delta_0$, $\Delta_1$ and $\Delta_2$. Their values can be read off from (\ref{Deltasformula}):
\begin{eqnarray}
   && \Delta_0(\mathbf{p})= \frac{\coth \left(\frac{\beta  \omega }{2}\right)}{2g(\mathbf{p})}\Delta_\Sigma(\mathbf{p}),\hspace{.4cm}
    \Delta_1(\mathbf{p}) = -\frac{1}{2} \Delta_\Sigma (\mathbf{p}),\hspace{.4cm}
    \Delta_2(\mathbf{p}) = -\frac{g(\mathbf{p})\coth \left(\frac{\beta  \omega }{2}\right)}{2}\Delta_\Sigma(\mathbf{p})\ .\nonumber
\end{eqnarray}
At finite temperature, the dissipative term of the ERG Lindbladian becomes
\begin{align}
\begin{split}
    \mathcal{D}_\beta(\rho)&= \frac{1}{2}\int_{\mathbf{p}} \Delta_\Sigma(\mathbf{p})\Bigg[g(\mathbf{p})\coth \left(\frac{\beta  \omega }{2}\right)\left(\hat{\varphi}(-\mathbf{p})\rho \hat{\varphi}(\mathbf{p}) - \frac{1}{2}\{\rho, \hat{\varphi}(\mathbf{p})\hat{\varphi}(-\mathbf{p})\}\right)\\
    &\hspace{2.5cm}+\frac{\coth \left(\frac{\beta  \omega }{2}\right)}{g(\mathbf{p})}\left(\hat{\pi}(-\mathbf{p})\rho \hat{\pi}(\mathbf{p}) - \frac{1}{2}\{\rho, \hat{\pi}(\mathbf{p})\hat{\pi}(-\mathbf{p})\}\right)\\
    &\hspace{2.5cm} -i\bigg(\hat{\varphi}(-\mathbf{p}) \rho \hat{\pi}(\mathbf{p}) -\hat{\pi}(-\mathbf{p})\rho \hat{\varphi}(\mathbf{p}) -  \frac{1}{2}\{\hat{\pi}(-\mathbf{p}) \hat{\varphi}(\mathbf{p}) - \hat{\varphi}(-\mathbf{p})\hat{\pi}(\mathbf{p}), \rho\}\bigg)\Bigg]
    \end{split}
\end{align}
Introducing the creation/annihilation operators which diagonalize the scale-dependent Hamiltonian (\ref{adef}) gives the Lindbladian 
\begin{align}\label{thermalLindblad}
    \mathcal{D}_\beta(\rho) 
    &= \int_{\mathbf{p}}\frac{\Delta_\Sigma(\mathbf{p})}{\sinh(\beta\omega/2)} \left[e^{\beta \omega/2}(a \rho a^\dagger  - \frac{1}{2}\{a^\dagger a, \rho\})+e^{-\beta \omega/2}(a^\dagger \rho a - \frac{1}{2}\{a a^\dagger, \rho\})\right]
\end{align}
We have suppressed the momentum/scale dependence of the operators here for notational clarity. Unlike the ``zero temperature'' case, the Lindblad generator above includes both the emission $a\rho a^\dagger$ and absorption $a^\dagger \rho a$ of modes weighted with thermal coefficients  $e^{\beta\omega/2}$ and $e^{-\beta \omega/2}$, respectively.\footnote{It is natural to interpret this as a quantum detailed balance condition, even though the thermal density matrix is not invariant under the flow.} Consider the contribution to the thermal state of a free theory from a single momentum mode $\mathbf{p}$:
\begin{eqnarray}
    \rho_\beta(\mathbf{p})\sim e^{-\beta (a(\mathbf{p})^\dagger a(\mathbf{p})+1/2)}\ .
\end{eqnarray}
Then, we have
\begin{eqnarray}\label{thermalcommut}
    a(\mathbf{p}) \rho_\beta(\mathbf{p})=\rho_\beta (\mathbf{p})a(\mathbf{p}) e^{-\beta \omega} 
\end{eqnarray}
Plugging this into (\ref{thermalLindblad}) we find that the diffusive part of the evolution of the thermal state vanishes
\begin{eqnarray}
    \mathcal{D}_\beta(\rho_\beta)=0\ .
\end{eqnarray}
To see how this changes the RG flows, it is instructive to consider an explicit example. Let us once again look at the Gaussian state flows. Written as a functional differential equation, the new bare flow equation becomes
\begin{align}
\begin{split}
\label{thermal-diffusion}
    -\Lambda \partial_\Lambda \rho(\varphi^-, \varphi^+) &= \frac{1}{2}\int_{\mathbf{p}} \Bigg(\Delta_\Sigma\coth(\beta\omega/2)\Bigg[\left(\frac{1}{g}D_\varphi^2 - \frac{g}{2}(\varphi^- -\varphi^+)^2\right)\rho(\varphi^-, \varphi^+)\\
    &\hspace{2cm} + \Delta_\Sigma\left[\left(\frac{\delta}{\delta \varphi^-}+\frac{\delta}{\delta \varphi^+}\right)((\varphi^-+\varphi^+)\rho(\varphi^-, \varphi^+) )\right]\Bigg).
\end{split}
\end{align}
where we have defined the functional differential operator
\begin{equation}
    D_\varphi^2 \equiv \frac{\delta^2}{\delta\varphi^-\delta\varphi^+} + \frac{1}{2}\frac{\delta^2}{\delta\varphi^-\delta\varphi^-}+\frac{1}{2}\frac{\delta^2}{\delta\varphi^+\delta\varphi^+}.
\end{equation}
The only difference from the zero temperature case in (\ref{matrix-convection-diffusion}) is the coefficient of $\coth(\beta \omega/2)$. Carrying this through the computation for a Gaussian density matrix, the flow equation becomes
\begin{equation}
    \Lambda\frac{d}{d\Lambda}c_+ =\Lambda\frac{d}{d\Lambda}\lb \frac{g^2}{c_-}\rb= \coth(\beta\omega/2)\Lambda\frac{d}{d\Lambda}g
\end{equation}
where $c_\pm$ are defined as before. Notice now that because $\coth(\beta\omega/2)$ is $\Lambda$ independent, we can define scaled kernels 
\begin{equation}
    \Tilde{c}_+ = \frac{c_+}{\coth(\beta\omega/2)}, \hspace{2cm}\Tilde{c}_- = \coth(\beta\omega/2) c_-
\end{equation}
such that the tilded kernels satisfy the same differential equation as the zero temperature case. The solutions are then of the same form, where one must take care in the implementation of the initial condition. For the purpose of comparison, consider again the sharp cutoff approximation, so that we may write the approximate solution as 
\begin{equation}
    \Tilde{c}_\pm(\mathbf{p};\Lambda) = \begin{cases}
        \Tilde{c}_\pm(\mathbf{p};\Lambda_0) & \abs{\mathbf{p}} < \Lambda \\
        g(\mathbf{p};\Lambda) & \abs{\mathbf{p}} > \Lambda
    \end{cases}
\end{equation}
Then, returning to the un-tilded variables, we conclude
\begin{equation}
    c_\pm(\mathbf{p};\Lambda) = \begin{cases}
        c_\pm(\mathbf{p};\Lambda_0) & \abs{\mathbf{p}} < \Lambda \\
        g(\mathbf{p};\Lambda)\coth(\beta\omega/2)^{(\pm 1)} & \abs{\mathbf{p}} > \Lambda
    \end{cases}
\end{equation}
The state below the cutoff is completely unaffected by the value of the temperature $\beta$. The new element of this solution is that, above the cutoff, rather than replacing the state with the massless ground state, we now replace it with the massless thermal state of temperature $\beta$.

\section{ERG monotones and approximate quantum error correction}\label{sec:monotones}

Consider again the rescaled flow equation for the density matrix with an $s$-dependent scaling dimension (\ref{rescaclingD}):
\begin{gather*}
    \frac{d}{ds}\rho=i[\hat{L}_s+\hat{K},\rho]+\mathcal{D}(\rho) \\
    \hat{L}_s=-\frac{1}{2}\int d^{d-1}\mathbf{x}\lb \mathbf{x}\cdot \nabla+\Delta_\phi(s))\hat{\varphi}(\mathbf{x}))\hat{\pi}(\mathbf{x})+h.c.\rb\ .
\end{gather*}
As we pointed out in Section \ref{sec:ERGdensitymatrix}, a key property of the ERG equation above is that it is memory-less. Physically, this means that knowledge of the state $\rho(s_0)$ is sufficient to determine its flow towards the infra-red at $s>s_0$. In the vicinity of a fixed point, the values of $\Delta(s)$ for scaling operators are set to their scaling dimensions. However, away from a fixed point, we are free to choose any $\Delta(s)$ that interpolates between the scaling dimensions at the fixed points. Different choices of functions $\Delta(s)$ correspond to different ERG schemes. Integrating the ERG flow, we obtain a one-parameter flow of quantum channels:
\begin{equation}\label{ERGflowint}
    \Phi_{s,s_0} = \mathcal{P}_{s'} e^{-\int_{s_0}^s ds' \mathcal{L}_{s'}}
\end{equation}
where $\mathcal{P}$ is scale-ordering. In this section, we explore two consequences of the ERG equations we have derived. 

First, we observe that for any ERG scheme we choose, the flow in (\ref{ERGflowint}) results in a large class of RG monotones, namely functions that monotonically decrease along the ERG flow.\footnote{Of course, our RG monotones are scheme-dependent.} As we discussed in Section \ref{sec:ERGdensitymatrix}, the integrated ERG map $\Phi_{s,s_0}$ is a quantum channel for any $s\geq s_0$. This implies that for any pair of states $\rho(s_0)$ and $\sigma(s_0)$ and any {\it distinguishability measure} $D(\rho,\sigma)$ we have the data processing inequality \cite{petz2003monotonicity}
\begin{eqnarray}
    D(\rho(s),\sigma(s))\leq D(\rho(s_0),\sigma(s_0))\ .
\end{eqnarray}
In fact, there exists a zoo of such multi-state multi-parameter distinguishability measures \cite{furuya2023monotonic}.
Physically, this implies that the distinguishability of states is non-increasing under the ERG, and hence is an RG monotone. Later in this section, we will explore two examples of such monotones, namely the quantum relative entropy and the (logarithmic) Uhlmann fidelity. We remind the reader that there is a distinction between RG monotones and $C$-theorems. To find a $C$-function, one needs an RG monotone that satisfies extra conditions such as stationarity at fixed points, and the requirement that $C(UV)-C(IR)$ is finite. We postpone the study of such requirements for our ERG monotones to future work. 

Second, we observe that our integrated ERG flow in (\ref{ERGflowint}) combined with the mathematical formalism of approximate quantum error correction can make the following intuition precise: the ERG flow erases the short-distance information of the correlators. It turns out that the Heisenberg picture (dual ERG map) is a more convenient language to connect with approximate quantum error correction \cite{Furuya:2020tzv,Furuya:2021lgx}.

Consider an ERG flow that starts by deforming a CFT at some UV scale $\Lambda$.
Deep in the IR $s\to \infty$ there are three possibilities: 
\begin{enumerate}
    \item {\bf Mass gap:} In the first case, at large enough $s$ the effective scale $e^{-s}\Lambda$ reaches the map gap. Flowing observables deeper into the IR  projects out all observables except for the ERG zero modes:
    \begin{eqnarray}
    \Phi^*_{s,s_0}(\mO_0)=\mO_0\ .\end{eqnarray}
Besides, the identity operator, any zero-mode of ERG results in vacuum degeneracies and topological order. In other words, deep in the IR, we have an exact quantum error correction code \cite{Furuya:2020tzv}. For theories with unique vacuum, below the mass gap, there is no non-trivial observable algebra and all our RG monotones (distinguishability measures) reach zero. 

\item {\bf An IR fixed point:} In some cases, deep in the IR, we can end up with a fixed point. In the vicinity of the IR CFT, we can choose the scheme $\Delta(s)=\Delta$ to get rid of the scale-dependence in the ERG equation and obtain 
\begin{eqnarray}
    \Phi_{s,s_0}=e^{-(s-s_0)\mathcal{L}}\ .
\end{eqnarray}
Such a flow is called a Markov semigroup. The eigen-operators of the Lindbladian are the IR conformal primary fields \cite{Furuya:2021lgx}
\begin{eqnarray}
    \Phi_{s,s_0}(\mO_\Delta)=e^{-(s-s_0)\Delta_\mO}\mO\ .
\end{eqnarray}
Using the spectral analysis of the Lindbladian $\mathcal{L}$ we can obtain bounds on how quickly the distinguishability decreases in scale \cite{gao2022complete}. In this case, we can view ERG as a scheme that prepares approximate quantum error correction codes, and decay of relative entropy can be used to write down approximate Knill-Laflamme equations \cite{Furuya:2021lgx}. In such an approximate quantum error correction code, the logical operators are the low-lying IR primary fields which are approximately protected against the erasure of short-distance information \cite{Furuya:2020tzv}.

\item {\bf Massless Particles:} In Abelian gauge theory (photons), and theories with spontaenous symmetry-breaking, deep in the IR, we end up with massless particles. In this case, we expect the spectrum of the Lindbladian to become continuous near zero to reflect the massless mode.
\end{enumerate}

In the remainder of this section, we make use of the solutions for Gaussian density matrices to compute exact results for the ERG flow of quantum fidelity and relative entropy of states as RG monotones.

\subsection{Ground state fidelity}\label{subsec:fidelity}

For a first example of an RG monotone, let us return to the flow of the massive ground state $\ket{\Omega_m}$. Although in the sharp cutoff limit each massive state remains approximately pure along the flow, this does not imply that the flow is approximately unitary, as we will see below. For a pair of pure states, quantum fidelity is the same as the square of the overlap of the wave-functions. Therefore, in the limit of a sharp momentum cut-off, our ERG monotone is $|\braket{\Omega_m}{\Omega_0}|^2$ where $\ket{\Omega_0}$ is the massless free ground state.

Denote by $g_m(s) = c_+(s) = c_-(s)$ the pure state Gaussian kernel for a state with initial mass $m$. A simple computation with Gaussian integrals yields the result
\begin{equation}
    |\braket{\Omega_m}{\Omega_0}|^2 = \frac{\det(2g_m)^{1/2}\det(2g_0)^{1/2}}{\det(g_m + g_0)} = \exp(\tr(\log(\frac{(4g_mg_0)^{1/2}}{(g_m + g_0)})))
\end{equation}
where $g_0=c_+(s=0)=c_-(0)$.
In evaluating the trace, we must account for the dilation scale $s$. As an integral over momenta, we therefore have
\begin{equation}
    \tr = V e^{-(d-1)s} \int \frac{d^{d-1}\mathbf{p}}{(2\pi)^{d-1}}
\end{equation}
where $V$ is the spatial volume. In the infinite volume limit, this becomes IR divergent, and so the fidelity always yields either zero or one. This is a standard problem for overlaps of QFT ground states, which is avoided by considering the IR finite quantity
\begin{equation}
   f(\ket{\Omega_m}, \ket{\Omega_0}) \equiv  -\frac{\log(|\braket{\Omega_m}{\Omega_0}|^2)}{V}.
\end{equation}
One can interpret $f$ as a ``fidelity density'' between the two states. It is also a monotone under quantum channels, and is zero only when the states are the same. 

\begin{figure}
    \centering
    \includegraphics[width = .7\linewidth]{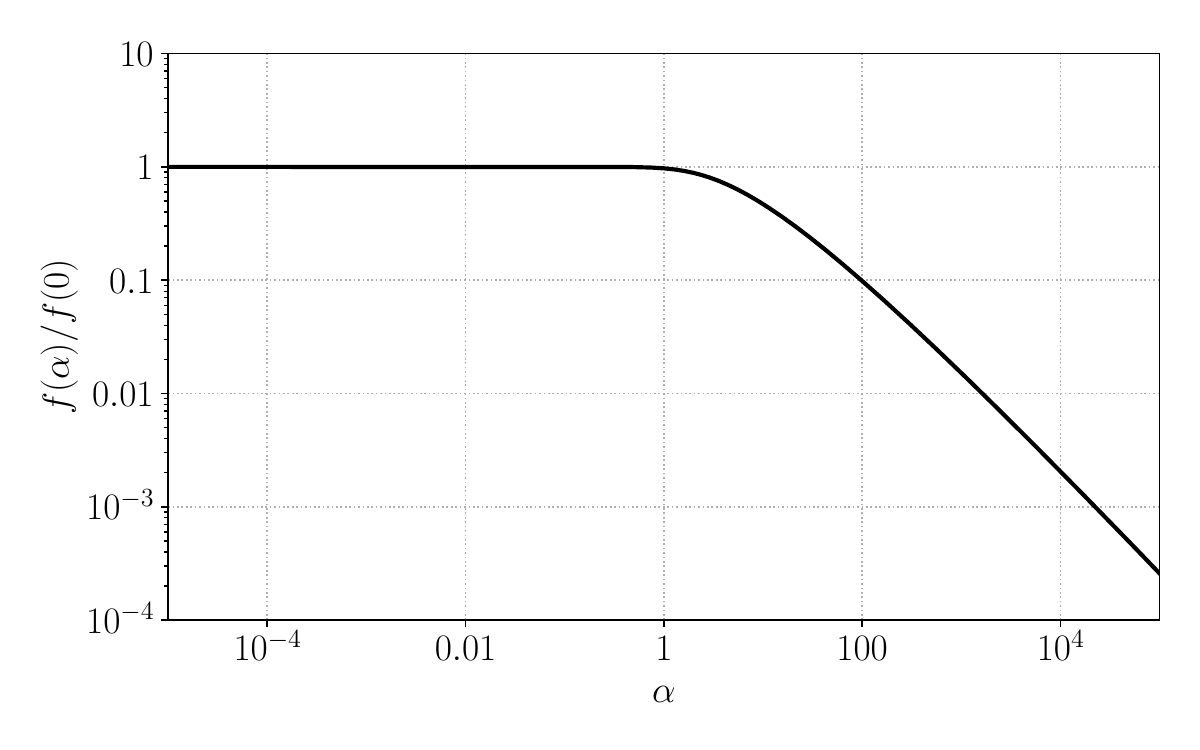}
    \caption{\small{The normalized log-fidelity $f = -\log(\mathcal{F})/V$ between a massive ground state and the massless fixed point ground state as a function of the normalized effective scale $\alpha = (m/\Lambda) e^s$. The UV behavior is consistent with a unitary RG flow, while the IR behavior indicates a loss of distinguishability between a massive and massless state. The crossover between these regimes occurs at $\alpha = 1$, when the effective scale is equal to the mass.}}
    \label{fig:fidelity}
\end{figure}

Using the piecewise approximation (\ref{piecewise-sol}) to $g_m$, we notice that for momenta above the cutoff, $g_m = g_0$, and so
\begin{equation}
    \log(\frac{(4g_mg_0)^{1/2}}{g_m + g_0}) = 0 \quad \text{for }|\mathbf{p}| > \Lambda
\end{equation}
Therefore, our fidelity density is
\begin{align}
    f(\ket{\Omega_m}, \ket{\Omega_0}) 
    &= e^{-(d-1)s}\frac{S_{d-2}}{(2\pi)^{d-1}}\int_0^\Lambda dp p^{d-2}\log(\frac{(4p\sqrt{p^2 + e^{2s}m^2} )^{1/2}}{\sqrt{p^2 + e^{2s}m^2} + p })
\end{align}
For simplicity, consider the case of $d = 2$.\footnote{The behavior of the result in arbitrary dimension is similar and can be expressed generally in terms of hypergeometric functions.} Define the ratio of the mass $m$ and the effective scale $e^{-s}\Lambda$ by 
\begin{equation}
    \alpha = \frac{m}{\Lambda e^{-s}}.
\end{equation}
Then we can exactly integrate the expression above
\begin{equation}
    f(\ket{\Omega_m}, \ket{\Omega_0}) = \frac{m}{2\pi}
    \left[\sqrt{1 + \frac{1}{\alpha^2}}- \frac{1}{\alpha} + \left(\frac{1}{2}\text{arccot}\left(\alpha\right)-1\right) + \frac{1}{2\alpha}\log(\frac{4\sqrt{1 + \alpha^2}}{\left(1 + \sqrt{1 + \alpha^2}\right)^2}) \right].
    \end{equation}
Up to an overall factor, the result depends only on the scale dependent ratio $\alpha$. The function is plotted in Figure \ref{fig:fidelity}. It is helpful to consider the asymptotic behavior of the function, which reads
\begin{align}
    &\frac{f(\alpha)}{f(0)} \sim\begin{cases} 1 + \mathcal{O}(\alpha^3) &   \alpha << 1\\
    \frac{2}{4-\pi} \frac{\log(\alpha/4)}{\alpha}  + \mathcal{O}(\alpha^{-2})&  \alpha >> 1
    \end{cases}
\end{align}
For small $\alpha$, the evolution is approximately unitary, and the fidelity density is amost constant. However, near $\alpha = 1$, the log-fidelity begins to drop, indicating non-unitary evolution. This particular value of $\alpha$ is special, as it represents the value of $s$ at which the effective scale $\Lambda e^{-s}$ and the original mass $m$ are equal. We interpret this as a threshold effect for the relevant perturbation to the massless ground state. That is, for values of $\alpha < 1$, the effective scale remains above the mass, and no information about the state is lost. For $\alpha >1$, the mass is greater than the effective scale, and so the state has essentially ``forgotten'' its initial condition.

\subsection{Quantum relative entropy}\label{subsec:relativeentropy}

As the second example let us consider the quantum relative entropy of two density matrices $\rho$ and $\sigma$,
\begin{equation}
    S_{rel}(\rho ||\sigma) = \tr (\rho \log\rho) - \tr(\rho \log\sigma)
\end{equation}
Relative entropy satisfies the data processing inequality, and is therefore guaranteed to be an RG monotone. As the relative entropy of any two pure states is infinite, the ground states we used in the previous section will not be of much use. Instead,  we compute the relative entropy of the thermal density matrices, $\rho_{\beta, m}$, where $\beta$ is the inverse temperature and $m$ is the mass. As usual, we write the Gaussian density matrix as
\begin{gather}
    \mel{\varphi_-}{\rho_{\beta, m}}{\varphi_+} = \mathcal{N}\exp(\int_{\mathbf{p}}\left[-\frac{1}{2}\varphi_- \cdot c_1 \cdot \varphi_- - \frac{1}{2}\cdot \varphi_+ \cdot c_1 \cdot \varphi_+ + \varphi_+ \cdot c_2\cdot \varphi_- \right]).
\end{gather}
For an initial thermal state in the sharp cut-off limit, we have
\begin{align}
     c_1(\mathbf{p})&= \begin{cases}\omega_m(\mathbf{p})\coth(\beta \omega_m(\mathbf{p})) & \abs{\mathbf{p}} < \Lambda, \\
    \frac{\abs{\mathbf{p}}}{K(\mathbf{p}^2/\Lambda^2)} & \abs{\mathbf{p}}>\Lambda
    \end{cases}\\
    \begin{split}
    \\
    c_2(\mathbf{p}) &=\begin{cases} \omega_m(\mathbf{p}) \csch(\beta \omega_m(\mathbf{p})) & \abs{\mathbf{p}} < \Lambda, \\
    \frac{\abs{\mathbf{p}}}{K(\mathbf{p}^2/\Lambda^2)} & \abs{\mathbf{p}}>\Lambda
    \end{cases}
    \end{split}
\end{align}
with $\omega_m(\mathbf{p}) = \sqrt{\mathbf{p}^2 + m^2}$. We can now use the exact solutions (\ref{lowercs}) to solve for the density matrix at a lower scale, and then perform rescaling to obtain the solution to the full flow equation. Just as in the massive ground state case, the flow is once again given by scaling $m$ and $\beta$ by their canonical mass dimensions (when the regulator is taken to be sharp). That is, the state at scale $s$ is given by the expressions for $c_{1,2}$ as written above, but with the replacements of $m\to e^sm$ and $\beta\to e^{-s}\beta$. 

Our goal is to compute the relative entropy
\begin{equation}
    S_{rel}(\rho_{\beta, m}(s)|| \rho_{\beta, 0}(s))
\end{equation}
which can be compute exactly for Gaussian states \cite{Casini_2009}. It is convenient to express the result purely in terms of the two point functions of the fields. In the case of a translationally invariant and UV regulated state, we can in fact express the relative entropy purely as a momentum space integral. As before, let $c_\pm = c_1 \pm c_2$ and define
\begin{equation}
    C = \frac{1}{2}\sqrt{\frac{c_+}{c_-}} \hspace{1.5cm} X = \frac{1}{2c_-} \hspace{1.5cm}P = \frac{c_+}{2}
\end{equation}
Let $C_{\rho}, X_\rho, P_\rho$ be the functions associated to the translationally invariant Gaussian state $\rho$ and likewise $C_{\sigma}, X_{\sigma}, P_{\sigma}$ for the state $\sigma$. Then, the relative entropy is given by
\begin{equation}
    \tr\left(-C_\rho \log(\frac{C_\rho + \frac{1}{2}}{C_\rho - \frac{1}{2}}) + \left(\frac{X_\rho}{X_\sigma} + \frac{P_\rho}{P_\sigma}\right)\frac{C_\sigma}{2}\log(\frac{C_\sigma + \frac{1}{2}}{C_\sigma - \frac{1}{2}}) + \frac{1}{2}\log(\frac{1 - 4C_\sigma^2 }{1 - 4C_\rho^2}) \right)
\end{equation}
It is now a matter of plugging in the explicit expression for the states and computing the momentum integral. Once again, the cutoff in the state causes the trace to vanish for momenta with magnitude greater than $\Lambda$. Working once again in 1+1 dimensions, the relative entropy is given by the integral 
\begin{align}
\begin{split}
    S_{rel}(\rho_{\beta, m}(s)|| \rho_{\beta, 0}(s)) &= \frac{1}{\pi}\int_0^\Lambda dp\Bigg[\frac{\beta  m^2 \coth \left(\frac{1}{2} \beta  \sqrt{m^2+e^{-2s}p^2}\right)}{4 \sqrt{e^{2s}m^2+p^2}}
    \\&\hspace{2cm}+e^{-s}\log \left(\frac{\coth \left(\frac{1}{2} \beta  \sqrt{m^2+e^{-2s}p^2}\right)+1}{\coth \left(\frac{e^{-s}\beta  p}{2}\right)+1}\right)\Bigg]
\end{split}
\end{align}
The above integral can be evaluated analytically in the $\beta \to 0$ limit.\footnote{The infinite temperature limit of bosonic states does not technically exist, but the relative entropy remains finite in the limit.} The result is given by
\begin{align}
    \lim_{\beta \to 0}S_{rel}(\rho_{\beta, m}(s)|| \rho_{\beta, 0}(s)) &= \frac{e^{-s}}{\pi}\int_0^\Lambda dp\left[ \log \left(\frac{p}{\sqrt{m^2 e^{2s}+p^2}}\right)+\frac{m^2 }{2 \left(m^2 +e^{-2s}p^2\right)}\right]\\
    &= \frac{m}{2\pi}\left[\frac{\log \left(\alpha ^2+1\right)}{\alpha }+\textrm{arccot}(\alpha )\right]
\end{align}
where we have once again defined $\alpha = e^{s}m/\Lambda$. As it turns out, this result has essentially the same qualitative behavior as the ground state fidelity density; both have the same UV and IR behavior, with the crossover occurring at $\alpha \approx 1$.

\section{Discussion}

In a seminal paper \cite{polchinski1984renormalization}, Polchinski provided a concrete realization of the ideas of Wilsonian renormalization, now known as the ERG.
By considering path integrals on manifolds with boundary, the authors of \cite{,fliss2017unitary} extended Polchinski's formalism to the ERG flow of wave-functionals. 
In this work, we applied this extended ERG formalism to derive the ERG flow of density matrices and discovered that, generically, it is described by a scale-dependent Lindblad equation:
\begin{equation}
    \frac{d}{ds}\rho =\mathcal{L}_s(\rho) = i[\hat{L}_s+\hat{K}, \rho] + \mathcal{D}(\rho)\ .
\end{equation}
The ``Hamiltonian" piece is a disentangler $\hat{K}$ followed by a scaling $\hat{L}_s$, as was already pointed out in \cite{fliss2017unitary}. We have found that, in order to satisfy the consistency condition of Figure \ref{fig:commutative-diagram}, an additional dissipative term appears, which is given by (\ref{Drho}) at zero temperature and (\ref{thermalLindblad}) at inverse temperature $\beta$.
We integrated this Lindbladian flow in the Heisenberg picture exactly, and worked out examples of this ERG flow in Gaussian states, and in perturbation theory for $\lambda \phi^4$.
An important implication of our result is that any distinguishability measure of states decreases monotonically under the ERG flow. We postpone the exploration of connections between our ERG monotones and $C$-functions to future work.

The main appeal of the ERG is that it provides a fully non-perturbative and systematic approach to the renormalization of states that connects with other proposals for renormalization of states such as tensor network renormalization \cite{haegeman2013entanglement,vidal2007entanglement,hu2018continuous,fliss2017unitary}, holographic renormalization \cite{heemskerk2011holographic,leigh2014holographic,leigh2015exact}, and optimal transport in statistical field theory \cite{Cotler_2023}. Here, we  comment on the connections between our results and tensor networks. We postpone a careful exploration of the  ERG in large $N$ models and holography, and the connections between our flow equation (\ref{thermal-diffusion}) and optimal transport in statistical field theory to future work.

It was already observed in \cite{fliss2017unitary}  that the ERG flow of states for particular choices of coarse-graining reduces to the unitary tensor network approaches to the RG of states. In particular, by choosing regulators that modify the relativistic dispersion in the UV, we explicitly showed that the ERG results in a continuous Multi-Scale Renormalization Ansatz (cMERA) \cite{goldman2023exact}. Our ERG scheme applied to free quantum fields closesly resembles the so-called continuous Tensor Network Renormalization (cTNR) scheme introduced in \cite{hu2018continuous}. However, as opposed to cMERA and cTNR, the ERG approach has the advantage that it can systematically include interactions, as we demonstrated in a sample calculation in Section \ref{subsec:interaction}.

In discrete tensor networks, applying TNR to discretized Euclidean path-integral yields a MERA network on the boundary \cite{evenbly2015tensor}. 
The authors of \cite{hu2018continuous} observed that in the continuum limit,  as opposed to the discrete case, applying cTNR to a Euclidean path-integral of a Gaussian free fields does not result  in a cMERA on the boundary. Our results suggest a resolution to their puzzle, as we observed that the non-localities in the ERG, generically, induce diffusive terms on the boundaries that need to be added to the cMERA flow.

\appendix

\section{Wegner-Morris Equation}\label{app:wegner}

In this appendix, we will review an alternative and more general approach to the ERG, as developed by Wegner and Morris. Although we only make use of the Polchinski equation in our paper, some of the conceptual discussion in Section \ref{sec:ERGdensitymatrix} is more directly inspired by the Wegner-Morris approach. The idea is to take seriously Kadanoff spin blocking in the continuum. In particular, much like blocked spins, one defines a coarse-grained field variable and re-expresses the partition function in terms of these new variables. The infinitesimal field redefinition is taken to be of the form
\begin{equation}
    \phi(x) \to \phi'(x) = \phi(x) + \frac{\delta \Lambda}{\Lambda}\Psi(x).
\end{equation}
Note the similarities with the cTNR proposal in \cite{hu2018continuous} for free fields.
By absorbing the change of the path integral into an effective action, one obtains a flow equation for effective Euclidean action $S_E$, which is simply the Schwinger-Dyson equation applied to this specific redefinition:
\begin{equation}
    \Lambda\frac{d}{d\Lambda}S_E = \Psi* \frac{\delta S_E}{\delta \phi} - \Tr\left(\frac{\delta \Psi}{\delta \phi}\right).
\end{equation}
The first term is the contribution of the change of the action under the redefinition while the second term accounts for the change of path integral measure. This result is known as the \textit{Wegner-Morris equation}. Of course, not every choice of $\Psi$ implements an RG flow. For example, one could take $\Psi$ to be the infinitesimal generator of some gauge transformation. At a conceptual level, one must require that $\Psi$ actually coarse-grains in some appropriate sense. In the literature, this is often done by parametrizing $\Psi$ as
\begin{equation}
    \Psi = \frac{1}{2}G'* \frac{\delta \Sigma}{\delta \phi}, \hspace{2cm}\Sigma = S_E - 2\hat{S}
\end{equation}
The particular scheme is specified by choosing the functions $G'$ and $\hat{S}$, referred to as the ERG kernel and the seed action, respectively. With this choice, it can be shown that the Wegner-Morris equation becomes a generalized heat equation for the probability density of field configurations. To recover the Polchinski equation, one splits the action in the form $S = S_0 + S_I$ and takes
\begin{equation}
G' = \frac{\Lambda\frac{d}{d\Lambda} K}{p^2},\qquad \hat{S} = S_0
\end{equation}
Note that using this parametrization, the field redefinition $\Psi$ always depends on the action at scale $\Lambda$, ultimately making the Wegner-Morris equation non-linear.

\section{Mixed state preparation in quantum mechanics}\label{app:mixed-states}

In this Appendix, we consider two examples of $0+1$ dimensional path-integrals that prepare mixed states.

\subsubsection*{Example 1: Two-state mixture}
Consider a quantum harmonic oscillator $x(t)$ with the standard Hamiltonian
\begin{equation}
    H = \frac{p^2}{2m} + \frac{1}{2}m\omega^2 x^2
\end{equation}
The ground state is prepared by the Euclidean path integral
\begin{equation}
    \ket{0} = \frac{1}{\sqrt{Z}} \int \mathcal{D}x e^{-\int_{-\infty}^0 dt L_E[x(t)]}.
\end{equation}
We can also consider a state prepared by a single Euclidean insertion, 
\begin{equation}
    \ket{\psi} = \frac{1}{\mathcal{N}}e^{-\delta \hat{H}}\hat{x}\ket{0} = \frac{1}{\sqrt{Z'}} \int \mathcal{D}x \left[x(-\delta) e^{-\int_{-\infty}^0 dt L_E[x(t)]}\right].
\end{equation}
with $\mathcal{N}^2 = \mel{0}{\hat{x} e^{-2\delta \hat{H}}\hat{x}}{0}$. Now we can construct a mixed state by taking the convex combination
\begin{equation}
    \rho = p \ket{0}\bra{0} + (1 - p) \ket{\psi}\bra{\psi}
\end{equation}
where $0< p <1$. Rewriting each state in its path integral representation, we have
\begin{equation}
    \rho = p\left(\frac{1}{Z}\int \mathcal{D}x^- \mathcal{D}x^+ e^{-\int_{-\infty}^\infty dt L_E[x(t)]}\right) + (1-p)\left(\frac{1}{Z'}\int \mathcal{D}x^- \mathcal{D}x^+ \left[ x(-\delta)x(\delta)e^{-\int_{-\infty}^\infty dt L_E[x(t)]}\right]\right)
\end{equation}
Here we have written the path integral measure explicitly as $\mathcal{D}x^- \mathcal{D}x^+$ to remind the reader that the $t = 0^\pm$ surface should be thought of as a cut in these path integrals, awaiting a boundary condition from a position basis bra or ket. Combining these path integrals, we have
\begin{align}
    \rho &= \int \mathcal{D}x^-\mathcal{D}x^+ \mathcal{F}[x] e^{-S_E}\\
    \mathcal{F}[x]&\equiv \frac{p}{Z} + \frac{1-p}{Z'} x(-\delta)x(\delta)
\end{align}
Clearly, $\mathcal{F}[x]$ fails to factorize into functionals of only $x^-$ and $x^+$, which was our stated condition for purity of a state. In this particularly simple example, it is clear that the mixed density matrix arises purely as a consequence of the linearity of the path integral. That is to say, if one can use path integrals to prepare pure density matrices, then one must also necessarily be able to prepare mixed states via convex combination of the integrand. 

\subsubsection*{Example 2: Non-local sources}

The previous example is essentially the full story for mixed state preparation, in its simplest form. However, in some cases one can use an alternative interpretation of the path integral as turning on non-local ``interactions'' between the upper and lower half planes which induce the mixed state. We will now discuss a simple example of such a case. Consider two coupled quantum harmonic oscillators with Hamiltonian 
\begin{equation}
    H = \frac{p_1^2}{2m} + \frac{p_2^2}{2m} + \frac{1}{2}m\omega^2 (x_1^2 + x_2^2) + \lambda x_1x_2
\end{equation}
Once again we denote the unique ground state by $\ket{0}$. If we then partial trace over the second particle's Hilbert space, we obtain the reduced density matrix corresponding to only the first particle in the ground state:
\begin{equation}
    \rho_1 = \tr_{2}(\ket{0}\bra{0}).
\end{equation}
In the path integral, we can obtain a representation of the reduced density matrix by integrating out the second particle. The global ground state density matrix is prepared by 
\begin{align}
\begin{split}
   &\braket{x_1^-, x_2^-}{0}\braket{0}{x_1^+, x_2^+} \\&\hspace{1cm}=\frac{1}{Z} \int \mathcal{D}x_1\mathcal{D}x_2 e^{-\int_{-\infty}^\infty dt\left[-\frac{m}{2}x_1\ddot{x}_1 - \frac{m}{2}x_2\ddot{x}_2 + \frac{1}{2}m\omega^2 (x_1^2 + x_2^2) + \lambda x_1 x_2\right] - \frac{m}{2}\left(x_1^- \dot{x}_1^- - x_1^+ \dot{x}_1^+ 
    + x_2^- \dot{x}_2^- - x_2^+ \dot{x}_2^+\right)}
\end{split}
\end{align}
The final terms in the exponent are boundary terms from integrating by parts in the kinetic term of the Euclidean action. The partial trace over $x_2$ removes the boundary terms and allows us to simply integrate the Gaussian path integral over $x_2$. Define the propagator $G$ as
\begin{equation}
    G(t-t') = \frac{1}{-\partial_t^2 + \omega^2} = \frac{1}{2\omega} e^{-\omega |t-t'|}
\end{equation}
Then it is not hard to show that the reduced density matrix is given by
\begin{equation}
    \mel{x^-}{\rho_1}{x^+} = \frac{1}{Z'} \int \mathcal{D}x e^{-\int_{-\infty}^\infty dt\left(-\frac{m}{2} x(t)\ddot{x}(t) + \frac{1}{2}m\omega^2 x^2 -\frac{\lambda^2}{m} \int_{-\infty}^\infty dt' x(t) G(t-t')x(t') \right) - \frac{m}{2}\left(x^- \dot{x}^- - x^+ \dot{x}^+\right)}
\end{equation}
Notice in the last term of the action that we have a non-local quadratic interaction of the form
\begin{equation}
    \int_{t, t'} x(t) G(t-t') x(t').
\end{equation}
Other than this non-local term, we simply have the usual elements for the  preparation of a harmonic oscillator ground state, and so without this non-local term we would find that $\rho$ is pure (and, in particular, the ground state of a single harmonic oscillator). 

Because this path integral is Gaussian, we can now explicitly compute the density matrix by performing a shift of variables of the form
\begin{equation}
    x \to x_{cl} + \eta
\end{equation}
where $x_{cl}$ satisfies the ``effective equation of motion'' 
\begin{equation}
    -m \ddot{x}_{cl}(t) + m\omega^2 x_{cl}(t) - \frac{\lambda^2}{m}\int dt' G(t-t')x_{cl}(t') = -m (x^+-x^-) \delta'(t) - m(\dot{x}^+ - \dot{x}^-) \delta(t)
\end{equation}
The two delta functions on the right hand side of the equation enforce the boundary conditions $\lim_{t\to 0^\pm} x(t) = x^\pm $. In particular, the first delta function enforces the strength of the discontinuity of $x(t)$ at $t = 0$ while the second determines the strength of the discontinuity in the first derivative. The derivative discontinuity is not determined by our boundary conditions, but will be given once we enforce the equations of motion. With this shift, $\eta$ no longer depends on the boundary conditions of the path integral, and so one finds
\begin{equation}\label{reduced-density-matrix-form}
    \mel{x^-}{\rho_1}{x^+} = \mathcal{N} e^{-\frac{m}{2}\left(x^- \dot{x}_{cl}^- - x^+ \dot{x}_{cl}^+\right)}
\end{equation}
It remains to compute $\dot{x}_{cl}$ at $t\to 0^{\pm}$. Fourier transforming from $t$ to $p$, the effective equation of motion becomes
\begin{equation}
    p^2 x_{cl}(p) +  \omega^2 x_{cl}(p) - \frac{\lambda^2}{m^2}\frac{1}{p^2+m^2}x_{cl}(p) = -ip (x^+-x^-) - (\dot{x}_{cl}^+ - \dot{x}_{cl}^-)
\end{equation}
and so we obtain the momentum space solution
\begin{equation}
    x_{cl}(p) = \frac{-ip (x^+-x^-) - (\dot{x}_{cl}^+ - \dot{x}_{cl}^-)}{p^2 + \omega^2 - \frac{\lambda^2}{m^2}\left(\frac{1}{p^2+m^2}\right)}.
\end{equation}
For convenience, let us define
\begin{equation}
    \Delta x = x^+ - x^-,\hspace{1cm} \qquad \Delta \dot{x} = \dot{x}_{cl}^+ - \dot{x}_{cl}^-,
\end{equation}
and
\begin{equation}
    \omega_\pm = \sqrt{\omega^2 \pm \lambda/m}.
\end{equation}
Then, the classical solution after inverse Fourier transforming is
\begin{equation}
    x_{cl}(t) = \frac{\text{sgn}(t)}{4}\Delta x\left(e^{-\omega_- \abs{t}} + e^{-\omega_+ \abs{t}}\right) - \frac{1}{4}\Delta \dot{x}\left(\frac{e^{-\omega_- \abs{t}}}{\omega_-} + \frac{e^{-\omega_+\abs{t}}}{\omega_+}\right)
\end{equation}
By enforcing the condition $\lim_{t\to 0^\pm} x_{cl}(t) = x^\pm$, we find that $\Delta \dot{x} = -\frac{2\omega_- \omega_+}{\omega_-+\omega_+}(x^- + x^+)$, and so as a function of the boundary conditions, we obtain
\begin{align}
    \dot{x}_{cl}^+ &= \frac{1}{4(\omega_+ + \omega_-)}\left[x^-\left(\omega_+ - \omega_-\right)^2 - x^+ \left(\frac{\omega_-^2 + 6 \omega_+\omega_- + \omega_+^2}{\omega_+ + \omega_-}\right)\right]\\
    \dot{x}_{cl}^- &= \frac{1}{4(\omega_+ + \omega_-)}\left[-x^+\left(\omega_+ - \omega_-\right)^2 + x^- \left(\frac{\omega_-^2 + 6 \omega_+\omega_- + \omega_+^2}{\omega_+ + \omega_-}\right)\right].
\end{align}
Plugging our results back into (\ref{reduced-density-matrix-form}), we obtain the density matrix
\begin{equation}
    \mel{x^-}{\rho_1}{x^+} = \mathcal{N} \exp[-\frac{m}{8(\omega_-+\omega_+)}\left((\omega_-^2 + 6\omega_+\omega_- + \omega_+^2)((x^-)^2 + (x^+)^2) - 2(\omega_+ - \omega_-)^2x^+ x^-\right)]
\end{equation}
This density matrix is indeed mixed unless $\lambda = 0$. Moreover, one can verify via standard canonical techniques that this is the correct reduced density matrix. 

In this calculation, the non-local term in the reduced density matrix path integral arose from the partial trace over the $x_2$ particle, and so it was clear that its contribution would lead to a mixed density matrix. However, we can now replace the particular kernel $G(t-t')$ with an arbitrary (reflection symmetric) bilocal function $\lambda(t, t')$, leading to the path integral
\begin{equation}
    \mel{x^-}{\rho_\lambda}{x^+} = \frac{1}{Z} \int \mathcal{D}x \exp(-\int dt \left(\frac{1}{2}m\dot{x}(t)^2+ \frac{1}{2}m\omega^2 x(t)^2 - \int dt' \lambda(t,t')x(t')\right)).
\end{equation}
This allows us to prepare a more general class of mixed states, and for all such path integrals it is convenient to regard $\lambda(t, t')$ as a non-local interaction of some effective action, rather than simply as the result of some convex combination of pure state path integrals. 

\section{Computation of Boundary Terms}\label{app:computationboundary}

In this appendix, we will explicitliy compute the boundary terms B.T defined implicitly in (\ref{implicit-bt-def}) and show that it is given by (\ref{bt}). Our starting point is the expression
\begin{equation}
    -\frac{1}{2}\phi * G^{-1}* G' * G^{-1}*\phi
\end{equation}
where we recall that $G^{-1}$ is the regulated Laplacian
\begin{eqnarray}
    G^{-1} = -K^{-1}(-\nabla^2/\Lambda^2) * \partial^2.
\end{eqnarray}
In this expression, each of the $G^{-1}$ factors is a differential operator acting to the right. In our derivation, we ``symmetrize'' this expression by moving the action of the leftmost $G^{-1}$ to the left. 
For now, we assume that the path integral has two boundaries located at $t =\pm \epsilon$. Later, we will take the limit $\epsilon \to 0$ left implicit in all expressions. Thus, only terms with time derivatives get non-trivial contributions from integrating by parts. To calculate the terms which appear, it is convenient to work in (spatial) momentum space, where our expression reads
\begin{eqnarray}
    &&-\frac{1}{2}\phi * G^{-1}* G' * G^{-1}*\phi \nn\\
    &&= -\int d^{d-1}\mathbf{p} dt dt' \frac{K^{-2}(\mathbf{p}^2/\Lambda^2)}{2(2\pi)^{d-1}}\left(\phi_{-\mathbf{p}}(t) (\mathbf{p}^2 - \partial_t^2) G'(t-t'; \mathbf{p})(\mathbf{p}^2 - \partial_{t'}^2)\phi_{\mathbf{p}}(t')  \right)
\end{eqnarray}
where we have used a (time-dependent) spatial mode decomposition
\begin{equation}
    \phi(\mathbf{x}, t) = \int \frac{d^{d-1}\mathbf{p}}{(2\pi)^{d-1}} \phi_{\mathbf{p}}(t) e^{-i\mathbf{p}\cdot \mathbf{x}}
\end{equation}
and the temporally non-local kernel is given by
\begin{equation}
    G'(t-t';\mathbf{p}) = \frac{\Lambda\frac{d}{d\Lambda}K(\mathbf{p}^2/\Lambda^2)}{2\sqrt{\mathbf{p}^2}} e^{-\sqrt{\mathbf{p}^2}|t-t'|}
\end{equation}
Because the integrand is local in momentum space, we can ignore the spatial integrals and focus only on the integrals over $t$ and $t'$. That is, for a single momentum mode, we have the expression
\begin{equation}
    -\frac{1}{2}K^{-2}(\mathbf{p}^2/\Lambda^2) \int dt dt' \left(\phi_{-\mathbf{p}}(t) (\mathbf{p}^2 - \partial_t^2) G'(t, t'; \mathbf{p})(\mathbf{p}^2 - \partial_{t'}^2)\phi_{\mathbf{p}}(t')  \right)\ .
\end{equation}
With this observation in mind, define
\begin{equation}
    \text{B.T.} = \int \frac{d^{d-1}\mathbf{p}}{(2\pi)^{d-1}} f(\mathbf{p})
\end{equation}
Integrating by parts twice in the $t$ integral then yields the explicit expression for $f$, 
\begin{align}
\begin{split}
    f(\mathbf{p})&=\frac{1}{2}K^{-2}(\mathbf{p}^2/\Lambda^2) \int dt' \Big[\big(\phi_{-\mathbf{p}}(-\epsilon) \partial_t G'(-\epsilon-t';\mathbf{p}) - \phi_{-\mathbf{p}}(\epsilon) \partial_t G'(\epsilon- t';\mathbf{p}) \\&\hspace{4cm}- \dot{\phi}_{-\mathbf{p}}(-\epsilon) G'(-\epsilon- t';\mathbf{p}) +\dot{\phi}_{-\mathbf{p}}(\epsilon) G'(\epsilon- t';\mathbf{p})\big) (\mathbf{p}^2 - \partial_{t'}^2) \phi_{\mathbf{p}}(t')\Big],
    \end{split}
\end{align}
where we have implemented above the notation $\dot{\phi}(t) = \partial_t \phi(t)$. We have generated four boundary terms from this procedure, since we have two boundaries and integrated by parts twice. However, this expression retains one ``bulk'' integral, from the coordinate $t'$. We can now integrate by parts twice on $t'$ to move the operator $\mathbf{p}^2 - \partial_{t'}^2$ onto $G'$. Because $G'$ is proportional to the propagator, this results in a delta function. The bulk terms are therefore proportional to the integral
\begin{align}
\begin{split}
    &\int dt' \Big[\big(\phi_{-\mathbf{p}}(-\epsilon) \partial_t \delta(-\epsilon-t') - \phi_{-\mathbf{p}}(\epsilon) \partial_t \delta(\epsilon-t')  - \dot{\phi}_{-\mathbf{p}}(-\epsilon) \delta(-\epsilon-t')  +\dot{\phi}_{-\mathbf{p}}(\epsilon) \delta(\epsilon-t') \big)  \phi_{\mathbf{p}}(t')\Big]\\
    &\hspace{2cm}=\phi_{-\mathbf{p}}(-\epsilon) \dot{\phi}_{\mathbf{p}}(-\epsilon) - \phi_{-\mathbf{p}}(\epsilon) \dot{\phi}_{\mathbf{p}}(\epsilon)  - \dot{\phi}_{-\mathbf{p}}(-\epsilon)\phi_{\mathbf{p}}(-\epsilon) +\dot{\phi}_{-\mathbf{p}}(\epsilon)\phi_{\mathbf{p}}(\epsilon)  =0
\end{split}
\end{align}
Note that the final expression is not technically zero, but will evaluate to zero inside of the remaining momentum integral, since all momentum space kernels in the expression are symmetric under a $\mathbf{p}\to -\mathbf{p}$ reflection.

So, as one might expect, we can remove all bulk time integrals and write B.T purely as a sum of boundary terms. However, before writing these terms out, we must address an order of limits issue. In obtaining boundary terms from the $t'$ integral, we will introduce additional limits of $t'\to 0$. To distinguish this limit from those coming from the first $t$ integral, we use $\epsilon'$ to denote the limit. Then, in expressions of the form $\phi(\pm \epsilon)\dot{\phi}(\pm\epsilon ')$ or $\phi(\pm \epsilon ')\dot{\phi}(\pm\epsilon)$ there is an ambiguity in the time order, depending on the order in which the two limits are taken\footnote{There is no ambiguity in the product $\phi(\pm\epsilon) \dot{\phi}(\mp\epsilon ')$ or similar insertions since the two factors live on different time slices.}. We single out these products in particular, because as operators they will ultimately correspond to products of $\hat{\varphi}$ and $\hat{\pi}$, and so we must make sure they are inserted in the correct order. To resolve the potential ambiguity, we must note that the order in which we perform the integration by parts provides a specification of the order in which we perform the limits. So, the limit should always be performed first on $\epsilon$ and then on $\epsilon'$. Using the notation of $0^\pm$ to denote an infinitesimal positive or negative quantity, we have, of course, that
\begin{equation}
    \pm\epsilon = \pm \epsilon' = 0^{\pm}
\end{equation}
The stated order of limits also resolves the difference of $\epsilon$ and $\epsilon'$, which should be taken as
\begin{eqnarray}
    \epsilon - \epsilon' = 0^-
\end{eqnarray}

In writing the result of the final two integration by parts, we will drop the momentum dependence of our functions, which can be straightforwardly added back in at the end of our computation. One then finds
\begin{align}
    \begin{split}
        -2f(\mathbf{p}) =& \phi(-\epsilon)D_tG'(0^+)D_{t'}\phi(-\epsilon') -  \phi(-\epsilon)D_tG'(0^-) D_t\phi(\epsilon)
    \\&- \phi(-\epsilon) D_tD_{t'}G'(0) \phi(-\epsilon) + \phi(-\epsilon) D_t D_{t'}G'(0)  \phi(\epsilon) \\
    &- \phi(\epsilon)D_tG'(0^+)D_t\phi(-\epsilon) +\phi(\epsilon) D_tG'(0^-)D_t\phi(\epsilon')
    \\&+\phi(\epsilon)D_tD_{t'}G'(0) \phi(-\epsilon) -\phi(\epsilon)  D_t D_{t'}G'(0) \phi(\epsilon) \\
    &- D_t\phi(-\epsilon) G'(0) D_t\phi(-\epsilon) +   D_t\phi(-\epsilon)G'(0) D_t\phi(\epsilon)
   \\& + D_t\phi(-\epsilon)  D_{t'}G'(0^+)  \phi(-\epsilon') -D_t\phi(-\epsilon)  D_{t'}G'(0^-)  \phi(\epsilon)\\
    &+ D_t\phi(\epsilon)G'(0) D_t\phi(-\epsilon) -  D_t\phi(\epsilon)G'(0) D_t\phi(\epsilon)
    \\&-D_t\phi(\epsilon) D_{t'}G'(0^+) \phi(-\epsilon) + D_t\phi(\epsilon)  D_{t'}G'(0^-) \phi(\epsilon')
    \end{split}
\end{align}
Note we have used the notation $D_t = K^{-1}\partial_t$. In terms where the time order of expressions in unambiguous or unimportant, we have simply set $\epsilon = \epsilon'$. Also, only the first time derivative of $G'$ depends on the direction of approach of $t\to 0$, and for all other terms we have simply taken the limit in the argument of $G'$. 

Finally, define the following momentum space functions:
\begin{align}\label{Deltasformula}
    \Delta_0 &= G'(0; \mathbf{p})\nn\\
    \Delta_1 &= K^{-1}(\mathbf{p}^2/\Lambda^2)\partial_tG'(0^+ ; \mathbf{p})\nn\\
    \Delta_2 &= K^{-2}(\mathbf{p}^2/\Lambda^2) \partial_t\partial_{t'}G'(0, ;\mathbf{p})
\end{align}
In identifying $\Delta_1$ in our expressions, we have the following identities,
\begin{equation}
    \partial_t G'(0^+; \mathbf{p})=-\partial_{t'} G'(0^+; \mathbf{p}) = -\partial_{t} G'(0^-; \mathbf{p}) = \partial_{t'} G'(0^-; \mathbf{p}).
\end{equation}
Then, $f(\mathbf{p})$ may be written as
\begin{align}
    \begin{split}
        -2f(\mathbf{p}) =& K^{-2}\Delta_0\left(2\dot{\phi}(\epsilon)\dot{\phi}(-\epsilon) - \dot{\phi}(-\epsilon)\dot{\phi}(-\epsilon) - \dot{\phi}(\epsilon)\dot{\phi}(\epsilon)\right)\\&+K^{-1}\Delta_1\Big(\phi(-\epsilon) \dot{\phi}(-\epsilon')+\phi(-\epsilon) \dot{\phi}(\epsilon)- \phi(\epsilon) \dot{\phi}(-\epsilon) - \phi(\epsilon) \dot{\phi}(\epsilon') \\&\hspace{2cm}- \dot{\phi}(-\epsilon) \phi(-\epsilon') - \dot{\phi}(-\epsilon) \phi(\epsilon) + \dot{\phi}(\epsilon) \phi(-\epsilon) + \dot{\phi}(\epsilon) \phi(\epsilon')\Big)\\
    & + \Delta_2\left(2\phi(\epsilon) \phi(-\epsilon) - \phi(-\epsilon) \phi(-\epsilon) - \phi(\epsilon) \phi(\epsilon)\right)
    \\
    & 
    \end{split}
\end{align}
B.T. is then readily recovered by integrating over all spatial momenta, resulting in equation (\ref{bt}).

\section{Massive ground state flow} \label{app:massive-flow}

In Section \ref{sec:gaussian-flows}, we claimed that the massive ground state flows are commensurate with the Polchinski equation for Gaussian sources. Here we will compute the flow of the massive free field ground state directly from the Polchinski equation and verify this claim. As in the computation of Gaussian wavefunctionals, we may ignore the flow of the field independent terms as they only contribute to the normalization, which is fixed. If we take an ansatz for $S_I$ of the form
\begin{equation}
    S_I = \frac{1}{2}\phi * B * \phi,
\end{equation}
then up to said field independent terms, the Polchinski equation becomes
\begin{equation}
    -\Lambda \frac{d}{d\Lambda}B = -B*G' * B.
\end{equation}
Let us also take the simplyfing assumption that $B$ is translation invariant and reflection symmetric, so that it may be written in momentum space as $B(p, \Lambda)$ with $B(-p, \Lambda) = B(p, \Lambda)$. Then, using the definition of $G'$, we have
\begin{equation}
    \frac{\frac{d}{d\Lambda}B(p, \Lambda)}{B(p, \Lambda)^2} = \frac{\frac{d}{d\Lambda}K(\mathbf{p}^2/\Lambda^2)}{p^2}.
\end{equation}
We can directly integrate this expression from $\Lambda_0$ to a final $\Lambda$ to obtain the solution
\begin{align}
    \frac{1}{B(p, \Lambda_0) } - \frac{1}{B(p, \Lambda)} = \frac{K(\mathbf{p}^2/\Lambda^2) - K(\mathbf{p}^2/\Lambda_0^2)}{p^2}
\end{align}
For convenience let us write $\delta K = K(\mathbf{p}^2/\Lambda^2) - K(\mathbf{p}^2/\Lambda_0^2)$. Then we may rearrange to find
\begin{align}
    B(p, \Lambda) = \frac{p^2 B(p, \Lambda_0)}{p^2 - B(p, \Lambda_0)\delta K}
\end{align}
If we start with the massive theory at scale $\Lambda_0$, we simply have $B(p, \Lambda_0) = m^2$, and so 
\begin{align}
    B(p, \Lambda) = \frac{p^2 m^2}{p^2 - m^2 \delta K} = m^2 + \frac{m^4\delta K}{p^2 - m^2 \delta K}
\end{align}
Now we use this solution to compute the ground state wavefunctional as a function of scale. The path integral is given by
\begin{align}
\begin{split}
    \mel{\varphi_-}{\rho}{\varphi_+} &= \int_{\phi(t= 0^-) = \varphi_-}^{\phi(t = 0^+) = \varphi_+} \mathcal{D}\phi \text{exp}\Bigg(\frac{1}{2}\Big[ \phi * K^{-1} * \partial^2 \phi - \phi*B(\Lambda)*\phi \\&\hspace{5cm}-\varphi_- \cdot K^{-1}\cdot \partial_t \phi_- +  \varphi_+ \cdot K^{-1}\cdot \partial_t \phi_+\Big]\Bigg)
    \end{split}.
\end{align}
Now we shift the field $\phi \to \phi_c + \eta$ where $\phi_c$ satisfies the classical equation of motion
\begin{align}\label{classical-eom}
    -K^{-1}\partial^2\phi_c(x) + \int d^dy B(x, y; \Lambda) \phi_c(y) = -K^{-1}\left((\varphi_+(\mathbf{x}) - \varphi_-(\mathbf{x}))\delta'(t) + c(\mathbf{x}) \delta(t)\right)
\end{align}
or more simply in momentum space,
\begin{align}
    K^{-1}p^2 \phi_c(p) + B(p, \Lambda) \phi_c(p) = -K^{-1}\left(ip_t (\varphi_+(\mathbf{p}) - \varphi_-(\mathbf{p}) + c(\mathbf{p})\right)
\end{align}
The right hand side in this expression enforces the discontinuous boundary conditions at $t = 0$; the first delta function sets the discontinuity of $\phi$ while the second sets the discontinuity of its derivative $\partial_t \phi$. We have left this in terms of an arbitrary function $c(\mathbf{p})$ to be determined later. With this shift, the density matrix becomes
\begin{align}\label{shifted-wf}
    \mel{\varphi_-}{\rho}{\varphi_+} = \mathcal{N} \exp(-\frac{1}{2}\int_{\mathbf{p}}\left(\varphi_- (\mathbf{p}) K^{-1}(\mathbf{p}^2/\Lambda^2) \partial_t \phi_{c,-}(\mathbf{p})-\varphi_+ (\mathbf{p}) K^{-1}(\mathbf{p}^2/\Lambda^2) \partial_t \phi_{c,+}(\mathbf{p})\right))
\end{align}
with $\mathcal{N}$ an overall normalization which is independent of the boundary conditions. It remains to solve the classical equations of motion and compute the first derivatives at the boundary. Fourier transforming equation (\ref{classical-eom}), we find
\begin{align}
    \phi_c(t, \mathbf{p}) &= -\int \frac{dp_t}{2\pi} \frac{ip_t(\varphi_+ - \varphi_-) + c}{K \left(K^{-1} p^2 + B(p, \Lambda)\right)} e^{i p_t t}\\
    &= -\int \frac{dp_t}{2\pi} \frac{\left(ip_t(\varphi_+ - \varphi_-) + c\right)(p^2 - m^2 \delta K)}{p^2\left(p^2 + K_0 m^2\right)} e^{i p_t t}.
\end{align}
where we have dropped the functional dependence on the momentum, to lighten the notation. We have also introduced $K_0 = K(\mathbf{p}^2/\Lambda_0^2)$ to denote the initial regulator. 

It is not hard to perform the integral over $p_t$ to obtain the explicit solution for $\phi_c(t)$. It is convenient to define $\omega_{m} = \sqrt{\mathbf{p}^2 + K_0 m^2},$ and as before write $\omega_0 = \sqrt{\mathbf{p}^2}$. Then the solution takes the form
\begin{equation}
    \phi_c(t, \mathbf{p}) = \frac{Ke^{-\omega_{m} |t|}}{2K_0}\left(\text{sgn}(t)(\varphi_+ -\varphi_-) - \frac{c}{\omega_{m}}\right) - \frac{(K - K_0)e^{-\omega_0 |t|}}{2K_0}\left(\text{sgn}(t)(\varphi_+ - \varphi_-)-\frac{c}{\omega_0} \right)
\end{equation}
Since the strength of the discontinuity in $\phi$ at $t = 0$ has already been set, $c(\mathbf{p})$ is determined by requiring 
\begin{equation}
    \lim_{t\to 0^+}\phi_c(t, \mathbf{p}) = \varphi_+(\mathbf{p}).
\end{equation}
which yields
\begin{align}
    c = \frac{\omega_0 (\varphi_+ + \varphi_-)}{\frac{K}{K_0}\left(1 - \frac{\omega_0}{\omega_{m}}\right) - 1}
\end{align}

With $c(\mathbf{p})$ determined we have the full solution for $\phi_c(t, \mathbf{p})$, and it only remains to compute the terms in the exponential of (\ref{shifted-wf}). The rather messy result is
\begin{align}
    &\frac{1}{2K}\left(\varphi_- \partial_t \phi_{c, -} - \varphi_+ \partial_t \phi_{c, +}\right)\\
    &\hspace{1cm}= \frac{K(K-K_0) (\varphi_- -\varphi_+)^2 \left(2 \omega_0 \left(\omega_0-\omega_{m}\right)+K_0 m^2\right)-2K_0^2 \omega_0 \left(\varphi_+^2+\varphi_-^2\right) \omega_{m}}{4 K K_0 \left(K\left(\omega_{m}-\omega_0\right)-K_0 \omega_{m}\right)}
\end{align}
We can clean things up a bit by writing the final answer in the form
\begin{align}
     \mel{\varphi_-}{\rho}{\varphi_+} = \mathcal{N} \exp(-\frac{1}{2}\left(\varphi_+ \cdot c_1 \cdot \varphi_+ + \varphi_- \cdot c_1 \cdot \varphi_-\right) + \varphi_- \cdot c_2 \cdot \varphi_+)
\end{align}
and defining the functions 
\begin{align}
    c_\pm = c_1\pm c_2
\end{align}
Then the final result of our computation yields
\begin{align}
    c_+ &= \frac{\omega_0}{K} - \frac{\omega_0}{K_0} + \frac{\omega_{m}}{K_0}\\
    c_- &= \frac{K_0\omega_0}{K\left(K_0 + K\left(\frac{\omega_0}{\omega_{m}}-1\right)\right)}
\end{align}
which matches precisely our result in (\ref{lowercs}). 

\section{Perturbative Calculations}\label{app:perturbative}

\subsection{Diagrammatic rules for states}

In this section, we will detail the ``spatial'' diagrammatic rules for a small coupling perturbative expansion to set the stage for the calculation of the ERG flow of the density matrix in perturbation theory. For our application, we only use the $\lambda \phi^4$ Lagrangian, but much of our discussion generalizes straightforwardly. Our starting point is the regulated Euclidean Lagrangian,
\begin{equation}
    \mathcal{L} = \int d^{d-1}\mathbf{x} \left[-\frac{1}{2}\phi(t, \mathbf{x})K^{-1}(-\nabla^2/\Lambda^2) \partial^2 \phi(t, \mathbf{x})  + \frac{1}{4!}\lambda \phi(t, \mathbf{x})^4\right].
\end{equation}
The ground state will be prepared by a half-space Euclidean path integral with corresponding action
\begin{equation}
    S_E = S_0 + \int_{-\infty}^0 V[\phi(t, \mathbf{x})] dt
\end{equation}
where we have defined the perturbative splitting
\begin{gather}
    S_0 = \int_{-\infty}^0dt \int d^{d-1}\mathbf{x} \left[-\frac{1}{2}\phi(t, \mathbf{x})K^{-1}(-\nabla^2/\Lambda^2) \partial^2 \phi(t, \mathbf{x})\right]\\
    V[\phi] = \frac{\lambda}{4!}\int d^{d-1}\mathbf{x} \phi(t, \mathbf{x})^4
\end{gather}
In order to get the correct normalization of the state, it is necessary to ``regulate'' the lower bound of the time integral of $V$, and so the path integral preparing the ground state may be written as 
\begin{equation}
    \ket{\Omega_{\lambda}} = \lim_{T\to \infty}\mathcal{N}_T\int \mathcal{D}\phi \, e^{-S_0-\int_{-T}^0 Vdt}
\end{equation}
where $\phi$ has an unspecified fixed boundary condition at $t = 0$. The normalization $\mathcal{N}_T$ is chosen so that the state has norm $1$ for all $T>0$\footnote{As we will see, some terms in the perturbative expansion are naively divergent in the $T\to \infty$ limit. However, these divergences are only intermediate quantities which ultimately cancel upon appropriately normalizing, and as such have no physical meaning. These divergences should not be confused with the UV divergences appearing from loop diagrams.}.
Expanding in small $\lambda$, one obtains
\begin{equation}
    \ket{\Omega_{\lambda}} = \mathcal{N} \int \mathcal{D}\phi \left(\sum_{n=0}^\infty \frac{(-1)^n\lambda^n}{4^n n!}\left(\int dt d^{d-1}\mathbf{x} \phi(t, \mathbf{x})^4\right)^n\right)e^{-S_0}
\end{equation}
where $S_0$ denotes the massless free field theory action. Each term in this sum can be treated as a Euclidean ordered operator insertion acting on the ground state, so we may write
\begin{align}
    \ket{\Omega_{\lambda}} &= \mathcal{N} \left(\sum_{n=0}^\infty \frac{(-1)^n\lambda^n}{(4!)^n n!} \int_{-T}^0 dt_1\cdots \int_{-T}^0  dt_n \mathcal{T}\left[\hat{\varphi}_1(t_1)^4 \cdots \hat{\varphi}_n(t_n)^4\right]\right)\ket{\Omega_0}\\
    &= \mathcal{N} \left( \sum_{n=0}^\infty \frac{(-1)^n\lambda^n}{(4!)^n} \int_{-T}^0 \int_{-T}^{t_n}\cdots \int_{-T}^{t_2} \left(\hat{\varphi}_n(t_n)^4 \hat{\varphi}_{n-1}(t_{n-1})^4 \cdots \hat{\varphi}_1(t_1)^4\right) dt_1\cdots dt_{n-1}dt_n \right)\ket{\Omega_0}
\end{align}
where we have used a subscript on $\hat{\varphi}$ to denote the spatial dependence of each insertion. We can remove the time dependence from each operator insertion by using the Euclidean time evolution operator. Writing $H$ for the regulated free Hamiltonian, we may rewrite 
\begin{equation}
    \hat{\varphi}_n(t_n)^4\cdots \hat{\varphi_n}(t_1)^4\ket{\Omega_0} = e^{-t_nH}\hat{\varphi}_n^4 e^{-(t_{n-1}-t_n)H} \hat{\varphi}_{n-1}^4\cdots\hat{\varphi}_{2}^4 e^{-(t_1-t_2)H}\hat{\varphi}_1^4 \ket{\Omega_0}
\end{equation}
Finally, each spatially local term can be written in momentum space where the spatial integrals may be traded for a momentum conserving delta function:
\begin{equation}
    \hat{\varphi}_i^4 = \int_{\mathbf{p}_1, \mathbf{p}_2, \mathbf{p}_3, \mathbf{p}_4}\hat{\varphi}(\mathbf{p}_1)\hat{\varphi}(\mathbf{p}_2)\hat{\varphi}(\mathbf{p}_3)\hat{\varphi}(\mathbf{p}_4) \delta^{(d-1)}(\mathbf{p}_1 + \mathbf{p}_2 + \mathbf{p}_3 + \mathbf{p}_4)
\end{equation}
This leads us to an expansion very similar to standard Feynman diagrams, but where notably we must take care to preserve the Euclidean time ordering, and each vertex conserves only the spatial momentum. To formalize this, let us now expand each field operator $\hat{\varphi}(\mathbf{p})$ in creation/annihilation operators via
\begin{equation}\label{field-expansion}
    \hat{\varphi}(\mathbf{p}) = \sqrt{\frac{K(\mathbf{p})}{2\omega(\mathbf{p})}}(a(\mathbf{p}) + a^\dagger(-\mathbf{p}))
\end{equation}
By expanding out all the resulting terms, we can construct diagrams by contracting operators using Wick's theorem. Any term with an excess of annihilation operators after contractions will evaluate to zero when acting on the unperturbed ground state, and so we only consider terms where the number of creation operators is greater than or equal to the number of annihilation operators. The only additional subtlety which must be addressed are the Hamiltonian factors $e^{Ht}$ appearing between each vertex insertion. To handle these, it is helpful to first consider an example.

Consider the following term which appears in the expansion at second  order:
\begin{equation}
    \left(\int_{-T}^{0} dt_2 e^{Ht_2}a^\dagger_8 a^\dagger_7 a_6 a_5 \left(\int_{-T}^{t_2}dt_1 e^{Ht_1} a_4 a^\dagger_3 a^\dagger_2 a^\dagger_1 \right)\right) \ket{\Omega_0}
\end{equation}
Here we are denoting the momentum dependence in the subscript. We commute $a_4$ all way to the right to produce the state:
\begin{equation}
    \left(\int_{-T}^{0} dt_2 e^{Ht_2}a^\dagger_8 a^\dagger_7 a_6 a_5 \left(\delta_{34}\int_{-T}^{t_2}dt_1 e^{Ht_1} a^\dagger_2 a^\dagger_1 \right)\right) \ket{\Omega_0} + \cdots
\end{equation}
where the $\cdots$ contain the other operator contractions. Now, the state to the right of $e^{Ht_1}$ is an eigenstate of $H$, and so we convert the expression to
\begin{equation}
    \left(\int_{-T}^{0} dt_2 e^{Ht_2}a^\dagger_8 a^\dagger_7 a_6 a_5 \left(\delta_{34}\int_{-T}^{t_2}dt_1 e^{E_{21}t_1} a^\dagger_2 a^\dagger_1 \right)\right) \ket{\Omega_0} + \cdots
\end{equation}
where $E_{21} = \omega(\mathbf{p}_1)+\omega(\mathbf{p}_2)$. We can now evaluate the $t_1$ integral, which yields 
\begin{equation}
    \int_{-T}^{t_1} dt_1 e^{E_{21}t_1} = \frac{1}{E_{21}} + \mathcal{O}(1/T).
\end{equation}
Note that because $E_{21}$ is non-zero, we can simply take the $T\to\infty$ limit immediately. However, in the case where the state is the vacuum and $E=0$, we must keep $T$ finite and carry it through the computation. The resulting term will be divergent for large $T$, but one can check order by order that all such terms cancel after normalizing the state. 

Finally, plugging back in, the form of the state becomes
\begin{equation}
    \cdots \frac{1}{E_{21}}\delta_{34}\left(\int_{-T}^{t_3} dt_2 e^{Ht_2}a^\dagger_8 a^\dagger_7 a_6 a_5  a^\dagger_2 a^\dagger_1 \right) \ket{\Omega_0}
\end{equation}
We are now free to commute $a_5$ and $a_6$ to the right, producing additional contractions. Finally, after this we can convert the factor of $e^{Ht_2}$ to its corresponding eigenvalue and perform the $t_2$ integral. For more general diagrams at higher order, this procedure then carries out iteratively to the left until we have exhausted all the time integrals. 

This procedure of contractions and evaluation of the time integrals from right to left can be summarized by the following diagrammatic rules:

\begin{enumerate}
    \item For a diagram at order $n$ in perturbation theory, first draw $n$ disconnected interaction vertices arranged from right to left. The diagrams represent operator insertions to be acted on the state $\ket{\Omega_0}$
    \item A particular diagram is generated by connecting pairs of legs to form a contraction. Not all legs have to be contracted. 
    \item To each leg, associate a spatial momentum. Each vertex enforces conservation of spatial momentum for its connected legs. For each internal leg, insert a factor of $\frac{K}{2\omega}$, and for each external leg, insert a factor of $\sqrt{\frac{K}{2\omega}}a^\dagger(\mathbf{p})$.
    \item To evaluate the energy factor, starting from the rightmost vertex and repeating from right to left for all vertices, consider the sub-diagram given by removing all vertices to the left. Any contraction to a removed vertex should be replaced with an external leg with the same momentum. Compute the energy $E_{(k)}$ of the state corresponding to the $k$-th sub-diagram. The diagram gets an overall factor of
    \begin{equation}
        \int_{-T}^0 dt_n \int_{-T}^{t_n}dt_{n-1} \cdots \int_{-T}^{t_2}dt_1\left[e^{t_n\left(E_{(n)}- E_{(n-1)}\right)}e^{t_{n-1}\left(E_{(n-1)}- E_{(n-2)}\right)}\cdots e^{t_2\left(E_{(2)} - E_{(1)}\right)} e^{t_1 E_{(1)}}\right]
    \end{equation}
    In the case that all the $E_{(k)}$ are non-zero, one may take the $T \to \infty$ limit immediately, in which case the above integral simplifies to
    \begin{equation}
       \prod_{k=1}^n\frac{1}{E_{(k)}}.
    \end{equation}
    However, if any of the $E_{(k)}$ are zero, the order of limits is to first take the energies to zero and then $T$ to infinity. Such terms will have power-law divergences in $T$, which are all ultimately cancelled by the normalization. 
    \item Multiply by a factor of $(-1)^n \lambda^n/(4!)^n$ as well as the overall multiplicity of the diagram. Integrate over all momenta (including external legs).
    \item To construct the ground state, sum over all diagrams to a given order in perturbation theory, and then compute the normalization $\mathcal{N_T}$ as a sum of vacuum diagrams. Finally, take the limit $T\to \infty$.
\end{enumerate}

The slicing procedure in step 4 retroactively implements the iterative replacement of energy factors we discussed above. The multiplicity of a given diagram may be determined via Wick's theorem.

Let us now use these rules for some explicit calculations. At first order in $\lambda$, there are only three diagrams contributing to the ground state, which are explicitly given by
\begin{align}\label{first-order-4pt-expression}
    \includegraphics[, valign = c]{figs/firstorder4pt.pdf} \quad &= -\frac{\lambda}{4!}\int_{\mathbf{p}_1, \mathbf{p_2}, \mathbf{p}_3, \mathbf{p}_4} \frac{\delta^{(d-1)}(\sum_{i=1}^4 \mathbf{p}_i)}{\sum_{i=1}^4\omega(\mathbf{p}_i)} \left(\prod_{i=1}^4\sqrt{\frac{K(\mathbf{p}_i)}{2\omega(\mathbf{p}_i)}} a^\dagger(\mathbf{p}_i)\right) \\\label{first-order-2pt-expression}
    \includegraphics[, valign = c]{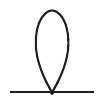}\quad&= -\frac{\lambda}{4} \int_{\mathbf{p}, \mathbf{q}} \left(\frac{1}{2\omega(\mathbf{p})}\right)\frac{K(\mathbf{q})}{2\omega(\mathbf{q})}\frac{K(\mathbf{p})}{2\omega(\mathbf{p})}a^\dagger(\mathbf{p})a^\dagger(-\mathbf{p})\\
    \includegraphics[, valign = c]{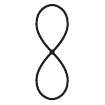}\quad &= -\frac{\lambda T}{6} \int_{\mathbf{q}_1, \mathbf{q}_2} \frac{K(\mathbf{q}_1)}{2\omega(\mathbf{q}_1)}\frac{K(\mathbf{q}_2)}{2\omega(\mathbf{q}_2)}
\end{align}
There are a few elements of these equations to be pointed out. The first two equations are non-vacuum states, and as such have a non-zero energy. For these diagrams, we have sent $\Lambda\to\infty$, which leaves behind an energy factor of $1/\sum_i \omega(\mathbf{p}_i)$. The final diagram, on the other hand, is a vacuum contribution and has a divergent factor of $T$ coming from the integral factor
\begin{equation}
    \int_{-T}^0 e^{-t(0)}dt = T.
\end{equation}
Now writing the normalization of the state as
\begin{equation}
    \mathcal{N}_T = 1 + \lambda \mathcal{N}^{(1)} +  \lambda^2 \mathcal{N}^{(2)} + \dots , 
\end{equation}
the ground state to first order in $\lambda$ is given by
\begin{equation}
    \ket{\Omega_{\lambda}} = \ket{\Omega_0} + \lambda\mathcal{N}^{(1)}\ket{\Omega_0} + \includegraphics[scale = .6, valign = c]{figs/firstorder4pt.pdf} + \includegraphics[scale = .6, valign = c]{figs/firstorder2pt.pdf} +\includegraphics[scale = .6, valign = c]{figs/firstorder0pt.pdf}
\end{equation}
and the normalization condition at this order reads
\begin{equation}\label{first-order-normalization}
    \lambda \mathcal{N}^{(1)} +\includegraphics[scale = .6, valign = c]{figs/firstorder0pt.pdf}= 0.
\end{equation}
As one expects, the normalization merely removes all first order contributions to the vacuum. Thus, the final expression for the ground state at first order is given by
\begin{equation}
    \ket{\Omega_{\lambda}} = \ket{\Omega_0} + \includegraphics[scale = .6, valign = c]{figs/firstorder4pt.pdf} + \includegraphics[scale = .6, valign = c]{figs/firstorder2pt.pdf}  
\end{equation}
which is indeed independent of the constant $T$, as claimed. 

\begin{figure}
    \centering
    \includegraphics[scale = 1]{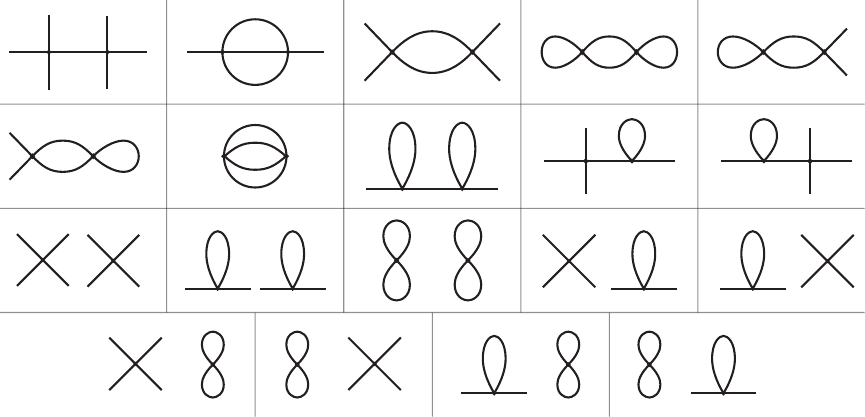}
    \caption{All diagrams contributing to the unnormalized ground state at second order in perturbation theory.}
    \label{fig:2ndorder-all}
\end{figure}

We now turn to the calculation of the state at second order. In Figure \ref{fig:2ndorder-all}, we have drawn all diagrams which contribute to the state before normalizing. Notice that at second order the non-commutativity of diagrams becomes relevant, and so diagrams which are not left-right reflection symmetric must be included in both orders. We will not explicitly compute each diagram, but we give special attention to the diagrams which have naive divergences in $T$. There are five such diagrams:
\begin{equation}
\begin{gathered}
\includegraphics[valign = c]{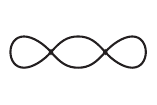}\hspace{1cm}
    \includegraphics[valign = c]{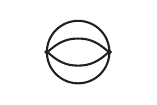}\hspace{1cm}
    \includegraphics[valign = c]{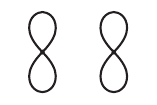}
    \\ \includegraphics[valign = c]{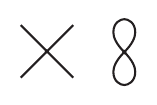}\hspace{1cm}
    \includegraphics[valign = c]{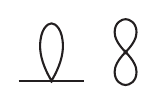}
\end{gathered}
\end{equation}
Computing the normalization at second order, we obtain the formula
\begin{align}\begin{split}
    &\lambda^2\left(2\mathcal{N}^{(2)} + ({\mathcal{N}^{(1)}})^2\right) + 4\lambda \mathcal{N}^{(1)}\includegraphics[scale = .5,valign =  c]{figs/firstorder0pt} + 2\left(\includegraphics[scale = .5,valign =  c]{figs/2-4.pdf}+\includegraphics[scale = .5,valign =  c,trim={.5cm 0 .5cm 0},clip]{figs/2-7.pdf}+\includegraphics[scale = .5,valign =  c,trim={.2cm 0 .2cm 0},clip]{figs/2-13.pdf}\right) \\&\hspace{2cm}+ \includegraphics[scale = .5,valign =  c]{figs/2-13.pdf}+\bra{\Omega_0}\left[\left(\includegraphics[scale = .4,valign =  c]{figs/firstorder2pt} \right)^\dagger\left(\includegraphics[scale = .4,valign =  c]{figs/firstorder2pt} \right) + \left(\includegraphics[scale = .4,valign =  c]{figs/firstorder4pt} \right)^\dagger\left(\includegraphics[scale = .4,valign =  c]{figs/firstorder4pt} \right)\right]\ket{\Omega_0} =0
    \end{split}
\end{align}
Using equation (\ref{first-order-normalization}), we see that
\begin{equation}
    \lambda^2 (\mathcal{N}^{(1)})^2 + 4 \lambda \mathcal{N}^{(1)}\includegraphics[scale = .5,valign =  c]{figs/2-13.pdf} + 3 \includegraphics[scale = .5,valign =  c]{figs/2-13.pdf} = 0
\end{equation}
and so all disconnected diagrams vanish from the condition for $\mathcal{N}^{(2)}$. The final result reads
\begin{equation}
    2\lambda^2 \mathcal{N}^{(2)} + 2\left(\includegraphics[scale = .5,valign =  c]{figs/2-4.pdf}+\includegraphics[scale = .5,valign =  c,trim={.5cm 0 .5cm 0},clip]{figs/2-7.pdf}\right) + \bra{\Omega_0}\left[\left(\includegraphics[scale = .4,valign =  c]{figs/firstorder2pt} \right)^\dagger\left(\includegraphics[scale = .4,valign =  c]{figs/firstorder2pt} \right) + \left(\includegraphics[scale = .4,valign =  c]{figs/firstorder4pt} \right)^\dagger\left(\includegraphics[scale = .4,valign =  c]{figs/firstorder4pt} \right)\right]\ket{\Omega_0}  = 0
\end{equation}

Plugging this result back into the expression for the state, one sees that all second order bubble diagrams vanish once again. This removes the first three divergent diagrams listed above. For the remaining two, we will now show that the following cancellations occur:
\begin{align}\label{cancel-1}
    \lambda \mathcal{N}^{(1)} \includegraphics[scale = .4, valign = c]{figs/firstorder2pt.pdf} + \includegraphics[scale = .5, valign = c]{figs/2-18}+ \includegraphics[scale = .5, valign = c]{figs/2-19} &= 0\\ \label{cancel-2}
    \lambda \mathcal{N}^{(1)} \includegraphics[scale = .4, valign = c]{figs/firstorder4pt.pdf} + \includegraphics[scale = .5, valign = c]{figs/2-16}+ \includegraphics[scale = .5, valign = c]{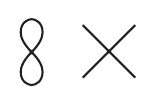} &= 0
\end{align}
This is our first true demonstration of the non-commutativity of our diagrammatic expansion. First, we calculate 
\begin{align}
    \includegraphics[valign = c]{figs/2-18.pdf} = \frac{9\lambda^2}{16}\int \left(\prod_{i=1}^{4}\frac{d^{d-1}\mathbf{p}_i}{(2\pi)^{d-1}}\frac{K(\mathbf{p}_i)}{2\omega(\mathbf{p}_i)}\right)\mathcal{E} a^\dagger(-\mathbf{p}_1)a^\dagger(\mathbf{p}_1)
\end{align}
Here $\mathcal{E}$ is the energy factor, which is given by
\begin{equation}
    \mathcal{E} = \int_{-T}^\infty dt_2 \int_{-T}^{t_2} dt_1 \left(e^{t_2(2\omega_m(\mathbf{p}_1))} e^{(0)t_1)}\right) \sim \frac{T}{2\omega(\mathbf{p}_1)} - \frac{1}{(2\omega(\mathbf{p}_1))^2}
\end{equation}
We obtain this expression because the first subdiagram is a zero-energy vacuum diagram, contributing to the divergence of the $t_1$ integral, while the second diagram is a finite energy two particle diagram. Now in contrast, the reflected diagram is given by
\begin{align}
    \includegraphics[valign = c]{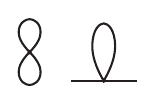} = \frac{9\lambda^2}{16}\int \left(\prod_{i=1}^{4}\frac{d^{d-1}\mathbf{p}_i}{(2\pi)^{d-1}}\frac{K(\mathbf{p}_i)}{2\omega_m(\mathbf{p}_i)}\right)\mathcal{E}' b^\dagger(-\mathbf{p}_1)b^\dagger(\mathbf{p}_1)
\end{align}
Since the leg structure is the same, the propagators and operator insertions are the same, but the energy factor is a different value, $\mathcal{E}'$, which is given by
\begin{equation}
    \mathcal{E}' = \int_{-T}^\infty dt_2 \int_{-T}^{t_2} dt_1 \left(e^{2t_2\omega_m(\mathbf{p}_1)} e^{2t_1 \omega_m(\mathbf{p}_1)}\right) = \frac{1}{(2\omega_m(\mathbf{p_1}))^2}
\end{equation}
We see that the finite contributions of the two diagrams cancel, leaving behind only a divergent contribution proportional to $T$. It is not hard to then check that the first term in (\ref{cancel-1}) cancels the remaining divergent term, and so we see that the claimed cancellation occurs. The calculation for (\ref{cancel-2}) follows analogously. 

Now we are prepared to write down the ground state to second order in perturbation theory, which we conclude this section with. Our full expression reads
\begin{align}
    \begin{split}\label{second-order-full}
        \ket{\Omega_{\lambda}} &= \left(1 - \frac{1}{2}\kappa\right)\ket{\Omega_0}   \\&\hspace{1cm}+ \Bigg(\includegraphics[scale = .6, valign = c]{figs/firstorder4pt.pdf} + \includegraphics[scale = .6, valign = c]{figs/firstorder2pt.pdf}   + \includegraphics[scale = .6, valign = c]{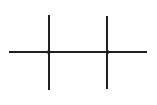} + \includegraphics[scale = .6, valign = c]{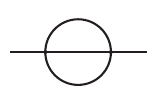} + \includegraphics[scale = .6, valign = c]{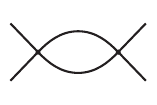} \\&\hspace{1cm}+ \includegraphics[scale = .6, valign = c]{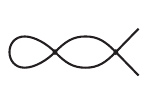} + \includegraphics[scale = .6, valign = c]{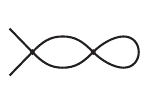} + \includegraphics[scale = .6, valign = c]{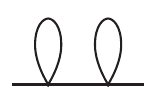} + \includegraphics[scale = .6, valign = c]{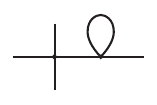}+ \includegraphics[scale = .6, valign = c]{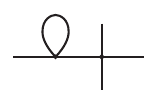} \\ &\hspace{1cm}+ \includegraphics[scale = .6, valign = c]{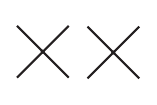} + \includegraphics[scale = .6, valign = c]{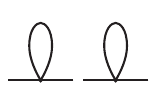} + \includegraphics[scale = .6, valign = c]{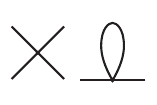} + \includegraphics[scale = .6, valign = c]{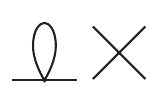} \Bigg)\ket{\Omega_0}
    \end{split}
\end{align}
where
\begin{equation}
    \kappa = \bra{\Omega_0}\left[\left(\includegraphics[scale = .4,valign =  c]{figs/firstorder2pt} \right)^\dagger\left(\includegraphics[scale = .4,valign =  c]{figs/firstorder2pt} \right) + \left(\includegraphics[scale = .4,valign =  c]{figs/firstorder4pt} \right)^\dagger\left(\includegraphics[scale = .4,valign =  c]{figs/firstorder4pt} \right)  \right]\ket{\Omega_0}
\end{equation}

\subsection{Proof of Equation (\ref{defineoperatorS})}
In the text, we made the claim
\begin{equation}
    -\Lambda\partial_\Lambda \rho = i[\hat{K}, \rho] - \mathcal{S}(\rho)
\end{equation}
Here we demonstrate this identity and define the quantity $\mathcal{S}(\rho)$ in the course of the calculation. The partial derivative is defined by
\begin{equation}
    -\Lambda \partial_\Lambda \rho\Big\vert_{\Lambda_0} \equiv -\Lambda \frac{\partial}{\partial \Lambda}\left(\int\mathcal{D}\phi e^{-S_0[\Lambda] - S_I[\Lambda_0] - S_{\textrm{bdry}}[\Lambda]}\right)\Bigg\vert_{\Lambda_0}.
\end{equation}
We expand the integrand as
\begin{equation}
     \int\mathcal{D}\phi e^{-S_0[\Lambda] - S_{\textrm{bdry}}[\Lambda]}\left( 1 + S_I[\Lambda_0] + \cdots\right).
\end{equation}
This is simply the perturbative expansion of the state at scale $\Lambda$, since $S_I[\Lambda_0]$ has no explicit $\Lambda$ dependence. After expressing the state as a sum of diagrams, each term will have $\Lambda$ dependence both through the unperturbed state $\ket{\Omega_0}$ as well as through the diagrams, which depend explicitly on functions of $\Lambda$. For an arbitrary term in the expansion, we therefore identify three sources of $\Lambda$ dependence: (a) the unperturbed state $\ket{\Omega_0}$, (b) the creation/annihilation operators $a(\mathbf{p};\Lambda), a^\dagger(\mathbf{p};\Lambda)$ coming from external legs and (c) the scalar coefficients of each propagator. First, for (a), it can be shown that
\begin{align}
    -\Lambda \partial_\Lambda \ket{\Omega_0}\bra{\Omega_0} &= i[\hat{K},\ket{\Omega_0}\bra{\Omega_0}].
\end{align}
Next we must account for the operator valued diagrams encompassing (b) and (c). Similarly to the state, we have
\begin{equation}
    -\Lambda\partial_\Lambda a^\dagger(\mathbf{p};\Lambda) = i[\hat{K}, a^\dagger(\mathbf{p};\Lambda)],
\end{equation}
and so the operator contribution to the derivative is just a commutator. As for the coefficients, notice that every propagator contributes a factor of either $K$ for internal legs or $\sqrt{K}$ for external legs. Since the full diagram is just the product over the contributions of each propagator, we can treat the propagators one by one and sum over the result. Using the fact
\begin{equation}
    -\Lambda \partial_\Lambda K^\alpha = -\alpha \Delta_\Sigma K^\alpha,
\end{equation}
we can account for the coefficient derivatives by multiplying with a factor of $\Delta_\Sigma$ for internal legs and $\frac{1}{2}\Delta_\Sigma$ for external legs. Recall the modified propagator 
\begin{equation}\label{prop-deriv}
    \includegraphics[valign = c, width = 2.5cm]{figs/prop-deriv.pdf} = \begin{cases}
        \Delta_\Sigma(\mathbf{p})\frac{K(\mathbf{p}^2/\Lambda^2)}{2\omega(\mathbf{p})} & \textrm{internal}\vspace{.2cm}\\
        \frac{1}{2}\Delta_\Sigma(\mathbf{p})\sqrt{\frac{K(\mathbf{p}^2/\Lambda^2)}{2\omega(\mathbf{p})}} a^\dagger(\mathbf{p};\Lambda) & \textrm{external}
    \end{cases}
\end{equation}
which accounts for precisely the logarithmic derivative of the diagram coefficient. Putting (b) and (c) together, the partial derivative of any diagram is given by a commutator and a replacement of propagators with modified propagators. For example, for the four-particle diagram we have
\begin{equation}
    -\Lambda \partial_\Lambda\, \includegraphics[scale = .7, valign = c]{figs/firstorder4pt.pdf} = i\left[\hat{K}, \includegraphics[scale = .7, valign = c]{figs/firstorder4pt.pdf}\right] - \left(\includegraphics[scale = .7, valign = c]{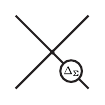}+\includegraphics[scale = .7, valign = c]{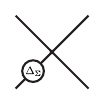}+\includegraphics[scale = .7, valign = c]{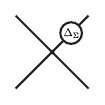}+\includegraphics[scale = .7, valign = c]{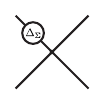}\right).
\end{equation}
More generally, we now define the linear function $\mathcal{S}$ as the function which computes the partial derivative of the diagram coefficients. Diagrammatically, this is given by the procedure of replacing propagators with modified propagators one by one in a diagram, and summing the result, as we saw in the example above. Formally, $\mathcal{S}$ has the properties:
\begin{gather}
    \mathcal{S}\left(D \ket{\Omega_0}\bra{\Omega_0}\right) = \mathcal{S}\left(D\right) \ket{\Omega_0}\bra{\Omega_0},\\
    \mathcal{S}\left(D_1\ket{\Omega_0}\bra{\Omega_0} D_2\right) = \mathcal{S}(D_1)\ket{\Omega_0}\bra{\Omega_0}D_2 +D_1\ket{\Omega_0}\bra{\Omega_0}\mathcal{S}(D_2).
\end{gather}
where $D, D_1, D_2$ are diagrams. The first property states that $\mathcal{S}$ ``ignores'' the unperturbed ground state, while the second is simply the Leibniz rule for diagrams on opposite sides of the density matrix.

Now we can put everything together. Since the $\Lambda$ derivative of the unperturbed state and the creation/annihilation operators are both generated by $\hat{K}$, the final expression for the partial derivative of the state is
\begin{equation}
    -\Lambda \partial_\Lambda \rho = i[\hat{K}, \rho] - \mathcal{S}(\rho).
\end{equation}
and the total derivative of the state under the flow of the Polchinski equation is
\begin{equation}
    -\Lambda \frac{d}{d\Lambda}\rho = \delta_{\scriptscriptstyle{S_I}}\rho + i[\hat{K}, \rho] - \mathcal{S}(\rho).
\end{equation}

\subsection{First order calculation}
In the main discussion, we considered the equality

\begin{equation}
    \mathcal{D}(\rho) = \delta_{\scriptscriptstyle{S_I}}\rho - \mathcal{S}(\rho)
\end{equation}
at second order in $\lambda$, where non-linear contributions of the Polchisnki equation became important. In this subsection, we will check this equation at first order in $\lambda$. For what follows, all equalities should be read as up to terms of order $\lambda^2$. First, consider the term $\delta_{\scriptscriptstyle{S_I}}\rho$. At first order in $\lambda$, the non-local term does not contribute and we simply have
\begin{equation}
    -\Lambda\frac{d}{d\Lambda}S_I\Big\vert_{\Lambda_0} = \frac{\lambda}{4} \int d^dx \phi(x)^2 G'(0) 
\end{equation}
Because this expression is local in Euclidean time, the purity of the state is preserved at this order in perturbation theory, and we may break its contribution into two terms
\begin{equation}
    \int d^{d}x\phi(x)^2 = \int_{-\infty}^0 dt\int  d^{d-1}\mathbf{x}\phi(x)^2 +  \int_{0}^\infty dt\int  d^{d-1}\mathbf{x}\phi(x)^2.
\end{equation}
The first term contributes to the ``ket'' and the second contributes to the ``bra'' of the pure ground state. Consider just the first term for the moment, as the second is just the Hermitian conjugate. Writing $H$ as the regulated free Hamiltonian associated to $S_0$, we have
\begin{equation}
    \int \mathcal{D}\phi \left[\int_{-\infty}^0 dt \int d^{d-x}\mathbf{x} \phi(x)^2 e^{-S_0- S_{\textrm{bdry}}}\right] = \int_{-\infty}^0 dt \left(e^{H t} \hat{\varphi}(\mathbf{x})^2\right) \ket{\Omega_0}
\end{equation}
Going to momentum space for the spatial arguments and expanding $\hat{\varphi}(\mathbf{x})$ in the creation/annihilation operators (\ref{adef}), we arrive at
\begin{align}
    \int_\mathbf{x}\int_{-\infty}^0 dt \left(e^{H t} \hat{\varphi}(\mathbf{x})^2\right) \ket{\Omega_0} &= \int_{\mathbf{p}}\left(\int_{-\infty}^0 e^{2\omega(\mathbf{p}) t}dt \right) \frac{K(\mathbf{p}^2/\Lambda^2)}{2\omega(\mathbf{p})}a^\dagger(\mathbf{p})a^\dagger(-\mathbf{p})\ket{\Omega_0} + \cdots\\
    &=  \int_{\mathbf{p}}\frac{K(\mathbf{p}^2/\Lambda^2)}{4\omega(\mathbf{p})^2}a^\dagger(\mathbf{p})a^\dagger(-\mathbf{p})\ket{\Omega_0} +\cdots 
\end{align}
The terms denoted by $\cdots$ are terms proportional to the unperturbed ground state, which at first order are always removed upon normalizing. So, we have that
\begin{equation}
    \delta_{\scriptscriptstyle{S_I}}\rho = -\frac{\lambda G'(0)}{4} \int_{\mathbf{p}}\frac{K(\mathbf{p}^2/\Lambda^2)}{4\omega(\mathbf{p})^2}a^\dagger(\mathbf{p})a^\dagger(-\mathbf{p})\ket{\Omega_0}\bra{\Omega_0} + h.c.
\end{equation}
We can compute the coefficient more explicitly using the definition of $G'$, which yields
\begin{equation}
    G'(0) = \int_{q} \frac{\Lambda \frac{d}{d\Lambda}K(\mathbf{q}^2/\Lambda^2)}{q^2} = \int_{\mathbf{q}} \frac{\Lambda \frac{d}{d\Lambda}K(\mathbf{q}^2/\Lambda^2)}{2\omega(\mathbf{q})}.
\end{equation}
Plugging this back in, we obtain the final expression:
\begin{align}
    \delta_{\scriptscriptstyle{S_I}}\rho &= -\frac{\lambda}{4}\int_{\mathbf{p}, \mathbf{q}} \Delta_\Sigma(\mathbf{p})\frac{K(\mathbf{p}^2/\Lambda^2)K(\mathbf{q}^2/\Lambda^2)}{8\omega(\mathbf{p})^2\omega(\mathbf{q})} a^\dagger(\mathbf{p})a^\dagger(-\mathbf{p}) \ket{\Omega_0}\bra{\Omega_0} + h.c.  + \mathcal{O}(\lambda^2)
\end{align}
Finally, comparing with the expression for the diagram in (\ref{first-order-2pt-expression}), we see that this result can be summarized as 
\begin{equation}
    \delta_{\scriptscriptstyle{S_I}}\rho = \left[\includegraphics[scale = .7, valign = c, trim = {0, 0, 0, .07cm}]{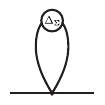} \right]\ket{\Omega_0}\bra{\Omega_0} + \ket{\Omega_0}\bra{\Omega_0}\left[\includegraphics[scale = .7, valign = c, trim = {0, 0, 0, .07cm}]{figs/2pt-modified-internal.pdf} \right].
\end{equation}
Now let us turn our attention to the operator $\mathcal{D}$. Acting on the initial state, we have three terms to consider:
\begin{align}
\begin{split}
    \mathcal{D}(\ket{\Omega_\lambda}\bra{\Omega_\lambda}) &= \int_{\mathbf{p}}\Delta_\Sigma(\mathbf{p}) \Big(a(\mathbf{p})\ket{\Omega_\lambda}\bra{\Omega_\lambda} a^\dagger(\mathbf{p})\\&\hspace{2cm}- \frac{1}{2}a^\dagger(\mathbf{p})a(\mathbf{p})\ket{\Omega_\lambda}\bra{\Omega_\lambda}  - \frac{1}{2}\ket{\Omega_\lambda}\bra{\Omega_\lambda} a^\dagger(\mathbf{p})a(\mathbf{p})\Big)
\end{split}
\end{align}
Schematically, the perturbed ground state takes the form $\ket{\Omega_\lambda} = \ket{\Omega_0} + \lambda \ket{\Omega_1} + \cdots $ and the density operator is
\begin{equation}
    \ket{\Omega_\lambda}\bra{\Omega_\lambda} = \ket{\Omega_0}\bra{\Omega_0} + \lambda\left(\ket{\Omega_1}\bra{\Omega_0} + \ket{\Omega_0}\bra{\Omega_1}\right)
\end{equation}
and so at this order in perturbation theory we have
\begin{equation}
    a(\mathbf{p})\ket{\Omega_\lambda}\bra{\Omega_\lambda}a^\dagger(\mathbf{p}) = 0.
\end{equation}
Now consider the term
\begin{equation}
    -\int_{\mathbf{p}}\frac{\Delta_\Sigma(\mathbf{p})}{2}a^\dagger(\mathbf{p})a(\mathbf{p}) \ket{\Omega_\lambda}\bra{\Omega_\lambda}
\end{equation}
Each diagram in the first order correction of the state has external legs, corresponding to a product of creation operators. For an arbitrary product of creation operators acting on the vacuum, we have
\begin{align}
    -\int_{\mathbf{p}}\frac{\Delta_\Sigma(\mathbf{p})}{2}a^\dagger(\mathbf{p}) a(\mathbf{p}) \left(\prod_i a^\dagger(\mathbf{q}_i)\right)\ket{\Omega_0} &= -\frac{1}{2}\int_\mathbf{p} \Delta_\Sigma(\mathbf{p}) a^\dagger(\mathbf{p})\sum_i \left(\prod_{j\neq i} a^\dagger(\mathbf{q}_j) \right)\delta^{(d-1)}(\mathbf{p} - \mathbf{q}_i)\ket{\Omega_0}\\
    &= -\frac{1}{2}\sum_i \Delta_\Sigma(\mathbf{q}_i) \left(\prod_{j}a^\dagger(\mathbf{q}_j)\right)\ket{\Omega_0}
\end{align}
For each term in the sum above, we are simply multiplying the diagram by $-\frac{1}{2}\Delta_\Sigma(\mathbf{q}_i)$ for a given external leg, which is precisely the result of replacing the external propagator with a modified propagator (with an overall minus sign). This applies to every diagram to the left of the unperturbed state. But, we can see that the remaining term in $\mathcal{D}$ performs precisely the same procedure to every diagram on the right of the unperturbed state. Motivated by this, define the splitting
\begin{equation}
    \mathcal{S} = \mathcal{S}_e + \mathcal{S}_i
\end{equation}
where $\mathcal{S}_{e/i}$ are defined similarly to $\mathcal{S}$, but where $\mathcal{S}_e$ applies the procedure of modifying diagrams only to external legs, and $\mathcal{S}_i$ to internal legs. If a diagram has no internal legs, then $\mathcal{S}_i$ evaluates to zero, and likewise for $\mathcal{S}_e$ and diagrams with no external legs. Then, what we have just shown is that
\begin{equation}
    \mathcal{D}(\ket{\Omega_\lambda}\bra{\Omega_\lambda})\Big\vert_{\Lambda_0} = -\mathcal{S}_e(\ket{\Omega_\lambda}\bra{\Omega_\lambda}) .
\end{equation}
Recall the first order state
\begin{equation}
    \ket{\Omega_\lambda}\bra{\Omega_\lambda}= \ket{\Omega_0}\bra{\Omega_0} +\left[ \left(\includegraphics[scale = .6, valign = c]{figs/firstorder4pt.pdf} + \includegraphics[scale = .6, valign = c]{figs/firstorder2pt.pdf} \right) \ket{\Omega_0}\bra{\Omega_0} + h.c.\right]
\end{equation}
The only diagram with an internal leg is the loop diagram. We therefore conclude that, 
\begin{align}
    \mathcal{S}(\ket{\Omega_\lambda}\bra{\Omega_\lambda}) &= \mathcal{S}_e(\ket{\Omega_\lambda}\bra{\Omega_\lambda}) +\mathcal{S}_i(\ket{\Omega_\lambda}\bra{\Omega_\lambda}) \\
    &=-\mathcal{D}(\ket{\Omega_\lambda}\bra{\Omega_\lambda})\Big\vert_{\Lambda_0}+ \left( \left[\includegraphics[scale = .6, valign = c, trim = {0, 0, 0, .07cm}]{figs/2pt-modified-internal.pdf} \right]\ket{\Omega_0}\bra{\Omega_0} + \ket{\Omega_0}\bra{\Omega_0}\left[\includegraphics[scale = .6, valign = c, trim = {0, 0, 0, .07cm}]{figs/2pt-modified-internal.pdf} \right]\right)\\
    &= \left[-\mathcal{D}(\ket{\Omega_\lambda}\bra{\Omega_\lambda}) + \delta_{\scriptscriptstyle{S_I}}\rho\right] \Big\vert_{\Lambda_0}
\end{align}
which is the desired identity. We can now clarify in what sense the first order calculation is ``trivial.'' Notice that, in solving the equation for $\mathcal{D}(\rho)$, all the contributions of $\delta_{\scriptscriptstyle{S_I}}\rho$ are cancelled by terms in $\mathcal{S}(\rho)$. So, at this order, $\mathcal{D}$ does not ``see'' the Polchinski equation at all. In general, at higher orders of $\lambda$, $\mathcal{D}$ can only see the \textit{non-local} contributions of the Polchinski equation.  

\section{Derivation of dilation action}\label{app:dilation}
In the second step of RG, we performed the Weyl scaling
\begin{eqnarray}
    \eta_{\mu\nu}\to e^{2s}\eta_{\mu\nu}, \qquad \phi(x)\to e^{-s(d-2)/2}\phi(x).
\end{eqnarray}
In this Appendix, we will derive how this transformation results in the action of the operator $\hat{L}$. First, note that effect of this transformation in the bulk of the path integral does not contribute any non-trivial boundary terms, so these terms can be entirely absorbed into the rescaled flow equation for $S_I$. So, we only need to address the change in the boundary term $S_{\textrm{bdry}}$. 

Under the rescalings above, $S_{\textrm{bdry}}$ takes the form
\begin{equation}
    -\frac{1}{2} \int_{\mathbf{x}}\varphi^-(\mathbf{x}) K^{-1}(-e^{-2s}\nabla^2/\Lambda^2) \partial_t \phi(\mathbf{x})^-
\end{equation}
This is only the $t = -\epsilon$ surface term, but the $t  = \epsilon$ surface term is completely symmetric. In the absence of the regulator $K$, the transformations above are a symmetry of the boundary terms, as one expects for the fixed point theory. Because of the regulator, we must now remove the $s$ dependence from its argument. By changing variables in the integral, we can rewrite this as
\begin{equation}
    -\frac{e^{-(d-1)s}}{2} \int_{\mathbf{x}} \varphi^-(e^{-s}\mathbf{x}) K^{-1}(-\nabla^2/\Lambda^2) \partial_t \phi(e^{-s}\mathbf{x})^-
\end{equation}
We must also take $t \to e^{-s}t$ to ensure consistency with the bulk terms, which results in additional factor 
\begin{equation}
    -\frac{e^{-(d-2)s}}{2}\int_{\mathbf{x}} \varphi^-(e^{-s}\mathbf{x}) K^{-1}(-\nabla^2/\Lambda^2) \partial_{t} \phi(e^{-s}\mathbf{x})^-
\end{equation}
For an infinitesimal $s = \epsilon$, this change in $S_{\textrm{bdry}}$ can be accounted for by the linear field redefinition
\begin{equation}
    \varphi(\mathbf{x}) \to \varphi(\mathbf{x}) -\epsilon\left[(\mathbf{x}\cdot \nabla)\varphi(\mathbf{x}) +\Delta_\phi\varphi(\mathbf{x})\right]
\end{equation}

With these observations, let us consider the change in the density matrix for an arbitrary linear field redefinition. In the configuration basis
 \begin{equation}
    \mel{\varphi^-}{\rho}{\varphi^+}=\rho(\varphi^-, \varphi^+ ),
 \end{equation}
we perform the transformation
 \begin{equation}
     \rho(\varphi^-, \varphi^+) \to  \rho'(\varphi^+,\varphi^-)=\mathcal{N}\rho(\varphi^- + \epsilon \eta^-, \varphi^+ + \epsilon \eta^+).
\end{equation}
 where $\mathcal{N}$ is a field-independent constant that restores the normalization of the density matrix and $\eta^\pm = \eta[\varphi^\pm]$ is a linear function of the boundary configurations. Expanding to linear order in the perturbation yields
 \begin{align}
     &\rho(\varphi^- + \epsilon \eta^-, \varphi^+ + \epsilon \eta^+) \\
     &\hspace{1cm}\sim \rho(\varphi^-, \varphi^+) +\epsilon \left[\eta^-\cdot \frac{\delta \rho}{\delta \varphi^-}+\eta^+\cdot \frac{\delta \rho}{\delta \varphi^+}\right]
 \end{align}
The normalization parameter $\mathcal{N}$ can be fixed using
 \begin{align}
     1 = \tr(\rho) &= \mathcal{N}\int \mathcal{D}\varphi \rho(\varphi + \epsilon \eta, \varphi + \epsilon \eta)\\
     &=\mathcal{N} \int \mathcal{D}\varphi' \left|\frac{\delta \varphi'}{\delta\varphi}\right|^{-1}\rho(\varphi', \varphi')\\
     &\sim \mathcal{N}\left(1 - \epsilon \tr(\frac{\delta \eta}{\delta \varphi})\right)
 \end{align}
 where in going to the second line we have made the field redefinition $\varphi' = \varphi + \eta$ and in going to the final line we used the normalization of the density operator $\rho$ along with the expansion for the determinant
 \begin{equation}
     |1 + \epsilon A| \sim 1 + \epsilon \tr(A).
 \end{equation}
Now specializing to our case, we have
\begin{equation}
    \eta[\varphi] = -(\mathbf{x}\cdot \nabla)\varphi(\mathbf{x}) - \Delta_\phi\varphi(\mathbf{x}),
\end{equation}
with $\Delta_\phi$ the canonical, or engineering, dimension of $\phi$. Applying this to the above expressions and using the functional representations of $\hat{\varphi}$ and $\hat{\pi}$, we deduce that the density matrix transforms as 
 \begin{align}
     \rho' &= \rho - \epsilon \int d^{d-1}\mathbf{x}\Big[i((\mathbf{x}\cdot \nabla +\Delta_\phi)\hat{\varphi}(\mathbf{x}) )\hat{\pi}(\mathbf{x})\rho - i\rho \hat{\pi}(\mathbf{x})(\mathbf{x}\cdot \nabla + \Delta_\phi) \hat{\varphi}(\mathbf{x})\\&\hspace{4cm} +  (\mathbf{x}\cdot \nabla + \Delta_\phi) \delta^{(d-1)}(0)\Big]\\
     &= \rho -\epsilon \int d^{d-1}\mathbf{x}\left[i((\mathbf{x}\cdot \nabla  +\Delta_\phi)\hat{\varphi}(\mathbf{x}) )\hat{\pi}(\mathbf{x})\rho - i\rho((\mathbf{x}\cdot \nabla+\Delta_\phi) \hat{\varphi}(\mathbf{x})) \hat{\pi}(\mathbf{x}) \right]
 \end{align}
 Comparing with the form of $\hat{L}$, we see that this is indeed the transformation
 \begin{equation}
     \rho' = \rho + i\epsilon[\hat{L}, \rho].
 \end{equation}

\section{Normal ordered correlators}\label{app:normal}

In Section \ref{sec:integrate}, we explicitly integrated the dual map to the ERG flow $\mathcal{L}^*_\Lambda$. In this appendix, we observe a property of the generator of the dual map that allows us to write simple flow equations for normal ordered operators. 

Define a normal ordered product of $n$ creation and $m$ annihilation operators
\begin{equation}
    \hat{N}_{n,m}(\mathbf{p}_1,\cdots, \mathbf{p}_{n+m};\Lambda) = a^\dagger(\mathbf{p}_1;\Lambda)\cdots a^\dagger(\mathbf{p}_n;\Lambda) a(\mathbf{p}_{n+1};\Lambda)\cdots a(\mathbf{p}_{n+m};\Lambda)
\end{equation}
By solving for the flow of expectation values of $\hat{N}_{n,m}$ we can obtain any correlator by linear combination. Thus, it is sufficient to consider the set of correlation functions
\begin{equation}
    \Gamma^{(n, m)}(\{\mathbf{p_i}\};\Lambda) = \tr(\rho(\Lambda) \hat{N}_{n,m}(\{\mathbf{p_i}\};\Lambda))
\end{equation}
where $\rho(\Lambda)$ is an arbitary density matrix. Taking the $\Lambda$ derivative, we have
\begin{equation}
    -\Lambda \frac{d}{d\Lambda}\Gamma^{(n, m)}(\Lambda) = \tr(\mathcal{L}_\Lambda (\rho(\Lambda)) \hat{N}_{n,m}(\Lambda)) -\Lambda \tr(\rho(\Lambda)\partial_\Lambda \hat{N}_{n,m}(\Lambda))
\end{equation}
Once again using the fact that $-\Lambda\partial_\Lambda g(\Lambda) = \frac{1}{2}\Delta_\Sigma g(\Lambda)$, one can show that
\begin{equation}
    -\Lambda\partial_\Lambda a(\mathbf{p};\Lambda) = i[\hat{K}, a(\mathbf{p};\Lambda)]
\end{equation}
from which it follows that $-\Lambda\partial_\Lambda \hat{N}_{n,m} = i[\hat{K}, \hat{N}_{n,m}]$. So, we have
\begin{align}
    -\Lambda \frac{d}{d\Lambda}\Gamma^{(n, m)}(\Lambda) &= i\tr([\hat{K}, \rho]\hat{N}_{n,m}) + \tr(\mathcal{D}(\rho)\hat{N}_{n,m}) + i\tr(\rho [\hat{K}, \hat{N}_{n,m}])\\
    &= \tr(\rho \mathcal{D}^*(\hat{N}_{n,m}))
\end{align}
The key observation is that $\hat{N}_{n, m}$ are eigenoperators of the superoperator $\mathcal{D}^*$, via the eigenvalue equation
\begin{equation}
    \mathcal{D}^*(\hat{N}_{n,m}) = \frac{1}{2}\left(\sum_{i=1}^{n+m}\Delta_\Sigma(\mathbf{p}_i)\right) \hat{N}_{n,m}
\end{equation}
where the $\mathbf{p}_i$ are the values of momenta appearing in the product. So, the correlator therefore satisfies the simple differential equation
\begin{equation}
    -\Lambda \frac{d}{d\Lambda}\Gamma^{(n, m)}(\Lambda) = \frac{1}{2}\left(\sum_{i=1}^{n+m}\Delta_\Sigma(\mathbf{p}_i)\right) \Gamma^{(n, m)}(\Lambda).
\end{equation}
Integrating from initial scale $\Lambda_0$ to final scale $\Lambda$, we have
\begin{equation}
    \Gamma^{(n, m)}(\Lambda) = \left(\prod_{i=1}^{n+m} \sqrt{\frac{K(\mathbf{p}_i^2/\Lambda^2)}{K(\mathbf{p}_i^2/\Lambda_0^2)}}\right) \Gamma^{(n, m)}(\Lambda_0)
\end{equation}
Keeping in mind the approximation for $\Lambda < \Lambda_0$ that
\begin{equation}
    \frac{K(\mathbf{p}^2/\Lambda^2)}{K(\mathbf{p}^2/\Lambda_0^2)} \approx K(\mathbf{p}^2/\Lambda^2), 
\end{equation}
we can simplify even further to the approximate solution
\begin{equation}
    \Gamma^{(n, m)}(\Lambda) \approx \left(\prod_{i=1}^{n+m} \sqrt{K(\mathbf{p}_i^2/\Lambda^2)}\right) \Gamma^{(n, m)}(\Lambda_0)
\end{equation}
Thus, we have a simple interpretation for the solution which is consistent with the Callan Symanzik equations: the normal ordered correlators are unaffected when $\abs{\mathbf{p}_i} <\Lambda$ for all the momenta in the correlator. The value of the correlator quickly drops to zero as soon as the scale $\Lambda$ drops below any of the momenta $\mathbf{p}_i$. For example, the expectation value of the number operator $N(\mathbf{p}) = a^\dagger(\mathbf{p})a(\mathbf{p})$ for modes at momentum $\mathbf{p}$ is 
given by $\expval{N(\mathbf{p})}_\Lambda = \Gamma^{(1, 1)}(\mathbf{p}, \mathbf{p};\Lambda)$, and we have
\begin{equation}
    \expval{N(\mathbf{p})}_\Lambda \approx K(\mathbf{p}^2/\Lambda^2) \expval{N(\mathbf{p})}_{\Lambda_0}
\end{equation}
The flow equation therefore removes the occupation of modes above the cutoff for any state.

\bibliographystyle{uiuchept}
\bibliography{erg_lindbladian}
\end{document}